\documentclass[a4paper, 12pt]{article}

\newcommand{\papertitle}{Cognitive Noise and Altruistic Preferences}

\usepackage{eurosym}
\newcommand{\mycaption}[2][]{\caption[#1]{#1 #2}}

\usepackage{ifxetex}  

\usepackage{caption}
\DeclareCaptionLabelFormat{andtable}{#1~#2  \&  \tablename~\thetable}
\DeclareCaptionLabelFormat{andfigure}{#1~#2  \&  \figurename~\thefigure}
\DeclareCaptionLabelFormat{threefigures}{#1~#2}

\usepackage{ifthen}

\usepackage[utf8]{inputenc}  

\ifxetex
	\usepackage[euenc]{fontspec}
\else
	\usepackage[LGR, T1]{fontenc}  
\fi

\usepackage[ngerman, american, USenglish]{babel}  
\usepackage[ngerman, USenglish]{isodate}

\usepackage[babel, german=quotes]{csquotes}  

\usepackage{calc}  
\usepackage{fp}  

\usepackage{etoolbox}  
\usepackage{xpatch}  
\usepackage{letltxmacro}
\usepackage{xparse}

\usepackage{geometry}  
\usepackage{ragged2e}

\usepackage{import}	 

\usepackage{placeins}  

\usepackage{amsmath}
\MakeRobust{\eqref}
\renewcommand{\eqref}[1]{(\ref{#1})}  
\usepackage{amsthm}	

\usepackage{mathtools}
\newcommand{\succeqq}{%
  \mathrel{%
    \vcenter{\offinterlineskip
      \ialign{##\cr$\succ$\cr\noalign{\kern 1pt}$=$\cr}%
    }%
  }%
}

\usepackage[
	colorlinks=true,
	linkcolor=black,
	citecolor=black,
	urlcolor=black,
	filecolor=black,
	bookmarks=true,
	bookmarksnumbered=true,
	bookmarksopenlevel=2,
	pdfstartview=Fit,
	pdfpagelayout=SinglePage,
	plainpages=false,
	pdfpagelabels=true
]{hyperref}
\newcommand{\email}[1]{\href{mailto:#1}{\nolinkurl{#1}}}
\addto\extrasUSenglish{%
}
\addto\extrasamerican{%
}
\addto\extrasngerman{%
}

\usepackage{verbatim}

\usepackage{graphicx}
\usepackage{epstopdf}  

\usepackage[table]{xcolor}
\definecolor{UBonnBlue}   {RGB}{0007,0082,0154}
\definecolor{darkblue}    {rgb}{0.00,0.20,0.40}
\definecolor{darkred}     {rgb}{0.80,0.00,0.00}
\colorlet   {darkred25}   {darkred!25!white}
\definecolor{customgreen} {rgb}{0.15,0.55,0.00}
\definecolor{custompurple}{rgb}{0.15,0.00,0.75}

\usepackage{pgf}
\usepackage{pgfplots}

\usepackage{tikz}
\usetikzlibrary{mindmap, trees, patterns, shapes, arrows, positioning, backgrounds, fit}

\usepackage{pdflscape}

\usepackage{afterpage}

\usepackage{soulutf8}
\usepackage{tabularx}
\usepackage{pdfpages}

\definecolor{lightor}{rgb}{1.0,0.93,0.8}

\usepackage{dsfont}

\usepackage[full]{textcomp}  
\usepackage{xfrac}	

\usepackage{soul}  

\usepackage{relsize}

\ifxetex
	\renewcommand{\caps}[1]{\textscale{0.96}{\addfontfeature{LetterSpace=5}\MakeUppercase{#1}}}
\else
	\renewcommand{\caps}[1]{\textscale{0.96}{\textls[35]{\MakeUppercase{#1}}}}
\fi

\usepackage{printlen}  

\usepackage[math]{blindtext}
\makeatletter
\def\blindtext@american{}
\renewcommand{\blindmathpaper}{%
	\blindtext
	\blindtext@formula\par
	\blindtext
	\blindtext@formula
	\blindtext
	\blindtext@formula\par
	\blindtext
	\blindtext@formula
	\blindtext
	\blindtext@formula\par
	\blindtext\relax%
}
\makeatother
\LetLtxMacro{\blindtextblindtext}{\blindtext}
\LetLtxMacro{\blindtextBlindlist}{\Blindlist}
\LetLtxMacro{\blindtextBlindtext}{\Blindtext}
\RenewDocumentCommand{\blindtext}{O{\value{blindtext}}}{%
	\begingroup%
	\iflanguage{USenglish}{\selectlanguage{american}}{}%
	\blindtextblindtext[#1]%
	\endgroup%
}
\RenewDocumentCommand{\Blindtext}{O{\value{blindtext}} O{\value{Blindtext}}}{%
	\begingroup%
	\iflanguage{USenglish}{\selectlanguage{american}}{}%
	\blindtextBlindtext[#1][#2]%
	\endgroup%
}
\RenewDocumentCommand{\Blindlist}{m O{\value{blindlist}}}{%
	\begingroup%
	\iflanguage{USenglish}{\selectlanguage{american}}{}%
	\blindtextBlindlist{#1}[#2]%
	\endgroup%
}

\makeatletter
\newcommand{\showfont}{{%
	\color{magenta}
	\textit{Encoding:} \f@encoding{},
	\textit{family:}   \f@family{},
	\textit{series:}   \f@series{},
	\textit{shape:}    \f@shape{},
	\textit{size:}     \f@size{}.
}}
\newcommand{\showfamily}{\f@family{}}
\makeatother

\makeatletter
\newcommand*{\checkgreekletters}{%
	\@for\@tempa:=%
	alpha,beta,gamma,delta,epsilon,varepsilon,zeta,eta,theta,vartheta,iota,kappa,lambda,mu,nu,xi,%
	omicron,pi,varpi,rho,varrho,sigma,varsigma,tau,upsilon,phi,varphi,chi,psi,omega,digamma,%
	Alpha,Beta,Gamma,Delta,Epsilon,Zeta,Eta,Theta,Iota,Kappa,Lambda,Mu,Nu,Xi,%
	Omicron,Pi,Rho,Sigma,Tau,Upsilon,Phi,Chi,Psi,Omega,Digamma%
	\do{$\csname\@tempa\endcsname,$ }%
}
\makeatother

\usepackage{fonttable}
\makeatletter
\renewcommand*\f@placedecimal[2]{#1\ {\color{gray}\tiny #2}}
\renewcommand*{\f@toct}[1]{\hbox{\color{gray}\rmfamily\'{}\kern-.2em\itshape#1\/\kern.05em}} 
\renewcommand*{\f@thex}[1]{\hbox{\color{gray}\rmfamily\H{}\ttfamily#1}} 
\makeatother
\hypersetup{
	pdftitle={\papertitle},
	pdfauthor={Niklas M. Witzig}
}

\newlength{\origbaselineskip}
\ifdef{\linesperpagedesired}
	{}
	{\newcommand{\linesperpagedesired}{39}}

\makeatletter
\newcommand{\newbaselinestretch}{1.1}
\makeatother
\newlength{\newbaselineskip}
\newlength{\newparindent}
\AtBeginDocument{%
	\setlength{\baselineskip}{\newbaselineskip}%
	\setlength{\parindent}{\newparindent}
	\setlength{\lineskiplimit}{0pt}
}
\newlength{\abovedisplayauxskip}
\newlength{\belowdisplayauxskip}
\newcommand{\predisplaycmd}{%
	\ifvmode\else\unskip\fi%
	\nopagebreak[2]%
	\vspace{\abovedisplayauxskip}%
}
\makeatletter
\def\@itemize@name{itemize}
\def\@enumerate@name{enumerate}
\def\@description@name{description}
\newcommand{\postdisplaycmd}{%
	\ifx\@currenvir\@itemize@name%
	\else%
		\ifx\@currenvir\@enumerate@name%
		\else%
			\ifx\@currenvir\@description@name%
			\else%
				\vskip\belowdisplayauxskip%
			\fi%
		\fi%
	\fi%
	\noindent%
}
\makeatother
\AtBeginEnvironment {align}      {\predisplaycmd}
\AfterEndEnvironment{align}      {\postdisplaycmd}
\AtBeginEnvironment {align*}     {\predisplaycmd}
\AfterEndEnvironment{align*}     {\postdisplaycmd}
\AtBeginEnvironment {alignat}    {\predisplaycmd}
\AfterEndEnvironment{alignat}    {\postdisplaycmd}
\AtBeginEnvironment {alignat*}   {\predisplaycmd}
\AfterEndEnvironment{alignat*}   {\postdisplaycmd}
\AtBeginEnvironment {displaymath}{\predisplaycmd}
\AfterEndEnvironment{displaymath}{\postdisplaycmd}
\AtBeginEnvironment {eqnarray}   {\predisplaycmd}
\AfterEndEnvironment{eqnarray}   {\postdisplaycmd}
\AtBeginEnvironment {eqnarray*}  {\predisplaycmd}
\AfterEndEnvironment{eqnarray*}  {\postdisplaycmd}
\AtBeginEnvironment {equation}   {\predisplaycmd}
\AfterEndEnvironment{equation}   {\postdisplaycmd}
\AtBeginEnvironment {equation*}  {\predisplaycmd}
\AfterEndEnvironment{equation*}  {\postdisplaycmd}
\AtBeginEnvironment {flalign}    {\predisplaycmd}
\AfterEndEnvironment{flalign}    {\postdisplaycmd}
\AtBeginEnvironment {flalign*}   {\predisplaycmd}
\AfterEndEnvironment{flalign*}   {\postdisplaycmd}
\AtBeginEnvironment {gather}     {\predisplaycmd}
\AfterEndEnvironment{gather}     {\postdisplaycmd}
\AtBeginEnvironment {gather*}    {\predisplaycmd}
\AfterEndEnvironment{gather*}    {\postdisplaycmd}
\AtBeginEnvironment {multiline}  {\predisplaycmd}
\AfterEndEnvironment{multiline}  {\postdisplaycmd}
\AtBeginEnvironment {multiline*} {\predisplaycmd}
\AfterEndEnvironment{multiline*} {\postdisplaycmd}

\newcommand{\thickmu}  {\thickmuskip=10mu \medmuskip=5mu}
\AtBeginEnvironment{align}      {\thickmu}
\AtBeginEnvironment{align*}     {\thickmu}
\AtBeginEnvironment{alignat}    {\thickmu}
\AtBeginEnvironment{alignat*}   {\thickmu}
\AtBeginEnvironment{displaymath}{\thickmu}
\AtBeginEnvironment{eqnarray}   {\thickmu}
\AtBeginEnvironment{eqnarray*}  {\thickmu}
\AtBeginEnvironment{equation}   {\thickmu}
\AtBeginEnvironment{equation*}  {\thickmu}
\AtBeginEnvironment{flalign}    {\thickmu}
\AtBeginEnvironment{flalign*}   {\thickmu}
\AtBeginEnvironment{gather}     {\thickmu}
\AtBeginEnvironment{gather*}    {\thickmu}
\AtBeginEnvironment{multiline}  {\thickmu}
\AtBeginEnvironment{multiline*} {\thickmu}

\usepackage{bm}

\makeatletter
\predisplaypenalty=\@medpenalty
\makeatother

\usepackage[all]{nowidow}

\AtBeginDocument{%
	
	\frenchspacing
	\sloppy
}

\ifxetex
	\usepackage[protrusion=true, expansion=false]{microtype}
\else
	\usepackage[protrusion=true, expansion=false, kerning=true]{microtype}
\fi

\spaceskip=\fontdimen2\font plus \fontdimen3\font minus 0.75\fontdimen4\font

  {\list{}{\leftmargin=\parindent \rightmargin=\parindent}%
   \linespread{\newbaselinestretch}\item[]\itshape\small}%
  {\endlist}
  {\list{}{\leftmargin=\parindent \rightmargin=\parindent
           \listparindent=\parindent \parsep=0pt}%
   \item[]}%
  {\endlist}

\usepackage{csquotes}

\usepackage{nameref}
\makeatletter
\newcommand*{\currentname}{\@currentlabelname}
\makeatother

\usepackage[multiple, bottom, hang]{footmisc}
\usepackage{etoolbox}

\makeatletter
\renewcommand{\@makefnmark}{%
	\ifnum\thefnsymbol=\@asteriskchar
	\else
	\rlap{\normalfont\@thefnmark.}%
	\fi}

\patchcmd{\@makefntext}{%
	\ifFN@hangfoot
	\bgroup}%
{%
	\ifFN@hangfoot
	\bgroup\def\@makefnmark{\rlap{\normalfont\@thefnmark}}}{}{}%
\patchcmd{\@makefntext}{%
	\ifdim\footnotemargin>\z@
	\hb@xt@ \footnotemargin{\hss\@makefnmark}}%
{%
	\ifdim\footnotemargin>\z@
	\hb@xt@ \footnotemargin{\@makefnmark\hss}}{}{}%
\makeatother

\DefineFNsymbols*{star}{%
	{$\mathrm{\star}$}{$\mathrm{\star\star}$}{$\mathrm{\ddagger}$}%
	{$\mathrm{\ddagger\ddagger}$}{\S}{\S\S}{\P}{\P\P}{$\mathrm{\|}$}{$\mathrm{\|\|}$}%
}
\setfnsymbol{star}

\usepackage[singlelinecheck=on]{caption}
\DeclareCaptionLabelSeparator{periodlargespace}{.\:\:}
\usepackage{subcaption}

\usepackage{booktabs}
\usepackage{tabularx}	
\newcolumntype{C}{>{\centering\arraybackslash}X}
\newcolumntype{J}{>{\arraybackslash}X}
\newcolumntype{L}{>{\RaggedRight\arraybackslash}X}
\newcolumntype{R}{>{\RaggedLeft\arraybackslash}X}

\LetLtxMacro{\oldtabular}{\tabular}
\LetLtxMacro{\endoldtabular}{\endtabular}
\RenewDocumentEnvironment{tabular}{O{c} m}{%
	\oldtabular[#1]{@{}#2@{}}%
}{%
	\endoldtabular%
}
\LetLtxMacro{\oldtabularx}{\tabularx}
\LetLtxMacro{\endoldtabularx}{\endtabularx}
\RenewDocumentEnvironment{tabularx}{m O{c} m}{%
	\oldtabularx{#1}[#2]{@{}#3@{}}%
}{%
	\endoldtabularx%
}

\usepackage{siunitx}[=v2]
\newcolumntype{T}[1]{@{}S[table-format = #1, table-space-text-pre = {***}, table-space-text-post = {***}]}
\ExplSyntaxOn\makeatletter
\newcommand{\thisseries}{\f@series}
\cs_set_protected:Npn \__siunitx_detect_font_weight_text: {%
	\let\origmdseries\mdseries@sf%
	\let\origbfseries\bfseries@sf%
	\let\currentseries\f@series%
	\edef\XcurrentseriesX{/\f@series/}%
	\tl_if_in:noTF
		{ /sb/ /b/ /bx/ /eb/ /ub/ /bold/ /extrabold/ /ultrabold/ /heavy/ /black/ /demibold/ /semibold/ }
		{ \XcurrentseriesX }
		{
			\cs_set:Nn \__siunitx_font_weight: {%
				\boldmath%
				\let\bfseries@sf\currentseries%
				\fontseries{\bfseries@sf}\selectfont%
			}%
			\let\bfseries@sf\origbfseries
		}
		{
			\cs_set:Nn \__siunitx_font_weight: {%
				\unboldmath
				\let\mdseries@sf\currentseries%
				\fontseries{\mdseries@sf}\selectfont%
			}%
			\let\mdseries@sf\origmdseries
		}%
}
\makeatother\ExplSyntaxOff

\usepackage[restart, breakwithin, indentafter]{parnotes}

\usepackage{makecell}

\usepackage{longtable}

\usepackage{multirow}

\makeatletter
\let\oldfigure\figure
\def\figure{\@ifnextchar[\figure@i \figure@ii}
\def\figure@i[#1]{\oldfigure[#1]\centering}
\def\figure@ii{\oldfigure\centering}
\makeatother

\usepackage[font={sf, small}]{floatrow}	
\usepackage{placeins} 

\usepackage{verbatim}

\usepackage{longtable}

\usepackage[inline]{enumitem}
\setlist{leftmargin=\parindent, listparindent=\parindent, itemsep=\smallskipamount, parsep=0pt}
\setlist[enumerate]{leftmargin=\parindent, labelsep=*}
\setlist[enumerate, 1]{label=(\arabic*), labelindent=-0.5pt}
\setlist[enumerate, 2]{label=\alph*., align=right}
\setlist[enumerate, 3]{label=\roman*., align=right, widest*=3, labelsep=0.3\parindent}
\setlist[itemize]{labelsep=0.435\parindent}
\setlist[description]{font=\rmfamily\normalsize}

\newtheoremstyle{Standard}
	{\topsep}    
	{\topsep}    
	{\itshape}   
	{}           
	{\bfseries}  
	{.}          
	{.5em}       
	{\thmname{#1}\thmnumber{\:#2}\thmnote{\bfseries\upshape\ (#3)}}

\theoremstyle{Standard}

\makeatletter
\newcounter{parenthypothesis}

\makeatother

\theoremstyle{definition}

\makeatletter
\@ifclassloaded{book}{%
	\numberwithin{theorem}{chapter}%
	\numberwithin{hypothesis}{chapter}%
	\numberwithin{result}{chapter}%
}
\makeatother

\usepackage{mfirstuc-english}
\gMFUnocap{about}
\gMFUnocap{at}
\gMFUnocap{against}
\gMFUnocap{around}
\gMFUnocap{between}
\gMFUnocap{by}
\gMFUnocap{from}
\gMFUnocap{on}
\gMFUnocap{over}
\gMFUnocap{per}
\gMFUnocap{to}
\gMFUnocap{versus}
\gMFUnocap{vs.}
\gMFUnocap{vis-\`a-vis}
\gMFUnocap{within}
\gMFUnocap{without}

\ifcase 0
	\renewcommand{\capitalisewords}[1]{#1}
	
	\renewcommand{\xcapitalisewords}[1]{#1}
\or
	\let\SavedContentsline\contentsline
	\renewcommand{\contentsline}[4]{%
		\SavedContentsline{#1}{\capitalisewords{#2}}{#3}{#4}%
	}
\else
	\let\SavedContentsline\contentsline
	\renewcommand{\contentsline}[4]{%
		\SavedContentsline{#1}{\capitalisewords{#2}}{#3}{#4}%
	}
	\LetLtxMacro{\SavedCaption}{\caption}
	\RenewDocumentCommand{\caption}{ O{\shortcaption} m }{%
		\def\shortcaption{%
			\xcapitalisewords{%
				#2%
			}%
		}%
		\SavedCaption[#1]{%
			\xcapitalisewords{%
				#2%
			}%
		}%
	}
\fi

\usepackage[title, titletoc]{appendix}  
\usepackage{chngcntr}  

\AtBeginEnvironment{appendices}{%
	\counterwithin{figure}{section}%
	\counterwithin{table}{section}%
	\counterwithin{equation}{section}%
}
\AtBeginEnvironment{subappendices}{%
	\counterwithin{figure}{section}%
	\counterwithin{table}{section}%
	\counterwithin{equation}{section}%
}

\makeatletter
\@ifclassloaded{book}{%
	\AfterEndEnvironment{appendices}{%
		\counterwithout{figure}{section}%
		\counterwithin{figure}{chapter}%
		\counterwithout{table}{section}%
		\counterwithin{table}{chapter}%
		\counterwithout{equation}{section}%
		\counterwithin{equation}{chapter}%
	}%
	\AfterEndEnvironment{subappendices}{%
		\counterwithout{figure}{section}%
		\counterwithin{figure}{chapter}%
		\counterwithout{table}{section}%
		\counterwithin{table}{chapter}%
		\counterwithout{equation}{section}%
		\counterwithin{equation}{chapter}%
	}%
}{}
\makeatother

\usepackage[sf, bf, raggedright, explicit]{titlesec}	

\let\dateorig\date
\renewcommand{\date}[1]{\dateorig{\small #1}}

\usepackage[style]{abstract}
\renewcommand{\abstitlestyle}[1]{\addcontentsline{toc}{section}{\abstractname}}

\usepackage[max6]{authblk}

\makeatletter
\renewcommand\AB@authnote[1]{\textsuperscript{{\kern.5pt}\textit{#1}}}
\renewcommand\AB@affilnote[1]{\textsuperscript{\textit{#1}}\,}
\makeatother

\newcommand{\appendixformat}[1][\currentname]{%
	\titleformat{\section}[hang]%
		{}{\textscale{1.3}{\textsf{\textbf{\appendixname~\thesection\quad}}}}{0pt}{\textscale{1.3}{\textsf{\textbf{#1}}}}[]%
}
\AtBeginEnvironment{appendices}{\appendixformat}

\titleformat{name=\section}[hang]%
	{}{\textscale{1.3}{\textsf{\textbf{\thesection\quad}}}}{0pt}{\textscale{1.3}{\textsf{\textbf{#1}}}}[]
\titleformat{name=\section, numberless}[hang]%
	{}{}{0pt}{\textscale{1.3}{\textsf{\textbf{#1}}}}[]
\titleformat{name=\subsection}[hang]%
	{}{\textscale{1.15}{\textsf{\textbf{\thesubsection\quad}}}}{0pt}{\textscale{1.15}{\textsf{\textbf{#1}}}}[]
\titleformat{name=\subsection, numberless}[hang]%
	{}{}{0pt}{\textscale{1.15}{\textsf{\textbf{#1}}}}[]
\titleformat{name=\subsubsection}[runin]%
	{}{}{0pt}{\textsf{\textbf{\thesubsubsection\quad #1\iftextterm{#1}{}{\unskip.}}}}[]
\titleformat{name=\subsubsection, numberless}[runin]%
	{}{}{0pt}{\textsf{\textbf{#1\iftextterm{#1}{}{\unskip.}}}}[]
\titleformat{name=\paragraph, numberless}[runin]%
	{}{}{0pt}{\textbf{#1\iftextterm{#1}{}{\unskip}}}[]
\titleformat{name=\subparagraph, numberless}[runin]%
	{}{}{0pt}{\textbf{#1\iftextterm{#1}{}{\unskip.}}}[]

\titlespacing*{\section}
	{0pt}{2\baselineskip plus 0.67\baselineskip}{1\baselineskip plus 0.5\baselineskip minus 0.33\baselineskip}
\titlespacing*{\subsection}%
	{0pt}{1.33\baselineskip plus 0.5\baselineskip minus 0.33\baselineskip}{0.67\baselineskip plus 0.33\baselineskip minus 0.17\baselineskip}
\titlespacing{\subsubsection}%
	{0pt}{1\baselineskip plus 0.67\baselineskip minus 0.33\baselineskip}{4\wordsep}[]
\titlespacing{\paragraph}
	{0pt}{0.5\baselineskip plus 0.5\baselineskip}{3\wordsep}[]
\titlespacing{\subparagraph}
	{\parindent}{0pt}{3\wordsep}[]

\usepackage[lining, scale=0.88]{FiraMono}  
\usepackage[book, semibold, lining, tabular, scale=0.88]{FiraSans}[2018/05/23]
\makeatletter
\@ifpackagelater{FiraSans}{2018/05/23}
	{}
	{\PackageError{FiraSans}
	     {Outdated 'FiraSans' package}
	     {Upgrade to FiraSans 2018/05/23 or newer!\MessageBreak
	      Otherwise the sans-serif font may not work properly, or the compilation might fail.\MessageBreak
	      This is a fatal error. I'm aborting now.}%
   }
\makeatother
\ifthenelse{\isundefined{\AfterPackage}}{\usepackage{afterpackage}}{}
\AfterPackage{siunitx}{%
  
}

\newcommand{\savesffamily}{\sfdefault}
\makeatletter
\newcommand{\savesfmdseries}{\mdseries@sf}
\newcommand{\savesfbfseries}{\bfseries@sf}
\makeatother

\usepackage[charter, greeklowercase=italicized, greekuppercase=italicized, sfscaled=false, ttscaled=false, euro=false]{mathdesign}
\let \savermfamily   \rmdefault
\let \savermmdseries \mddefault
\let \savermbfseries \bfdefault

\makeatletter
\@for\@tempa:=%
	alpha,beta,gamma,delta,epsilon,varepsilon,zeta,eta,theta,vartheta,iota,kappa,lambda,mu,nu,xi,%
	omicron,pi,varpi,rho,varrho,sigma,varsigma,tau,upsilon,phi,varphi,chi,psi,omega,digamma,%
	Alpha,Beta,Gamma,Delta,Epsilon,Zeta,Eta,Theta,Iota,Kappa,Lambda,Mu,Nu,Xi,%
	Omicron,Pi,Rho,Sigma,Tau,Upsilon,Phi,Chi,Psi,Omega,Digamma%
	\do{%
		\expandafter\let\csname up\@tempa\expandafter\endcsname\csname\@tempa up\endcsname%
	}%
\makeatother

\ifx\nequiv\undefined
	\newcommand{\nequiv}{\not\equiv}
\fi

\usepackage[scaled=1, lining, sups]{XCharter}  

\renewcommand{\textellipsis}{\mbox{.{\kern.09em}.{\kern.09em}.}}
\makeatletter
\newcommand{\@makefnmarkorig}{%
	\hbox{\sufigures\hspace*{.04em}\@thefnmark\hspace*{.04em}}%
}  
\makeatother

\ifxetex
\else
	\DisableLigatures[f]{family = {rm*, tt*}}
	\SetExtraKerning[unit=space]
		{encoding=*, family=*, series=*, size={*, normalsize, footnotesize}, font = */*/*/*/*}
		{\textemdash = {325, 325},
		           / = {100, 100},
		           : = { 50,   0},
		           ; = { 50,   0}}
\fi

\usepackage{xfrac}	

\AtBeginDocument{%
	\providecommand{\euro}[1]{\relax\ifmmode\text{\texteuro}#1\else\texteuro #1\fi}%
}

\makeatletter

\def\mweights@init{%
\ifdefined\mdseries@rm\else\edef\mdseries@rm{\mddefault}\fi
\ifdefined\bfseries@rm\else\edef\bfseries@rm{\bfdefault}\fi
\ifdefined\mdseries@sf\else\edef\mdseries@sf{\mddefault}\fi
\ifdefined\bfseries@sf\else\edef\bfseries@sf{\bfdefault}\fi
\ifdefined\mdseries@tt\else\edef\mdseries@tt{\mddefault}\fi
\ifdefined\bfseries@tt\else\edef\bfseries@tt{\bfdefault}\fi
\edef\rmdef@ult{\rmdefault}%
\edef\sfdef@ult{\sfdefault}%
\edef\ttdef@ult{\ttdefault}%
\edef\bfdef@ult{\bfdefault}%
\edef\mddef@ult{\mddefault}%
\edef\famdef@ult{\familydefault}%
}

\DeclareRobustCommand\bfseries{%
\mweights@init
\not@math@alphabet\bfseries\mathbf
\ifx\f@family\rmdef@ult\fontseries\bfseries@rm
\else\ifx\f@family\sfdef@ult\fontseries\bfseries@sf
\else\ifx\f@family\ttdef@ult\fontseries\bfseries@tt
\else\fontseries\bfdefault\fi\fi\fi\selectfont}%

\DeclareRobustCommand\mdseries{%
\mweights@init
\not@math@alphabet\mdseries\relax
\ifx\f@family\rmdef@ult\fontseries\mdseries@rm
\else\ifx\f@family\sfdef@ult\fontseries\mdseries@sf
\else\ifx\f@family\ttdef@ult\fontseries\mdseries@tt
\else\fontseries\mddefault\fi\fi\fi\selectfont}

\DeclareRobustCommand\rmfamily{%
\mweights@init
\not@math@alphabet\rmfamily\mathrm
\ifx\f@family\sfdef@ult
    \ifx\f@series\mdseries@sf\fontseries\mdseries@rm
    \else\ifx\f@series\bfseries@sf\fontseries\bfseries@rm
    \else\ifx\f@series\mddef@ult\fontseries\mdseries@rm
    \else\ifx\f@series\bfdef@ult\fontseries\bfseries@rm
    \else\fontseries\mdseries@rm
    \fi\fi\fi\fi
\else\ifx\f@family\ttdef@ult
    \ifx\f@series\mdseries@tt\fontseries\mdseries@rm
    \else\ifx\f@series\bfseries@tt\fontseries\bfseries@rm
    \else\ifx\f@series\mddef@ult\fontseries\mdseries@rm
    \else\ifx\f@series\bfdef@ult\fontseries\bfseries@rm
    \else\fontseries\mdseries@rm
    \fi\fi\fi\fi
\else\ifx\f@family\rmdef@ult
    \ifx\f@series\mdseries@rm\fontseries\mdseries@rm
    \else\ifx\f@series\bfseries@rm\fontseries\bfseries@rm
    \else\ifx\f@series\mddef@ult\fontseries\mdseries@rm
    \else\ifx\f@series\bfdef@ult\fontseries\bfseries@rm
    \else\fontseries\mdseries@rm
    \fi\fi\fi\fi
\else\fontseries\mdseries@rm
\fi\fi\fi\fontfamily\rmdefault\selectfont}

\DeclareRobustCommand\sffamily{%
\mweights@init
\not@math@alphabet\sffamily\mathsf
\ifx\f@family\rmdef@ult
    \ifx\f@series\mdseries@rm\fontseries\mdseries@sf
    \else\ifx\f@series\bfseries@rm\fontseries\bfseries@sf
    \else\ifx\f@series\mddef@ult\fontseries\mdseries@sf
    \else\ifx\f@series\bfdef@ult\fontseries\bfseries@sf
    \else\fontseries\mdseries@sf
    \fi\fi\fi\fi
\else\ifx\f@family\ttdef@ult
    \ifx\f@series\mdseries@tt\fontseries\mdseries@sf
    \else\ifx\f@series\bfseries@tt\fontseries\bfseries@sf
    \else\ifx\f@series\mddef@ult\fontseries\mdseries@sf
    \else\ifx\f@series\bfdef@ult\fontseries\bfseries@sf
    \else\fontseries\mdseries@sf
    \fi\fi\fi\fi
\else\ifx\f@family\sfdef@ult
    \ifx\f@series\mdseries@sf\fontseries\mdseries@sf
    \else\ifx\f@series\bfseries@sf\fontseries\bfseries@sf
    \else\ifx\f@series\mddef@ult\fontseries\mdseries@sf
    \else\ifx\f@series\bfdef@ult\fontseries\bfseries@sf
    \else\fontseries\mdseries@sf
    \fi\fi\fi\fi
\else\fontseries\mdseries@sf
\fi\fi\fi\fontfamily\sfdefault\selectfont}

\DeclareRobustCommand\ttfamily{%
\mweights@init
\not@math@alphabet\ttfamily\mathtt
\ifx\f@family\rmdef@ult
    \ifx\f@series\mdseries@rm\fontseries\mdseries@tt
    \else\ifx\f@series\bfseries@rm\fontseries\bfseries@tt
    \else\ifx\f@series\mddef@ult\fontseries\mdseries@tt
    \else\ifx\f@series\bfdef@ult\fontseries\bfseries@tt
    \else\fontseries\mdseries@tt
    \fi\fi\fi\fi
\else\ifx\f@family\sfdef@ult
    \ifx\f@series\mdseries@sf\fontseries\mdseries@tt
    \else\ifx\f@series\bfseries@sf\fontseries\bfseries@tt
    \else\ifx\f@series\mddef@ult\fontseries\mdseries@tt
    \else\ifx\f@series\bfdef@ult\fontseries\bfseries@tt
    \else\fontseries\mdseries@tt
    \fi\fi\fi\fi
\else\ifx\f@family\ttdef@ult
    \ifx\f@series\mdseries@tt\fontseries\mdseries@tt
    \else\ifx\f@series\bfseries@tt\fontseries\bfseries@tt
    \else\ifx\f@series\mddef@ult\fontseries\mdseries@tt
    \else\ifx\f@series\bfdef@ult\fontseries\bfseries@tt
    \else\fontseries\mdseries@tt
    \fi\fi\fi\fi
\else\fontseries\mdseries@tt
\fi\fi\fi\fontfamily\ttdefault\selectfont}

\AtBeginDocument{\mweights@init
\ifx\famdef@ult\rmdef@ult\rmfamily
\else\ifx\famdef@ult\sfdef@ult\sffamily
\else\ifx\famdef@ult\ttdef@ult\ttfamily
\fi\fi\fi}

\makeatother

\makeatletter
\newif\ifkp@upRm
\newif\ifkp@osm
\newif\ifkp@vosm
\makeatother

\DeclareMathVersion{normalup}
\DeclareMathVersion{boldup}
\DeclareMathVersion{sans}

\SetSymbolFont{operators}   {sans}{OT1}{mdbch}{m}{n}
\SetSymbolFont{letters}     {sans}{OML}{jkpss}{m}{it}
\SetSymbolFont{symbols}     {sans}{OMS}{jkp}  {m}{n}
\DeclareSymbolFont{extrasymbols}  {OMS}{cmbrs}{m}{n}
\SetSymbolFont{extrasymbols}{sans}{OMS}{cmbrs}{m}{n}

\SetMathAlphabet{\mathit} {sans}{T1}{\savesffamily}{\savesfmdseries}{it}
\SetMathAlphabet{\mathbf} {sans}{T1}{\savesffamily}{\savesfbfseries}{n}
\SetMathAlphabet{\mathtt} {sans}{OT1}{cmtl}{m}{n}
\SetMathAlphabet{\mathcal}{sans}{OMS}{ntxsy}{m}{n}
\SetSymbolFont{largesymbols} {sans}{OMX}{mdbch}{m}{n}

\DeclareMathVersion{sansup}
\SetSymbolFont{letters}  {sansup}{OML}{jkpss}{m}{it}
\SetSymbolFont{symbols}  {sansup}{OMS}{jkp}  {m}{n}

\DeclareMathVersion{boldsans}
\SetSymbolFont{operators}{boldsans}{OT1}{mdbch}{bx}{n}
\SetSymbolFont{letters}  {boldsans}{OML}{jkpss}{bx}{it}
\SetSymbolFont{symbols}  {boldsans}{OMS}{jkp}  {bx}{n}
\SetMathAlphabet{\mathit} {boldsans}{T1}{\savesffamily}{\savesfbfseries}{it}
\SetMathAlphabet{\mathtt} {boldsans}{T1}{cmtl}{b}{n}
\SetMathAlphabet{\mathcal}{boldsans}{OMS}{ntxsy}{b}{n}
\SetSymbolFont{largesymbols}{boldsans}{OMX}{mdbch}{bx}{n}

\DeclareMathVersion{boldsansup}
\SetSymbolFont{letters}{boldsansup}{OML}{jkpss}{bx}{it}
\SetSymbolFont{symbols}{boldsansup}{OMS}{jkp}  {bx}{n}

\DeclareSymbolFont{uprightglyphs}{T1}{\savermfamily}{\savermmdseries}{n}
\SetSymbolFont{uprightglyphs}{normal}    {T1}{\savermfamily}{\savermmdseries}{n}
\SetSymbolFont{uprightglyphs}{normalup}  {T1}{\savermfamily}{\savermmdseries}{n}
\SetSymbolFont{uprightglyphs}{bold}      {T1}{\savermfamily}{\savermbfseries}{n}
\SetSymbolFont{uprightglyphs}{boldup}    {T1}{\savermfamily}{\savermbfseries}{n}
\SetSymbolFont{uprightglyphs}{sans}      {T1}{\savesffamily}{\savesfmdseries}{n}
\SetSymbolFont{uprightglyphs}{sansup}    {T1}{\savesffamily}{\savesfmdseries}{n}
\SetSymbolFont{uprightglyphs}{boldsans}  {T1}{\savesffamily}{\savesfbfseries}{n}
\SetSymbolFont{uprightglyphs}{boldsansup}{T1}{\savesffamily}{\savesfbfseries}{n}
\DeclareSymbolFont{italicglyphs} {T1}{\savermfamily}{\savermmdseries}{it}
\SetSymbolFont{italicglyphs} {normal}    {T1}{\savermfamily}{\savermmdseries}{it}
\SetSymbolFont{italicglyphs} {normalup}  {T1}{\savermfamily}{\savermmdseries}{n}
\SetSymbolFont{italicglyphs} {bold}      {T1}{\savermfamily}{\savermbfseries}{it}
\SetSymbolFont{italicglyphs} {boldup}    {T1}{\savermfamily}{\savermbfseries}{n}
\SetSymbolFont{italicglyphs} {sans}      {T1}{\savesffamily}{\savesfmdseries}{it}
\SetSymbolFont{italicglyphs} {sansup}    {T1}{\savesffamily}{\savesfmdseries}{n}
\SetSymbolFont{italicglyphs} {boldsans}  {T1}{\savesffamily}{\savesfbfseries}{it}
\SetSymbolFont{italicglyphs} {boldsansup}{T1}{\savesffamily}{\savesfbfseries}{n}

\DeclareMathSymbol{0}{\mathalpha}{uprightglyphs}{`0}
\DeclareMathSymbol{1}{\mathalpha}{uprightglyphs}{`1}
\DeclareMathSymbol{2}{\mathalpha}{uprightglyphs}{`2}
\DeclareMathSymbol{3}{\mathalpha}{uprightglyphs}{`3}
\DeclareMathSymbol{4}{\mathalpha}{uprightglyphs}{`4}
\DeclareMathSymbol{5}{\mathalpha}{uprightglyphs}{`5}
\DeclareMathSymbol{6}{\mathalpha}{uprightglyphs}{`6}
\DeclareMathSymbol{7}{\mathalpha}{uprightglyphs}{`7}
\DeclareMathSymbol{8}{\mathalpha}{uprightglyphs}{`8}
\DeclareMathSymbol{9}{\mathalpha}{uprightglyphs}{`9}
\DeclareMathSymbol{+}{\mathbin}  {operators}    {`+}
\DeclareMathSymbol{=}{\mathrel}  {operators}    {`=}
\DeclareMathSymbol{.}{\mathord}  {uprightglyphs}{`.}
\DeclareMathSymbol{,}{\mathpunct}{uprightglyphs}{`,}
\DeclareMathSymbol{;}{\mathpunct}{uprightglyphs}{`;}
\DeclareMathSymbol{/}{\mathord}  {uprightglyphs}{`/}
\DeclareMathSymbol{\prime}{\mathord}{extrasymbols}{"30}
\DeclareMathDelimiter{(}      {\mathopen} {uprightglyphs}{`(} {largesymbols}{"00}
\DeclareMathDelimiter{)}      {\mathclose}{uprightglyphs}{`)} {largesymbols}{"01}
\DeclareMathDelimiter{[}      {\mathopen} {uprightglyphs}{`[} {largesymbols}{"02}
\DeclareMathDelimiter{]}      {\mathclose}{uprightglyphs}{`]} {largesymbols}{"03}
\DeclareMathDelimiter{\lbrace}{\mathopen} {uprightglyphs}{`\{}{largesymbols}{"08}
\DeclareMathDelimiter{\rbrace}{\mathclose}{uprightglyphs}{`\}}{largesymbols}{"09}
\DeclareMathSymbol{A}{\mathalpha}{italicglyphs}{`A}
\DeclareMathSymbol{B}{\mathalpha}{italicglyphs}{`B}
\DeclareMathSymbol{C}{\mathalpha}{italicglyphs}{`C}
\DeclareMathSymbol{D}{\mathalpha}{italicglyphs}{`D}
\DeclareMathSymbol{E}{\mathalpha}{italicglyphs}{`E}
\DeclareMathSymbol{F}{\mathalpha}{italicglyphs}{`F}
\DeclareMathSymbol{G}{\mathalpha}{italicglyphs}{`G}
\DeclareMathSymbol{H}{\mathalpha}{italicglyphs}{`H}
\DeclareMathSymbol{I}{\mathalpha}{italicglyphs}{`I}
\DeclareMathSymbol{J}{\mathalpha}{italicglyphs}{`J}
\DeclareMathSymbol{K}{\mathalpha}{italicglyphs}{`K}
\DeclareMathSymbol{L}{\mathalpha}{italicglyphs}{`L}
\DeclareMathSymbol{M}{\mathalpha}{italicglyphs}{`M}
\DeclareMathSymbol{N}{\mathalpha}{italicglyphs}{`N}
\DeclareMathSymbol{O}{\mathalpha}{italicglyphs}{`O}
\DeclareMathSymbol{P}{\mathalpha}{italicglyphs}{`P}
\DeclareMathSymbol{Q}{\mathalpha}{italicglyphs}{`Q}
\DeclareMathSymbol{R}{\mathalpha}{italicglyphs}{`R}
\DeclareMathSymbol{S}{\mathalpha}{italicglyphs}{`S}
\DeclareMathSymbol{T}{\mathalpha}{italicglyphs}{`T}
\DeclareMathSymbol{U}{\mathalpha}{italicglyphs}{`U}
\DeclareMathSymbol{V}{\mathalpha}{italicglyphs}{`V}
\DeclareMathSymbol{W}{\mathalpha}{italicglyphs}{`W}
\DeclareMathSymbol{X}{\mathalpha}{italicglyphs}{`X}
\DeclareMathSymbol{Y}{\mathalpha}{italicglyphs}{`Y}
\DeclareMathSymbol{Z}{\mathalpha}{italicglyphs}{`Z}
\DeclareMathSymbol{a}{\mathalpha}{italicglyphs}{`a}
\DeclareMathSymbol{b}{\mathalpha}{italicglyphs}{`b}
\DeclareMathSymbol{c}{\mathalpha}{italicglyphs}{`c}
\DeclareMathSymbol{d}{\mathalpha}{italicglyphs}{`d}
\DeclareMathSymbol{e}{\mathalpha}{italicglyphs}{`e}
\DeclareMathSymbol{f}{\mathalpha}{italicglyphs}{`f}
\DeclareMathSymbol{g}{\mathalpha}{italicglyphs}{`g}
\DeclareMathSymbol{h}{\mathalpha}{italicglyphs}{`h}
\DeclareMathSymbol{i}{\mathalpha}{italicglyphs}{`i}
\DeclareMathSymbol{\imath}{\mathalpha}{italicglyphs}{"19}
\DeclareMathSymbol{j}{\mathalpha}{italicglyphs}{`j}
\DeclareMathSymbol{\jmath}{\mathalpha}{italicglyphs}{"1A}
\DeclareMathSymbol{k}{\mathalpha}{italicglyphs}{`k}
\DeclareMathSymbol{l}{\mathalpha}{italicglyphs}{`l}
\DeclareMathSymbol{m}{\mathalpha}{italicglyphs}{`m}
\DeclareMathSymbol{n}{\mathalpha}{italicglyphs}{`n}
\DeclareMathSymbol{o}{\mathalpha}{italicglyphs}{`o}
\DeclareMathSymbol{p}{\mathalpha}{italicglyphs}{`p}
\DeclareMathSymbol{q}{\mathalpha}{italicglyphs}{`q}
\DeclareMathSymbol{r}{\mathalpha}{italicglyphs}{`r}
\DeclareMathSymbol{s}{\mathalpha}{italicglyphs}{`s}
\DeclareMathSymbol{t}{\mathalpha}{italicglyphs}{`t}
\DeclareMathSymbol{u}{\mathalpha}{italicglyphs}{`u}
\DeclareMathSymbol{v}{\mathalpha}{italicglyphs}{`v}
\DeclareMathSymbol{w}{\mathalpha}{italicglyphs}{`w}
\DeclareMathSymbol{x}{\mathalpha}{italicglyphs}{`x}
\DeclareMathSymbol{y}{\mathalpha}{italicglyphs}{`y}
\DeclareMathSymbol{z}{\mathalpha}{italicglyphs}{`z}

\makeatletter
\@for\@tempa:=%
	alpha,beta,gamma,delta,epsilon,varepsilon,zeta,eta,theta,vartheta,iota,kappa,lambda,mu,nu,xi,%
	omicron,pi,varpi,rho,varrho,sigma,varsigma,tau,upsilon,phi,varphi,chi,psi,omega,digamma,%
	Alpha,Beta,Gamma,Delta,Epsilon,Zeta,Eta,Theta,Iota,Kappa,Lambda,Mu,Nu,Xi,%
	Omicron,Pi,Rho,Sigma,Tau,Upsilon,Phi,Chi,Psi,Omega,Digamma%
	\do{%
		\expandafter\let\csname\@tempa orig\expandafter\endcsname\csname\@tempa\endcsname%
		\expandafter\let\csname\@tempa uporig\expandafter\endcsname\csname\@tempa up\endcsname%
	}%
\makeatother

\makeatletter

\@for\@tempa:=%
	epsilon,theta,pi,rho,sigma,phi%
\do{%
	\expandafter\let\csname var\@tempa orig\expandafter\endcsname\csname var\@tempa\endcsname%
}

\newcommand*{\sansmath}{%
	\@for\@tempa:=%
		alpha,beta,gamma,delta,epsilon,zeta,eta,theta,iota,kappa,lambda,mu,nu,xi,%
		omicron,pi,rho,sigma,varsigma,tau,upsilon,phi,chi,psi,omega,digamma,%
		Alpha,Beta,Gamma,Delta,Epsilon,Zeta,Eta,Theta,Iota,Kappa,Lambda,Mu,Nu,Xi,%
		Omicron,Pi,Rho,Sigma,Tau,Upsilon,Phi,Chi,Psi,Omega,Digamma%
	\do{%
		\expandafter\let\csname\@tempa\expandafter\endcsname\csname\@tempa LGR\endcsname%
		\expandafter\let\csname\@tempa up\expandafter\endcsname\csname\@tempa upLGR\endcsname%
		\expandafter\let\csname up\@tempa\expandafter\endcsname\csname\@tempa upLGR\endcsname%
	}%
	\@for\@tempa:=%
		epsilon,theta,pi,rho,phi%
	\do{%
		\expandafter\let\csname var\@tempa\expandafter\endcsname\csname\@tempa\endcsname%
		\expandafter\let\csname var\@tempa up\expandafter\endcsname\csname\@tempa up\endcsname%
		\expandafter\let\csname upvar\@tempa\expandafter\endcsname\csname up\@tempa\endcsname%
	}%
}

\newcommand*{\unsansmath}{%
	\@for\@tempa:=%
		alpha,beta,gamma,delta,epsilon,zeta,eta,theta,iota,kappa,lambda,mu,nu,xi,%
		omicron,pi,rho,sigma,varsigma,tau,upsilon,phi,chi,psi,omega,digamma,%
		Alpha,Beta,Gamma,Delta,Epsilon,Zeta,Eta,Theta,Iota,Kappa,Lambda,Mu,Nu,Xi,%
		Omicron,Pi,Rho,Sigma,Tau,Upsilon,Phi,Chi,Psi,Omega,Digamma%
	\do{%
		\expandafter\let\csname\@tempa\expandafter\endcsname\csname\@tempa orig\endcsname%
		\expandafter\let\csname\@tempa up\expandafter\endcsname\csname\@tempa uporig\endcsname%
		\expandafter\let\csname up\@tempa\expandafter\endcsname\csname\@tempa uporig\endcsname%
	}%
	\@for\@tempa:=%
		epsilon,theta,pi,rho,phi%
	\do{%
		\expandafter\let\csname var\@tempa\expandafter\endcsname\csname var\@tempa orig\endcsname%
		\expandafter\let\csname var\@tempa up\expandafter\endcsname\csname var\@tempa uporig\endcsname%
		\expandafter\let\csname upvar\@tempa\expandafter\endcsname\csname var\@tempa uporig\endcsname%
	}%
}

\newcommand*{\upgreekletters}{%
	\@for\@tempa:=%
		alpha,beta,gamma,delta,epsilon,varepsilon,zeta,eta,theta,vartheta,iota,kappa,lambda,mu,nu,xi,%
		omicron,pi,varpi,rho,varrho,sigma,varsigma,tau,upsilon,phi,varphi,chi,psi,omega,digamma,%
		Alpha,Beta,Gamma,Delta,Epsilon,Zeta,Eta,Theta,Iota,Kappa,Lambda,Mu,Nu,Xi,%
		Omicron,Pi,Rho,Sigma,Tau,Upsilon,Phi,Chi,Psi,Omega,Digamma%
	\do{%
		\expandafter\let\csname\@tempa\expandafter\endcsname\csname\@tempa up\endcsname%
	}%
}
\newcommand*{\itgreekletters}{%
	\@for\@tempa:=%
		alpha,beta,gamma,delta,epsilon,varepsilon,zeta,eta,theta,vartheta,iota,kappa,lambda,mu,nu,xi,%
		omicron,pi,varpi,rho,varrho,sigma,varsigma,tau,upsilon,phi,varphi,chi,psi,omega,digamma,%
		Alpha,Beta,Gamma,Delta,Epsilon,Zeta,Eta,Theta,Iota,Kappa,Lambda,Mu,Nu,Xi,%
		Omicron,Pi,Rho,Sigma,Tau,Upsilon,Phi,Chi,Psi,Omega,Digamma%
	\do{%
		\expandafter\let\csname\@tempa\expandafter\endcsname\csname\@tempa orig\endcsname%
	}%
}

\makeatother

\let \bmorig \bm
\renewcommand{\bm}[1]{%
	\IfInSansMode%
		\textbf{\mathversion{boldsans}\(#1\)}%
	\else%
		\bmorig{#1}%
	\fi\relax%
}
\renewcommand{\mathbf}[1]{\bm{#1}}

\renewcommand{\mathcal}[1]{\mathscr{#1}}

\newif\IfInSansMode
\newif\IfInBoldMode
\newif\IfInUpMode
\let \oldsf \sffamily
\renewcommand*{\sffamily}{%
	\oldsf\sansmath\InSansModetrue%
	\IfInBoldMode\mathversion{boldsans}\else\mathversion{sans}\fi\relax%
}
\let \oldbf \bfseries
\renewcommand*{\bfseries}{%
	\oldbf\InBoldModetrue%
	\IfInSansMode\sansmath\mathversion{boldsans}\else\mathversion{bold}\fi\relax%
}
\let \oldmd \mdseries
\renewcommand*{\mdseries}{%
	\oldmd\InBoldModefalse%
	\IfInSansMode\sansmath\mathversion{sans}\else\mathversion{normal}\fi\relax%
}
\let \oldnorm \normalfont
\renewcommand*{\normalfont}{%
	\oldnorm\InSansModefalse\InBoldModefalse\mathversion{normal}%
	\unsansmath%
}
\let \oldrm \rmfamily
\renewcommand*{\rmfamily}{%
	\oldrm\InSansModefalse%
	\IfInBoldMode\mathversion{bold}\else\mathversion{normal}\fi\relax%
	\unsansmath%
}

\let \mathnormalorig \mathnormal
\renewcommand{\mathnormal}[1]{%
	\IfInSansMode%
		\IfInBoldMode%
			\mathversion{boldsans}%
			{\textbf{\(#1\)}}%
		\else%
			\mathversion{sans}%
			{\textmd{\(#1\)}}%
		\fi\relax%
	\else%
		\mathnormalorig{#1}%
	\fi\relax%
}

\let \mathrmorig \mathrm
\renewcommand{\mathrm}[1]{%
	\IfInSansMode%
		{\textrm{%
			\IfInBoldMode%
				\mathversion{bold}%
				\(\mathrmorig{#1}\)%
			\else%
				\mathversion{normal}%
				\(\mathrmorig{#1}\)%
			\fi\relax%
		}}%
	\else%
		\mathrmorig{#1}%
	\fi\relax%
}

\newcommand{\mathup}[1]{%
	\IfInSansMode%
		{\textup{%
			\InUpModetrue%
			\IfInBoldMode%
				\mathversion{boldsansup}%
				\(#1\)%
			\else%
				\mathversion{sansup}%
				\(#1\)%
			\fi\relax%
		}}%
	\else%
		{\upgreekletters\mathrm{#1}\itgreekletters}%
	\fi\relax%
}

\let \operatornameorig \operatorname
\renewcommand{\operatorname}[1]{%
	\operatornameorig{\mathup{#1}}%
}
\makeatletter
\@for\@tempa:=%
	arccos,arccot,arccsc,arcsec,arcsin,arctan,arg,cos,cosh,cot,coth,csc,%
	deg,det,dim,exp,gcd,hom,inf,ker,lg,lim,liminf,limsup,ln,log,max,min,%
	Pr,sec,sin,sinh,sup,tan,tanh%
	\do{%
		\expandafter\let\csname\@tempa\endcsname\relax%
	}%
\makeatother

\DeclareMathOperator*{\lim}   {\mathup{lim}}

\DeclareMathOperator {\ln}    {\mathup{ln}}
\DeclareMathOperator {\log}   {\mathup{log}}

\usepackage[style=apa, backend=biber, natbib=true, backref=true, uniquename=false]{biblatex}

\renewcommand{\cite}{\citet}

\AtBeginDocument{%
	\setlength{\bibindent}{\parindent}%
}
\DefineBibliographyStrings{english}{%
	backrefpage = {},	
	backrefpages = {}	
}
\DefineBibliographyStrings{ngerman}{%
	backrefpage  = {},	
	backrefpages = {}	
}

\renewbibmacro*{pageref}{%
	\addperiod
	\iflistundef{pageref}
	{}
	{\printtext[brackets]{
			\ifnumgreater{\value{pageref}}{1}
			{\bibstring{backrefpages}}
			{\bibstring{backrefpage}}%
			\printlist[pageref][-\value{listtotal}]{pageref}%
		}%
	}%
}

\DeclareFieldFormat{doi}{%
	\Urlmuskip = 0mu\relax%
	\mathchardef\UrlBigBreakPenalty=1\relax%
	\mathchardef\UrlBreakPenalty=2\relax%
	\url{https://doi.org/#1}%
}
\DeclareFieldFormat{url}{%
	\Urlmuskip = 0mu\relax%
	\mathchardef\UrlBigBreakPenalty=1\relax%
	\mathchardef\UrlBreakPenalty=2\relax%
	\url{#1}%
}
\AtEveryBibitem{\iffieldundef{doi}{}{\clearfield{eprint}}}  
\AtEveryBibitem{\iffieldundef{doi}{}{\clearfield{url}}}  

\let\comment\undefined  

\usepackage[todonotes={textsize=scriptsize}, authormarkuptext=name, authormarkup=none]{changes}

\makeatletter
\tikzstyle{notestyleraw} = [
	draw=\@todonotes@currentbordercolor,
	fill=\@todonotes@currentbordercolor,
	text=white,
	line width = 0.35pt,
	text width = \@todonotes@textwidth - 1.5ex - 1pt,
	inner sep = 0.75ex,
	rounded corners = 0pt
]
\colorlet{authorcolor}{magenta}
\@namedef{Changes@AuthorColor}{magenta}
\colorlet{Changes@Color}{magenta}
\newcommand{\changesfontsettings}{\sffamily\scriptsize\baselineskip=2.25ex}
\renewcommand{\@todonotes@useSizeCommand}{\changesfontsettings}
\xpatchcmd{\@todonotes@drawLineToLeftMargin} {connectstyle}{connectstyle, line width=0.6pt}{}{}
\xpatchcmd{\@todonotes@drawLineToRightMargin}{connectstyle}{connectstyle, line width=0.6pt}{}{}
\makeatother

\definecolor{electricultramarine}{rgb}{0.25, 0.0, 1.0}
\definecolor{alizarin}{rgb}{0.82, 0.1, 0.26}
\definecolor{dartmouthgreen}{rgb}{0.05, 0.5, 0.06}
\definecolor{goldenpoppy}{HTML}{FCC200}
\definecolor{internationalorange}{rgb}{1.0, 0.31, 0.0}
\definechangesauthor[name={Holger}, color=alizarin]           {HG}
\definechangesauthor[name={Lou~E.}, color=internationalorange]{LV}
\definechangesauthor[name={U.~R.},  color=electricultramarine]{UR}

\LetLtxMacro{\todoorig}{\todo}
\makeatletter
\renewcommand{\todo}[2][]{%
	\todoorig[color=Changes@Color, bordercolor=Changes@Color, #1, tickmarkheight=0.1cm, linecolor=authorcolor]{#2}%
}
\makeatother

\LetLtxMacro{\commentorig}{\comment}
\newsavebox{\commentbox}
\newlength{\commentboxwidth}
\makeatletter
\newcommand{\commentpatched}[2][]{%
	\savebox{\commentbox}[\commentboxwidth][t]{%
		\parbox{\commentboxwidth}{%
			\changesfontsettings\color{white}\RaggedRight%
			\hrule\smallskip%
			#2%
		}%
	}%
	\commentorig[#1]{%
		\strut\newline%
		\usebox{\commentbox}%
	}%
}
\makeatother
\renewcommand{\comment}[2][]{%
	\commentpatched[#1]{#2}%
}

\setdeletedmarkup{}

\makeatletter
\newcommand{\lineabovecomment}[1]{%
	\setbox0=\hbox{#1\unskip}%
	\ifdim\wd0=0pt%
	\else%
		\hrule\smallskip\strut%
	\fi%
}
\newcommand{\insertchangesadded}{%
	\smallskip\hrule\smallskip%
	\strut\textit{\mbox{\changesaddedname}}\\[\smallskipamount]
	\lineabovecomment{\Changes@added@comment}%
}
\newcommand{\insertchangesdeleted}[1]{%
	\smallskip\hrule\smallskip%
	\strut\textit{\mbox{\changesdeletedname:}} #1\\[\smallskipamount]
	\lineabovecomment{\Changes@deleted@comment}%
}
\newcommand{\insertchangesreplaced}[1]{%
	\smallskip\hrule\smallskip%
	\strut\textit{\mbox{\changesreplacedname:}} #1\\[\smallskipamount]
	\lineabovecomment{\Changes@replaced@comment}%
}
\newcommand{\insertchangeshighlight}{%
	\strut\lineabovecomment{\Changes@highlight@comment}%
}
\Changes@set@commandname{added}
\expandafter\newcommand\expandafter{\csname\Changes@commandname\endcsname}[2][\@empty]{%
	\setkeys{Changes@added}{#1}%
	\Changes@output%
		{added}%
		{\Changes@added@id}%
		{#2}%
		{}%
		{\insertchangesadded\Changes@added@comment}%
		{\changesaddedname}%
		{#2}%
}
\Changes@set@commandname{deleted}
\expandafter\newcommand\expandafter{\csname\Changes@commandname\endcsname}[2][\@empty]{%
	\setkeys{Changes@deleted}{#1}%
	\Changes@output%
		{deleted}%
		{\Changes@deleted@id}%
		{}%
		{}
		{\insertchangesdeleted{#2}\Changes@deleted@comment}
		{\changesdeletedname}%
		{#2}%
}
\Changes@set@commandname{replaced}
\expandafter\newcommand\expandafter{\csname\Changes@commandname\endcsname}[3][\@empty]{%
	\setkeys{Changes@replaced}{#1}%
	\Changes@output%
		{replaced}%
		{\Changes@replaced@id}%
		{#2}%
		{}  
		{\insertchangesreplaced{#3}\Changes@replaced@comment}
		{\changesreplacedname}%
		{#2}%
}
\Changes@set@commandname{highlight}
\expandafter\newcommand\expandafter{\csname\Changes@commandname\endcsname}[2][\@empty]{%
	\setkeys{Changes@highlight}{#1}%
	\Changes@output%
		{highlight}%
		{\Changes@highlight@id}%
		{#2}%
		{}%
		{\insertchangeshighlight\Changes@highlight@comment}%
		{\changeshighlightname}%
		{#2}%
}
\makeatother

\newcommand{\coloruline}[2]{%
	\newcommand{\tempuline}{%
		\bgroup\markoverwith{\textcolor{#1}{\rule[-1ex]{1.5pt}{0.25ex}}}%
		\ULon%
	}%
	\tempuline{#2}%
}
\sethighlightmarkup{%
	\colorlet{authorcoloraux}{authorcolor!33}%
	\IfIsColored{%
		\sethlcolor{authorcoloraux}%
		\hl{#1\strut}%
	}{}%
}

\usepackage{pdfcomment}
	{\end{pdfsidelinecomment}%
}

 \definecolor{lightor}{rgb}{1.0,0.93,0.8}
\definecolor{lightgreen}{RGB}{210, 247, 173}

\let\pgfimage=\includegraphics
\graphicspath{{../graphs/paper/}}

\pgfdeclarelayer{background}
\pgfdeclarelayer{foreground}
\pgfsetlayers{background,main,foreground}

\providecommand{\keywords}[1]
{  \vspace{1em}
	{\small
		\noindent\textbf{\textit{Keywords:  }} #1
	}
}

\providecommand{\JEL}[1]
{
	{\small
			\noindent\textbf{\textit{JEL-Codes:}} #1
	}
}

\newcommand{\sigstar}{\raisebox{0.66ex}{\scalebox{0.95}{$\star$}}}

\newcommand{\balCL}[1][1]{\mbox{\caps{BAL}$_{\mathup{CL}}$}\xspace}
\newcommand{\unbalCLA}[1][1]{\mbox{\caps{UNBAL}$^{\mathup{I}}_{\mathup{CL}}$}\xspace}
\newcommand{\unbalCLB}[1][1]{\mbox{\caps{UNBAL}$^{\mathup{II}}_{\mathup{CL}}$}\xspace}

\usetikzlibrary{calc}

\bibliography{meta_references.bib}

\begin{document}

\selectlanguage{USenglish} \frenchspacing



\title{\sffamily\bfseries%
	\papertitle%
	\thanks{\protect
I thank Alexander Dzionara, Markus Eyting, Ben Grodeck, Katharina Hartinger, Florian Hett, Marc Kaufmann, Sebastian Olschewski, Daniel Schunk, Ferdinand Vieider and Isabell Zipperle for helpful comments and discussion. I owe a special thanks to Dominik Straub. I gratefully acknowledge funding from the Gutenberg Academy Fellows Program and the interdisciplinary research unit IPP at Johannes Gutenberg-University of Mainz.}%
}

\author[a]{Niklas M. Witzig\thanks{Corresponding author (\email{niklas.witzig@uni-mainz.de}).}}

\affil[a]{Johannes Gutenberg University of Mainz}

\date{
	\smallskip
	{\small \sffamily \textit{This version: December, 2024 \\
}}%
}

\maketitle

\vfill

\begin{abstract}%

{\small \noindent
I study altruistic choices through the lens of a cognitively noisy decision-maker. I introduce a theoretical framework that demonstrates how increased cognitive noise can directionally affect altruistic decisions and put its implications to the test: In a laboratory experiment, participants make a series of binary choices between taking and giving monetary payments. In the treatment, to-be-calculated math sums replace straightforward monetary payments, increasing the cognitive difficulty of choosing. The Treatment group exhibits a lower sensitivity towards changes in payments and decides significantly more often in favor of the other person, i.e., is more altruistic. I explore the origins of this effect with Bayesian hierarchical models and a number-comparison task, mirroring the "mechanics" of the altruism choices absent any altruistic preference. The treatment effect is similar in this task, suggesting that the perception of numerical magnitudes drives treatment differences. The probabilistic model supports this interpretation. A series of additional results show a negative correlation between cognitive reflection and individual measures of cognitive noise, as well as associations between altruistic choice and number comparison. Overall, these results suggest that the expression of altruistic preferences -- and potentially social preferences more generally -- is affected by the cognitive difficulty of their implementation.
}

\keywords{{Cognitive Noise, Altruism, Bayesian Hierarchical Models, Experiment}}

\JEL{{C91, D91}}
\end{abstract}

\vfill

\clearpage

\hypertarget{introduction}{%
\section{Introduction}\label{introduction}}

Theories of social and other-regarding preferences characterize a crucial advancement in economics and help to explain the results of various laboratory and field outcomes irreconcilable with traditional assumptions of pure self-interest \autocite{levineModelingAltruismSpitefulness1998, andreoniGivingAccordingGARP2002,fehrTheoryFairnessCompetition1999a,boltonERCTheoryEquity2000,charnessUnderstandingSocialPreferences2002,fismanIndividualPreferencesGiving2007}, with reviews in \textcite{fehrSocialPreferencesFundamental2023a, fehrChapterEconomicsFairness2006, cooperOtherRegardingPreferencesSelective2016}. Quantifying the underlying motivations of prosocial behavior, a growing body of research estimates population- and individual-level parameters of different social preference frameworks (e.g., \nptextcite{bellemarePreferencesIntentionsExpectation2011, klockmannArtificialIntelligenceEthics2022, carpenterMeasuringSociallyAppropriate2024, bruhinManyFacesHuman2019a, fismanIndividualPreferencesGiving2007} with a meta-analysis of inequality-aversion estimates available in \nptextcite{nunnariMetaAnalysisDistributionalPreferences2024}). 

While functional forms and parameter values differ, what these approaches share is an (implicit) assumption that social preferences are (i) a stable and fixed quantity and (ii) ``fundamental'', i.e., that -- in a standard individual utility-maximizing framework -- only differences in social preferences explain differences in behavior. While the first assumption is at odds with within-person inconsistencies typically observed in experiments, the second assumption contrasts with the advent of a ``cognitive turn'' \autocite{enkeCognitiveTurnBehavioral2024} in behavioral economics. There, a growing body of evidence shows how cognitive imprecision, e.g., in the mental representations of objective decision problem features such as lottery payoffs and probabilities, can generate risk aversion, probability weighting and hyperbolic discounting as the result of an optimal adaptation to imprecise perceptions. In addition, this literature micro-founds inconsistencies in behavior beyond ad-hoc solutions as an immediate consequence of such noisy perceptions \autocite{woodfordProspectTheoryEfficient2012, khawCognitiveImprecisionSmallStakes2021a, woodfordModelingImprecisionPerception2020, vieiderDecisionsUncertaintyBayesian2024,frydmanEfficientCodingRisky2022}. Similarly, the complexity of deciding according to one's preference and ``cognitive uncertainty'' can produce behavior previously understood as a choice anomaly and bias \autocite{enkeComplexityHyperbolicDiscounting2023, opreaDecisionsRiskAre2024,enkeCognitiveUncertainty2023}.

It is only natural to assume that social preferences are affected by cognitive imprecisions and the complexity of their implementation. Tasks involving social preference require that a decision-maker assesses the (non-trivial) value of different options before deciding, rendering such operations ``complex''.\footnote{\textcite[3]{opreaDecisionsRiskAre2024} writes: ``When we say a lottery is ``complex,'' we mean only that its value is not transparent to the decision maker because the procedure required to optimally aggregate its disaggregated components into a value is costly or difficult.''} If past experiences shape social preferences, a noisy recollection of these experiences will also affect choices (see \nptextcite{polaniaEfficientCodingSubjective2019} for the original argument). Additionally, as social preferences are usually identified via monetary trade-offs, imprecise perceptions of numerical magnitudes -- as in previous work -- are a candidate, but not-yet considered driver of choices related to social preferences. 

First indicative evidence that noisy cognition or complexity-related processes can guide prosocial choice is now starting to amass: For example, \textcite{enkeBehavioralAttenuation2024} consider the dictator game as one example of how cognitive uncertainty moderates reactions to changes in objective problem features across over 30 experiments. Similarly, \textcite{baoCognitiveUncertaintyGPT2024} find that higher cognitive uncertainty is associated with higher contributions in the public goods game. Beyond that, empirical evidence for noisy cognition (or complexity-related effects more generally) on social preferences is still lacking, however, particularly, in domains with no clear ``default-action''.


 

In this paper, I investigate altruistic choices -- a simple form of social preference-related decisions -- through the lens of a cognitively noisy decision-maker. Based on \textcite{vieiderDecisionsUncertaintyBayesian2024}, I develop a model of altruistic choice and show how an increase in cognitive noise can directionally affect altruistic choices. The core intuition is that higher cognitive noise -- either in perceiving monetary payments or altruistic preferences -- will lead an optimal Bayesian decision-maker away from acting upon true preferences and monetary stakes and instead towards simpler mental default representations (i.e., their prior beliefs). With increased noise, the decision maker reacts less strongly to changes in underlying problem features and, depending on the mental default, also chooses systematically differently. 

To test these implications, I implement a laboratory experiment that consists of two parts: In the first part, each of 300 participants makes a series of binary choices between taking a payment $\text{self}$ for themselves or giving a payment $\text{other}$ to another person. The Treatment group faces the same decision but has the values of $\text{self}$ and $\text{other}$ replaced by to-be-calculated sums, i.e., decide between $\text{self}_1 + \text{self}_2 \; (=\text{self})$ and $\text{other}_1 + \text{other}_2 \; (=\text{other})$. Encasing the stakes in to-be-calculated sums increases the cognitive difficulty of perceiving the monetary payments and deciding on this task. In the second part of the experiment, participants face the \textit{identical} numerical values as previously but have to judge which of two numbers $\text{A}$ (previously $\text{self}$), or $\text{B}$ (previously $\text{other}$) $\times \sfrac{1}{2}$ is numerically larger. This task aims to mirror the ``mechanics'' of the altruism decisions (as participants have to compare two numbers), yet abstracts from any subjective altruistic preference with $\sfrac{1}{2}$ replacing the individual-specific and subjective altruistic-preference-dependent decision threshold with an objective and fixed term.

The main results of the experiment are as follows: In the altruism task, participants in the Treatment group exhibit (i) a flatter association between changes in payments and behavior (are less sensitive) and (ii) decide significantly more often for $\text{other}$, i.e., are more altruistic. The theoretical framework offers multiple explanations for this effect, which I begin to investigate using a probabilistic (Bayesian hierarchical) model of participants' choices. The model indicates a considerable degree of uncertainty around the mechanism of the treatment effect, suggesting that additional data beyond altruistic decisions is necessary to make a precise statement about the origin of the treatment effect. Herein lies the main contribution of the number comparison task: In this task, although abstracting from altruistic preferences, the \textit{treatment effect} remains qualitatively similar: The Treatment group again is less sensitive towards changes in numerical values and decides significantly less often for $\text{A}$ (previously $\text{self}$). Interpreted together, this implies that the perception of numerical magnitudes, i.e., some intuitive prior default that $\widehat{\text{self}} < \widehat{\text{other}}$ and $\widehat{\text{A}} < \widehat{\text{B}}$, is a candidate driver for the treatment effect in both tasks. This conclusion is supported by probabilistic models based on both the number comparison data and on a \textit{combined} dataset of behavior in both tasks, indicating a high probability of such an ``intermediate'' prior belief of numerical magnitudes.


In additional analyses, I further investigate associations between cognition and altruistic preferences more generally. Given identical numerical magnitudes in the altruism and number comparison task, I can closely examine potential relationships between behavior across domains: Correlation analyses show that choosing $\text{self}$ correlates with choosing $\text{A}$ and identifying the $correct$ answer in the number comparison task, suggesting that numerical cognition can play a role when measuring altruistic behavior more generally. Further, individual parameter estimates (based on the hierarchical models) of cognitive noise correlate with performance on the Cognitive Reflection Test and Berlin Numeracy Task, showing that more cognitively able persons are also less cognitively noisy, providing support for the cognitive motivation of the general framework. More exploratory analyses show how measures of meta-cognition (e.g., self-reported confidence and attention) and response times -- both key informants of choice processes -- are both more strongly affected by the treatment variation and more closely related to behavior in the number comparison versus the altruism task. This, in turn, suggests that metacognitive processes could ``play out'' differently in domains of purely subjective preference versus domains with more objective benchmarks of choice.




With these findings, this paper predominantly speaks to three strands of literature: Primarily, the results relate to the recent work of the cognitive turn in behavioral economics. Most of this work so far focuses on the domain of risk, ambiguity, belief updating and intertemporal choice \autocite{khawCognitiveImprecisionSmallStakes2021a, woodfordModelingImprecisionPerception2020, vieiderBayesianEstimationDecision2024,frydmanEfficientCodingRisky2022,enkeComplexityHyperbolicDiscounting2023,vieiderCognitiveFoundationsDelayDiscounting2023}. This paper shows that the core theoretical postulate, of a Bayesian decision maker optimally integrating noisy perceptions with prior knowledge, is applicable to the domain of social preferences, too and offers a potential avenue for future work into the direction of cognitive noise and subjective valuations more generally. This paper also makes a methodological contribution by showing how to \textit{causally test} the impact of increased noise beyond standard (and arguably ad-hoc) treatments of cognitive load or time pressure. The to-be-calculated sums proposed here, inspired by treatments in \textcite{enkeComplexityHyperbolicDiscounting2023}, provide an easy-to-implement method of increasing uncertainty in the perception of objective problem features that also have proven to be suitable in a more extensive repeated-trials experiment. Employing exogenous manipulations further speaks to a broader ongoing discussion in this literature: \textcite[57]{enkeCognitiveTurnBehavioral2024} outlines how it is often unclear which assumptions to make about the (locations of the) prior distributions in the Bayesian models. Here, I show that typical ignorance assumptions are not necessarily valid (see also \nptextcite[p.\ 33][]{opreaMindingGapOrigins2024}) and demonstrate how a combination of experimental variations increasing noise, ``mirror'' tasks isolating parts of the decision-making process and probabilistic modeling allow inferring the parameters of prior distribution and likelihood in the Bayesian models.


This paper also relates to the literature on structural estimations of social preference parameters \autocite{bellemarePreferencesIntentionsExpectation2011, klockmannArtificialIntelligenceEthics2022, carpenterMeasuringSociallyAppropriate2024, bruhinManyFacesHuman2019a, fismanIndividualPreferencesGiving2007,nunnariMetaAnalysisDistributionalPreferences2024, echeverryStructuralIdentificationSocial2023}. Here, I show how altruistic behavior (and thereby ``revealed altruistic preferences'') can be affected by an increase in the cognitive difficulty of choosing. In turn, social preference parameter estimates are thus likely to be \textit{biased} due to the presence of unaccounted-for cognitive noise. Accordingly, classifying subjects into distinct preference types (e.g., as done in \nptextcite{bruhinManyFacesHuman2019a,vanleeuwenEstimatingSocialPreferences2024,carpenterMeasuringSociallyAppropriate2024}) or using estimated social preferences are used to predict or related to real-world outcomes (e.g., as in \nptextcite{grafParametersSocialPreference2013}) potentially suffers from biases. Furthermore, this paper makes an additional contribution to this literature: In an additional analysis, I show that the proposed theoretical model of a noisy Bayesian decision maker outperforms a standard random utility benchmark \autocite{mcfaddenEconometricModelsProbabilistic1981} commonly used in this literature. The ``noisy cognition'' model proposed here thus offers both a theoretically more micro-founded model of altruistic choice and provides empirical arguments in its favor, motivating its application to social preference modeling more generally.


 
 
Lastly, this paper relates to an interdisciplinary literature on dual-process models of cognition, altruism, and social preferences. This literature studies differences in the level of pro-sociality between fast (more intuitive) and slow (more deliberate) decisions. For example, \textcite{randSpontaneousGivingCalculated2012} show how cooperation is largest when participants are put under time pressure, which in turn sparked a debate about whether ``fairness is intuitive'' \autocite{cappelenFairnessIntuitive2016} (also ``social heuristics hypothesis''). The theoretical model and empirical evidence presented here add two insights to this literature: First (i), the model demonstrates that depending on the intuition in a given context, more intuitive (i.e., more prior-based) decision-making may also lead to more selfish choices, e.g., if monetary payments are intuitively perceived to be the same in less-for-me vs.\ more-for-other types of decisions. Next, while the treatment effect towards more altruism goes in a similar direction as in \textcite{randSpontaneousGivingCalculated2012}, the fact that (ii), the perception of monetary payments is a likely driver for more altruistic choices in the Treatment group highlights how experimental manipulations (e.g., including time pressure) may drive (pro-)social choices through channels other than via a genuine impact on social preferences per se. \textcite{hutchersonNeurocomputationalModelAltruistic2015} put forward a comparable argument and highlight how -- in light of a drift-diffusion model -- individual differences in decision thresholds (which are related to decision noise) can lead to differences in altruistic choices independent of altruistic preferences. 
This paper provides additional evidence in favor of this line of argument.





The remainder of this paper is structured as follows: Section \ref{sec:theory} describes the theoretical model that illustrates how an increase in cognitive noise can directionally affect altruistic choices. Section \ref{sec:experiment} details the between-subject experimental design and differences in implementation for the Baseline and Treatment group. Section \ref{sec:results} introduces the results of the experiment, focusing on the main group differences in altruistic choices and number comparison, accompanied by details on the structural estimations. Additional analyses on cognitive ability, noise and the relationship between altruism and number comparison as well as metacognition and response times, follow. Section \ref{sec:discussion} discusses the main results of the paper, outlining potential avenues for future research while Section \ref{sec:conclusion} briefly concludes, highlighting the limitations of the current paper.

\section{Theoretical Framework}\label{sec:theory}

The theoretical model modifies models of noisy Bayesian cognition by \textcite{vieiderDecisionsUncertaintyBayesian2024} and \textcite{khawCognitiveImprecisionSmallStakes2021a} and applies them to choices involving altruistic preferences.

\paragraph{Altruistic Preferences}

Imagine a decision maker (DM) who has to choose between taking a monetary payment $\text{self}$ for themselves or giving an amount $\text{other}$ to another person. They choose $\text{self}$ if

\begin{equation}\label{eq:equation1}
(1-\beta) \times \text{self} > \beta \times \text{other}	
\end{equation}

where $\beta$ is the weight the DM places on the material well-being of the other person (i.e., an altruism parameter) and its complement, $1-\beta$, is the weight the DM places on their own well-being (see e.g., \nptextcite{bernheimAltruismFamilyReconsidered1988, levineModelingAltruismSpitefulness1998}). While the value of $\beta$ can, in principle, be any real number, a sensible restriction is to expect $\beta \in (0,0.5)$, i.e., that the DM places a positive weight on the other person's payment yet still cares more strongly about their own payment. 
 This choice rule abstracts from many important notions relevant to social preferences, such as (dis-)advantageous inequality aversion \autocite{fehrTheoryFairnessCompetition1999a}, reciprocity concerns \autocite{falkTheoryReciprocity2006, bellemarePreferencesIntentionsExpectation2011}, or social norms \autocite{carpenterMeasuringSociallyAppropriate2024} and also does not distinguish between (non-)warm-glow giving \autocite{andreoniGivingImpureAltruism1989}. Instead, this rule focuses on the core trade-off akin to many types of social preference decisions: Trading off one's own vs.\ another person's material wealth. This rule, in turn, is similar to notions of ``pure altruism'' \autocite{levineModelingAltruismSpitefulness1998}, ``preferences for giving'' \autocite{fismanIndividualPreferencesGiving2007}, and ``social welfare preferences'' \autocite{andreoniGivingAccordingGARP2002, charnessUnderstandingSocialPreferences2002} assuming a strict positive weight on the payment of the other person.
 
 Rearranging equation \ref{eq:equation1} and applying the natural log\footnote{This follows the original model by \textcite{vieiderDecisionsUncertaintyBayesian2024} who demonstrates that logging the choice rule does not alter the results in a meaningful qualitative manner. See there for a derivation for the un-logged later (probabilistic) choice rule.} to both sides gives:

\begin{equation}\label{eq:equation2}
\ln \left(\frac{\text{self}}{\text{other}}\right) > \ln \left(\frac{\beta}{1-\beta} \right)
\end{equation}

which states that the DM assesses whether the (log) ratio of monetary payments, $\ln \frac{\text{self}}{\text{other}}$, is larger than their (log) altruism preference threshold $\ln \frac{\beta}{1-\beta}$. This structure predominantly makes the (computation of the) later model more tractable, yet expressing the payments and the preference threshold as (logs of) ratios also has a natural interpretation: $\frac{\beta}{1-\beta}$ is the weight a DM places on the other person's payment relative to their own. For example in the case of $\beta = 0.2$, which implies $\frac{\beta}{1-\beta} = 0.25$, the DM values each euro for the other person one-fourth as much compared to a euro for themselves. Judging monetary payments as ratios further aligns with evidence from cognitive psychology about numerical judgments (a feature discussed more extensively below) which in turn will be relevant to many choice rules featuring a comparison of monetary payments. 

\paragraph{Noisy Bayesian Decision Maker}

Following \textcite{vieiderDecisionsUncertaintyBayesian2024} and \textcite{khawCognitiveImprecisionSmallStakes2021a}, I apply a Bayesian perspective to equation \ref{eq:equation2} to allow cognitive noise to affect altruistic choices based on an intuition of ``perceptual uncertainty'', i.e., that the perception of problem features -- $\frac{\text{\text{self}}}{\text{other}}$ and $\frac{\beta}{1-\beta}$ -- gives rise to a noisy mental representation of both.\footnote{The exact origins of this cognitive noise are beyond the scope of this paper. The general motivation can be linked to the idea of a ``Bayesian Brain'' \autocite{doyaBayesianBrainProbabilistic2006} from neuroscience, i.e., that the brain optimally combines uncertain sensory evidence with prior knowledge. 
} 

Noisily representing monetary payments is a feature well grounded in research from cognitive psychology: Ample evidence suggests that humans possess an ``approximate number sense'' for mental representations of numerosity, e.g., judging which of two boxes on a screen contains more dots \autocite{feigensonCoreSystemsNumber2004}. Such approximate behavior is also likely to be at play for symbolic characterizations of numbers, including that of Arabic numerals \autocite{niederRepresentationNumberBrain2009, dehaeneNumberSenseHow2011}.\footnote{Electrophysiological recordings of monkeys can single out specific neurons favoring the mental representation of specific numbers. Crucially, the activations of these neurons are \textit{bell-shaped}: They activate strongest at their designated neuron and less pronounced at other numbers while the activation declines in numerical difference \autocite{diesterSemanticAssociationsSigns2007}.} This e.g., manifests in the ``numerical ratio effect'': People's performance in distinguishing between two Arabic numerals strongly depends on the numerical ratio between both numbers \autocite{dehaeneSymbolsQuantitiesParietal1993}. This effect is evident in neuroimaging data and materializes during ``passive viewing of numerical stimuli without an explicit behavioral task'' \autocite[2219]{cantlonNeuralDevelopmentAbstract2009}. These observations suggest that the famous ``Weber's-Law'', which states that the necessary increase to detect a difference to a base stimulus is proportional to the base stimulus, also holds for numerical stimuli. Supporting this conjecture, studies aiming to map the ``mental number line'' also find evidence for a non-linear compressed mental representation of Arabic numerals \autocite{longoSpatialAttentionMental2007}. Furthermore, \textcite{prat-carrabinEfficientCodingNumbers2022} show that the relationship between discriminability and bias -- a core law of human perception \autocite{weiLawfulRelationPerceptual2017} and originally formulated for sensory domains -- also holds for numerical cognition.



 A noisy mental representation of the preference threshold is plausible as well: Given that the true preference $\beta$ remains an entirely \textit{subjective} quantity, the DM must rely on introspection to form a belief about their preference. If past experiences shape $\beta$, an imperfect memory could introduce uncertainty around the true preference for the DM, i.e., introduce noise (see \textcite{polaniaEfficientCodingSubjective2019} for the original argument for subjective valuations).

To formalize the noisy mental representations of $\frac{\text{self}}{\text{other}}$ and $\frac{\beta}{1-\beta}$, assume that the DM obtains mental signals about the true values from a distribution of possible representations:

\begin{equation}\label{eq:likelihoods}
s_{\frac{\text{self}}{\text{other}}} \mid \ln \frac{\text{self}}{\text{other}} \sim \mathcal{N}\left(\ln \left(\frac{\text{self}}{\text{other}}\right), \nu_{\frac{\text{self}}{\text{other}}}^2 \right),s_{\frac{\beta}{1-\beta}}\mid \ln \frac{\beta}{1-\beta} \sim \mathcal{N}\left(\ln \left(\frac{\beta}{1-\beta}\right), \nu_{\frac{\beta}{1-\beta}}^2\right)
\end{equation}

where $s_{\frac{\text{self}}{\text{other}}}$ and $s_{\frac{\beta}{1-\beta}}$ are the mental signals. Importantly, I do not assume that the mental signals share a common variance, but instead that $\nu_{\frac{\text{self}}{\text{other}}}$ characterizes ``noise in monetary payments'' and $\nu_{\frac{\beta}{1-\beta}}$ characterizes ``noise in altruistic preferences''. Cognitive noise, as understood here, thus consists of two different sources of noise in perceiving problem features. This departs from previous work, where a common noise variance is a typical assumption \autocite{vieiderDecisionsUncertaintyBayesian2024}. The main argument for separately modeling the noise terms in the present setting is that $s_{\frac{\text{self}}{\text{other}}}$ and $s_{\frac{\beta}{1-\beta}}$ do not both refer to an explicitly stated numerical quantity, like e.g., a lottery payoff and probability, but to monetary payments and a \textit{subjective} preference threshold. Assuming that the same (cognitive) process underlies the perception of both features is thus less justified in the present setting. 

What remains common to both noise terms is the overall log-normal noise structure. With both means as logarithms, the noise terms become signal-dependent (as the variance of the exponentiated values increases in the mean of the original distribution), matching the intuition of Weber's Law (see also \nptextcite{barretto-garciaIndividualRiskAttitudes2023}).


In addition to the mental signals, the Bayesian DM has \textit{prior beliefs} about both the ratio of monetary payments and the preference threshold.
\begin{equation}\label{eq:priors}
\ln \frac{\text{self}}{\text{other}} \sim \mathcal{N} \left(\ln \mu_{\hat{r}}, \sigma_{\hat{r}}^2\right),\ln \frac{\beta}{1-\beta} \sim \mathcal{N}\left( \ln \mu_{\hat{b}}, \sigma_{\hat{b}}^2\right)
\end{equation}

where $\ln \mu_{\hat{r}} = \ln \frac{\widehat{\text{self}}}{\widehat{\text{other}}}$ and $\ln \mu_{\hat{b}} = \ln \frac{\widehat{\beta}}{1-\widehat{\beta}}$, the default representations of the problem features (the hat indicating prior values). A common assumption about the prior means is that they are equal to 0, i.e., $\ln \mu_{\hat{r}} = 0 \Leftrightarrow \frac{\widehat{\text{self}}}{\widehat{\text{other}}} = 1$ and similarly $\ln \mu_{\hat{b}} = 0 \Leftrightarrow \widehat{\beta} = 0.5$. These prior means imply that the DM intuitively does not distinguish between payments $\widehat{\text{self}} = \widehat{\text{other}}$ and treats the importance of both their own and the other person's well-being alike $1-\widehat{\beta} = \widehat{\beta}$. This ``ignorance assumption'' fits a possible interpretation of prior means by \textcite[266]{gabaixBehavioralInattention2019a} as ``the value that spontaneously comes to mind with no thinking''. 



Given likelihoods in equation \ref{eq:likelihoods} and priors in equation \ref{eq:priors}, a Bayesian DM will arrive at the following posterior distributions:

\begin{equation*}
\ln \left(\frac{\text{self}}{\text{other}}\right) \mid s_{\frac{self}{\text{other}}} \sim \mathcal{N}\left(\frac{\sigma_{\hat{r}}^2}{\sigma_{\hat{r}}^2 + \nu_{\frac{\text{self}}{\text{other}}}^2} \times s_{\frac{self}{\text{other}}}+\frac{\nu_{\frac{\text{self}}{\text{other}}}^2}{\sigma_{\hat{r}}^2+\nu_{\frac{\text{self}}{\text{other}}}^2} \times \ln \mu_{\hat{r}}, \frac{\nu_{\frac{\text{self}}{\text{other}}}^2 \sigma_{\hat{r}}^2}{\nu_{\frac{\text{self}}{\text{other}}}^2+\sigma_{\hat{r}}^2}\right)
\end{equation*}

\begin{equation*}
\ln \left(\frac{\beta}{1-\beta}\right) \mid s_{\frac{\beta}{1-\beta}} \sim \mathcal{N}\left(\frac{\sigma_{\hat{b}}^2}{\sigma_{\hat{b}}^2+\nu_{\frac{\beta}{1-\beta}}^2} \times s_{\frac{\beta}{1-\beta}}+\frac{\nu_{\frac{\beta}{1-\beta}}^2}{\sigma_{\hat{b}}^2+\nu_{\frac{\beta}{1-\beta}}^2} \times \ln \mu_{\hat{b}}, \frac{\nu_{\frac{\beta}{1-\beta}}^2 \sigma_{\hat{b}}^2}{\nu_{\frac{\beta}{1-\beta}}^2+\sigma_{\hat{b}}^2}\right)
\end{equation*}
with the following expected values:

\begin{equation*}
E\left[\ln \left(\frac{\text{self}}{\text{other}}\right) \mid s_{\frac{\text{self}}{\text{other}}}\right]=\alpha \times s_{\frac{\text{self}}{\text{other}}}+(1-\alpha) \times\ln \mu_{\hat{r}}
\end{equation*}
\begin{equation*}
E\left[\ln \left(\frac{\beta}{1-\beta}\right) \mid s_{\frac{\beta}{1-\beta}}\right]=\gamma \times s_{\frac{\beta}{1-\beta}}+(1-\gamma) \times \ln \mu_{\hat{b}}
\end{equation*}

where $\alpha = \frac{\sigma_{\hat{r}}^2}{\sigma_{\hat{r}}^2+\nu_{\frac{\text{self}}{\text{other}}}^2}$ and $\gamma = \frac{\sigma_{\hat{b}}^2}{\sigma_{\hat{b}}^2+\nu_{\frac{\beta}{1-\beta}}^2}$, the Bayesian evidence weights. The lower $\gamma$ and $\alpha$, the more the DM relies on the ``intuitive'' prior values, treating payments and persons alike, and the closer $\gamma$ and $\alpha$ are to $1$, the more the DM relies on the (noisy signals of the) true values of $\frac{\text{self}}{\text{other}}$ and $\frac{\beta}{1-\beta}$. 

So far, this setup follows a common structure in the noisy cognition literature. Incorporating a prior belief for monetary payments -- with the values $\frac{\text{self}}{\text{other}}$ varying from trial to trial in a typical experiment -- leads to a regularization in the posterior belief as shown above and typically assumed. However, altruistic preferences are again conceptually somewhat different: First, $\frac{\beta}{1-\beta}$ is usually assumed to be a fixed quantity for a given DM, which in turn makes it difficult (although not impossible) to distinguish between true and prior preferences. From an empirical perspective, this translates into identification challenges and risks of overly parameterizing the later choice model. Therefore, I adjust this common setup and abstract from any additional influence of the prior belief over preferences on choices. Therefore, throughout, I assume that the prior belief is maximally uninformative in terms of inference over true preferences. Importantly, noise in the mental signals $s_{\frac{\beta}{1-\beta}}$ still impacts choices (discussed in more detail below) and this assumption does not imply that perception of preferences in noiseless, only that noise in perceiving preferences does not lead to a bias towards some mental default preference. In terms of the theoretical framework, by setting $\sigma_{\hat{b}} \rightarrow \infty$, this implies that $\lim_{\sigma_{\hat{b}} \to \infty} \gamma = 1$ and, in turn $E\left[\ln \left(\frac{\beta}{1-\beta}\right) \mid s_{\frac{\beta}{1-\beta}}\right]= s_{\frac{\beta}{1-\beta}}$. Therefore, the prior belief regarding altruistic preferences does not affect the posterior distribution (or expectation); only the monetary payment prior fulfills the typical regularizing role.

Notwithstanding, the expectations of both posterior distributions form the basis of the choice rule. Mirroring equation \ref{eq:equation2}, the Bayesian DM will decide for $\text{self}$ if

\begin{equation}
E\left[\ln \left(\frac{\text{self}}{\text{other}}\right) \mid s_{\frac{\text{self}}{\text{other}}}\right]> 	E\left[\ln \left(\frac{\beta}{1-\beta}\right) \mid s_{\frac{\beta}{1-\beta}}\right] 
\end{equation}

and plugging in the above expressions for the posterior expectations results in

\begin{equation}\label{eq:choice_rule_signals}
\alpha \times s_{\frac{\text{self}}{\text{other}}} - s_{\frac{\beta}{1-\beta}} > \ln\delta
\end{equation}

where $\delta = \frac{1}{\mu_{\hat{r}}^{1-\alpha}}$. The DM decides for $\text{self}$ if the difference between the weighted signal of monetary payments and the signal of their altruistic preference is larger than a prior-induced threshold. To arrive at a probabilistic choice rule, subtract the $z$-score of the random variable $\alpha \times s_{\frac{\text{self}}{\text{other}}} - s_{\frac{\beta}{1-\beta}} \sim \mathcal{N}(\alpha \times \ln \frac{\text{self}}{\text{other}} - \ln \frac{\beta}{1-\beta},\nu_{\frac{\text{self}}{\text{other}}}^2\alpha^2 + \nu_{\frac{\beta}{1-\beta}}^2)$ from the equivalent $z$-score of equation \ref{eq:choice_rule_signals}, which results in the following Probit equation (see \textcite{vieiderDecisionsUncertaintyBayesian2024} for the original proof):

\begin{equation}\label{eq:choice_function}
{Pr}([\text{self} \succ \text{other}])=\Phi\left(\frac{\alpha \times \ln \left(\frac{\text{self}}{\text{other}}\right)- \ln \left(\frac{\beta}{1-\beta}\right) -\ln (\delta)}{\sqrt{\nu_{\frac{\text{self}}{\text{other}}}^2\alpha^2 + \nu_{\frac{\beta}{1-\beta}}^2}}\right)
\end{equation}

Choosing $\text{self}$ is thus the outcome of a \textit{probabilistic} process in which both noise in monetary payments -- and a potentially invoked bias of a prior belief due to increased noise -- as well as noise in altruistic preferences guide choices.


\paragraph{The Impact of Cognitive Noise}

Equipped with the probabilistic choice rule, I can more closely investigate the (numerical) impact of an increase in noise -- both in monetary payments and altruistic preferences -- on altruistic choices. Figure \ref{fig:theory_noise_plot} simulates both the impact of increasing noise $\nu_{\frac{\text{self}}{\text{other}}}$ and $\nu_{\frac{\beta}{1-\beta}}$ on the probability of choosing $\text{self}$ as a function of the ratio $\frac{\text{self}}{\text{other}}$ (equation \ref{eq:choice_function}) for varying values of $\mu_{\hat{r}}$. Throughout all panels, I fix $\beta = 0.3$ and $\sigma_{\hat{r}} = 1$, whereas $\nu_{\frac{\beta}{1-\beta}} = 0.25$ in the top and $\nu_{\frac{\text{self}}{\text{other}}} = 0.25$ in the bottom row. 
 
 

\begin{figure}[H]
{\centering
\resizebox{1\linewidth}{!}{%
\input{graphs_export_restricted/theory/main_theory_plot_manual.tex}}
\mycaption[Impact of Cognitive Noise on Altruistic Choices]{\quad This figure plots the impact of changes in cognitive noise $\nu_{\frac{\text{self}}{\text{other}}}$ (top) and $\nu_{\frac{\beta}{1-\beta}}$ (bottom) on the probabilistic choice function (equation \ref{eq:choice_function}) depending on different values of $\mu_{\hat{r}}$. Throughout all panels $\beta = 0.30$ and $\sigma_{\hat{r}} = 1$. The plot also includes average values of choosing $\text{self}$. 
} \label{fig:theory_noise_plot}}
\end{figure}

Consider the first row of Figure \ref{fig:theory_noise_plot}. There, I vary  $\nu_{\frac{\text{self}}{\text{other}}} \in [0.25, 0.5, 1]$ with fixed values of $\nu_{\frac{\beta}{1-\beta}} = 0.25$ and $\mu_{\hat{r}} = 1$ (i.e., $\ln \delta = 0$). An increase in noise in perceiving monetary payments \textit{increases} choices for $\text{self}$. With increased noise, the DM relies more strongly on their prior knowledge (i.e., $\alpha$ decreases) and $\ln \frac{\text{self}}{\text{other}}$ is attenuated towards $0$. Given the log space of the choice rule, an attenuation towards $0$ (of $\ln \frac{\text{self}}{\text{other}}$) implies an increase towards $1$ on the original scale. In other words, smaller values of $\frac{\text{self}}{\text{other}}$ are perceived to be larger. 

However, differences in the values of the prior mean will lead to different effects on choices: Consider the second graph in the first row, where $\mu_{\hat{r}} = 0.5$, i.e., an ``intermediate'' intuitive perception of monetary payments. Now, an increase in $\nu_{\frac{\text{self}}{\text{other}}}$ \textit{decreases} choices for $\text{self}$ as larger values of $\frac{\text{self}}{\text{other}}$ will be downwards adjusted due to the impact of the prior. Conversely, if $\mu_{\hat{r}} = 1.5$ (third graph), an increase in $\nu_{\frac{\text{self}}{\text{other}}}$ (again) \textit{increases} choices for $\text{self}$, which is quantitatively larger compared to the first instance. Overall -- in this exercise -- an increase in $\nu_{\frac{\text{self}}{\text{other}}}$ will therefore increase choices for $\text{self}$, unless $\mu_{\hat{r}} < 1$ (i.e., an ``intermediate'' payment perception).

In the second row, $\nu_{\frac{\text{self}}{\text{other}}}$ remains fixed at $0.25$, but noise in preferences varies between $\nu_{\frac{\beta}{1-\beta}} \in [0.25,0.5,1]$. Focusing on the first graph, note that an increase in noise in preferences \textit{decreases} choices for $\text{self}$. The origin of this effect -- recall that equation \ref{eq:choice_function} abstracts from the impact of a prior over preferences -- lies in the log-normal noise structure of the mental signals $s_{\frac{\beta}{1-\beta}}$: Due to the concavity of the log-transform, increasing signal noise leads to stronger attenuation for larger values of the $z$-score of the difference between a given trials' payment ratio $\frac{\text{self}}{\text{other}}$ and the preference-induced threshold $\frac{\beta}{1-\beta}$, which in turn, translates into \textit{fewer} choices for $\text{self}$. Varying $\mu_{\hat{r}}$ (second and third graph in the second row) leads to changes in the overall level of choices for $\text{self}$, as $\alpha < 1$, yet the effect of $\nu_{\frac{\beta}{1-\beta}}$ on choices remains largely unaffected. Note further that -- across all instances -- an increase in $\nu_{\frac{\beta}{1-\beta}}$ does not shift the \textit{indifference values}, as $\nu_{\frac{\beta}{1-\beta}}$ is absent from the numerator of equation \ref{eq:choice_function}. The difference in average choices is thus driven by the asymmetric effect of the log-normal noise structure.\footnote{Accordingly, a linear version of equation \ref{eq:choice_function} with linear encoding and Gaussian priors, i.e., where $Pr(self) = \Phi \left(\frac{\alpha \times \frac{\text{self}}{\text{other}} - \frac{\beta}{1-\beta} - \delta}{\sqrt{\nu_{\frac{\text{self}}{\text{other}}}^2\alpha^2 + \nu_{\frac{\beta}{1-\beta}}^2}} \right)$ with $\delta = 1 - (1-\alpha)\mu_{\hat{r}}$ does not feature such an asymmetrical effect of noise.}
 
Overall, the simulations show that the impact of increasing noise will depend on whether increased cognitive noise is primarily in perceiving monetary payments or altruistic preferences -- circling back to the discussion in Section \ref{sec:theory}. Due to the Bayesian regularization towards a prior belief for monetary payments, the impact of noise in monetary payments will additionally depend on the moments of that prior, as illustrated.

\paragraph{Hypotheses}

Based on these discussions, I can formulate hypotheses that center on the potential mechanisms of a treatment effect due to increased noise. These hypotheses remain stylized in nature and should be understood as examples that do not necessarily apply to the entire parameter range (given the non-linear nature of the model) but help in organizing the mechanisms of the model nonetheless.\footnote{The hypotheses further illustrate the considerable degree of flexibility of the Bayesian model, which requires a careful interpretation of the empirical results and the exact mechanisms of a potential treatment effect in the experimental data later on.} Similar to the exercise above, they should also be understood \textit{ceteris paribus}.

\bigskip

\textbf{Hypothesis 1$_a$ (Noise in Payments $\bm{a}$):} An increase in $\nu_{\frac{\text{self}}{\text{other}}}$ given $\mu_{\hat{r}} \geq 1$ \textbf{increases} average choices for $\text{self}$.

\textbf{Hypothesis 1$_b$ (Noise in Payments $\bm{b}$):} An increase in $\nu_{\frac{\text{self}}{\text{other}}}$ given $\mu_{\hat{r}} < 1$ \textbf{decreases} average choices for $\text{self}$.

\textbf{Hypothesis 1$_c$ (Noise in Preferences):} An increase in $\nu_{\frac{\beta}{1-\beta}}$ \textbf{decreases} average choices for $\text{self}$, whereas the indifference value remains unchanged.

\paragraph{Additional Hypotheses} Outside the impact of an increase in cognitive noise emanating from the theoretical framework, other hypotheses emerge following a ``cognitive lens'' to altruistic choices more generally. Throughout, I do not formulate assumptions on whether $\nu_{\frac{\text{self}}{\text{other}}}$ or $\nu_{\frac{\beta}{1-\beta}}$ is the more appropriate measure of cognitive noise in a particular instance, but understand both to be measures of different aspects of cognitive noise. I therefore refrain from discussing them separately for the additional hypotheses.

A key assumption of the noisy cognition literature is that noisy mental representations drive choice variability and bias. Crucially, these noisy representations, e.g., of numerical magnitudes, should thus have comparable effects on behavior across domains with similar ``mechanics'' of choice irrespective of the \textit{subject} of the decision. Further, if perceiving numerical values (and subjective preferences) is person-specific, individual measures of cognitive noise should be positively correlated across domains within a person. Supporting evidence in this direction is presented by \textcite{frydmanEfficientCodingRisky2022, frydmanCoordinationCognitiveNoise2023}, who show how lottery choice and behavior in a coordination game correlates with choices in a ``perceptual'' number discrimination task. 

For altruistic choices, this implies that individual measures of cognitive noise and overall behavior, more generally, should be positively related to data from a comparable choice task, e.g., a number comparison task (considered in the experiment).\\

\textbf{Hypothesis 2:} There is a positive correlation between measures of cognitive noise and behavior in altruism choices and choices in a number comparison task. \\

Further, noisy mental representations of problem features are generally assumed to stem from \textit{cognitive} processes. In line with this argument, a broad class of work shows how performance in the Cognitive Reflection Test \autocite{frederickCognitiveReflectionDecision2005a} -- a popular tool to measure reflective thinking -- empirically correlates with various biases and mistakes in choices: For instance, \textcite{augenblickOverinferenceWeakSignals2022} find that subjects who score high on the CRT infer more (less) from strong (weak) signals, \textcite{opreaDecisionsRiskAre2024} finds that lower CRT performance is associated with more prospect-theoretic behavior (i.e., probability weighting and loss-aversion). \textcite{assenzaPerceivedWealthCognitive2019} report a negative correlation between CRT performance and misjudgments in a portfolio valuation task and \textcite{chewMultipleswitchingBehaviorChoicelist2022} show a negative relationship between CRT performance and multiple switching behavior in choice lists. For \textit{altruistic choices}, this implies that an association between measures of cognitive ability and individual measures of cognitive noise $\nu_{\frac{\text{self}}{\text{other}}}$ and $\nu_{\frac{\beta}{1-\beta}}$ -- key drivers of choice inconsistency and bias -- should emerge, with more cognitively able people exhibiting lower values of noise.\\

\textbf{Hypothesis 3:} Individual measures of cognitive noise negatively correlate with measures of cognitive ability. \\

\section{Experiment}\label{sec:experiment}

In this section, I describe the setup and implementation of the experiment, which fulfills four objectives: (i) eliciting altruistic decisions in terms of the choice rule in equation \ref{eq:equation1}. (ii) Exogenously manipulating cognitive noise during altruistic decisions. (iii) Eliciting choices in a number comparison task similar to the altruistic decisions, and (iv) gathering additional personal characteristics, especially regarding subjects' cognitive ability. Accordingly, the experiment consists of three parts: Part 1 entails the altruistic decisions, where cognitive noise is manipulated in a \textit{between-subject} treatment condition. Part 2 introduces the number comparison task, and Part 3 elicits additional behavioral and survey data. All three parts are described in detail below, and a graphical overview of the experiment outline is depicted in Figure \ref{fig:session_outline}.

\subsection{Part 1: Altruistic Choice}

\paragraph{Altruistic Choice}

\begin{figure*}
        \centering
        \begin{subfigure}[b]{0.475\textwidth}
        \setcounter{subfigure}{0}
            \centering
   \includegraphics[width=1\linewidth]{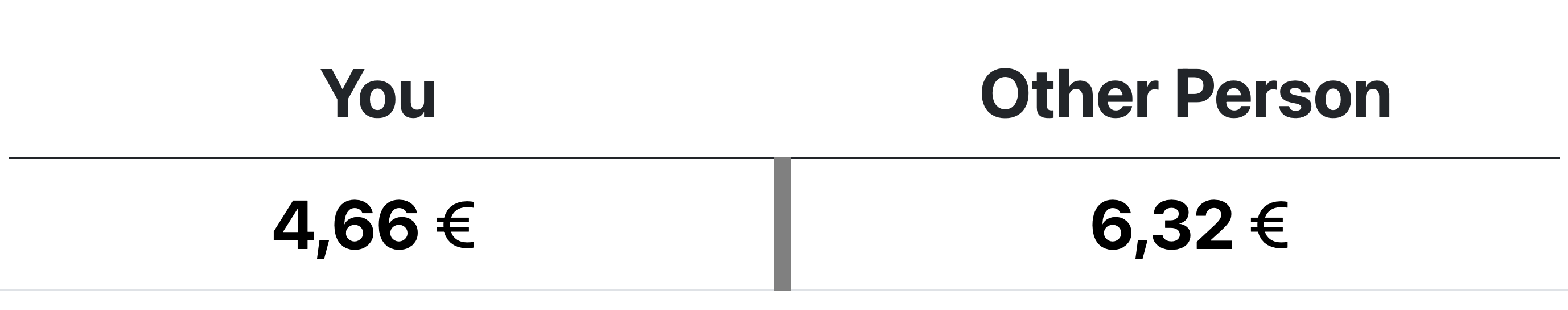}
		\centering\footnotesize
		\textbf{(a)} Altruism Baseline
        \end{subfigure}
        \hfil
        \begin{subfigure}[b]{0.475\textwidth}  
            \centering 
   \includegraphics[width=1\linewidth]{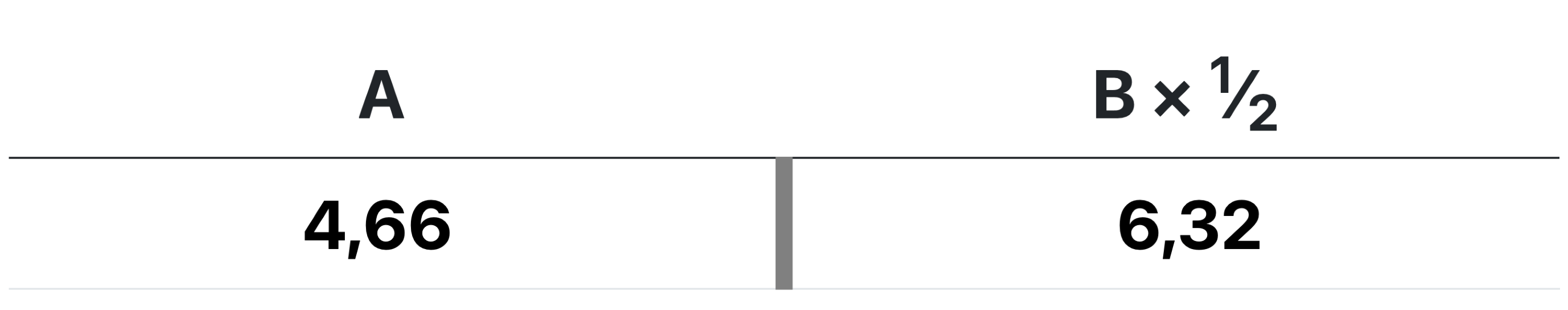}
		\centering\footnotesize
		\textbf{(c)} Number Comparison Baseline
        \end{subfigure}
        \vskip\baselineskip
        \begin{subfigure}[b]{0.475\textwidth}   
            \centering 
   \includegraphics[width=1\linewidth]{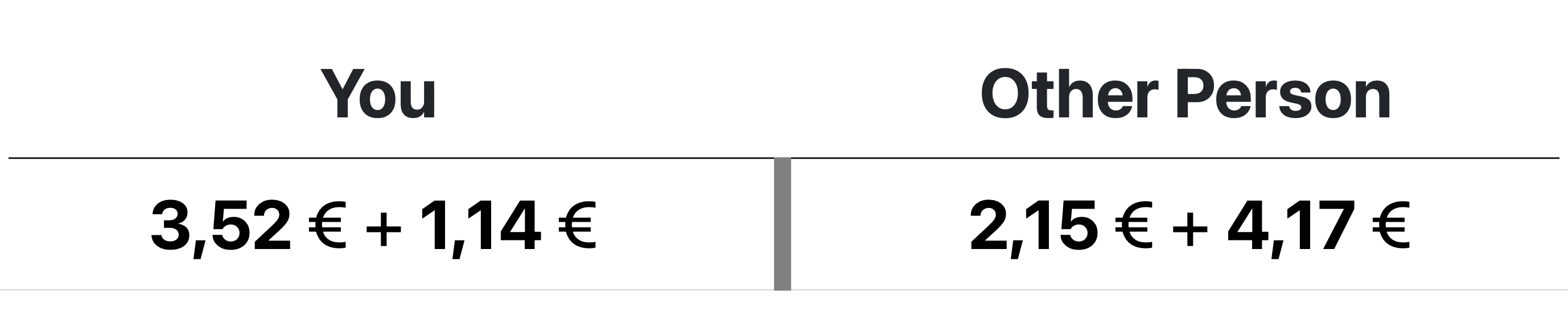}
		\centering\footnotesize
		\textbf{(b)} Altruism Treatment
        \end{subfigure}
        \hfill
        \begin{subfigure}[b]{0.475\textwidth}   
            \centering 
   \includegraphics[width=1\linewidth]{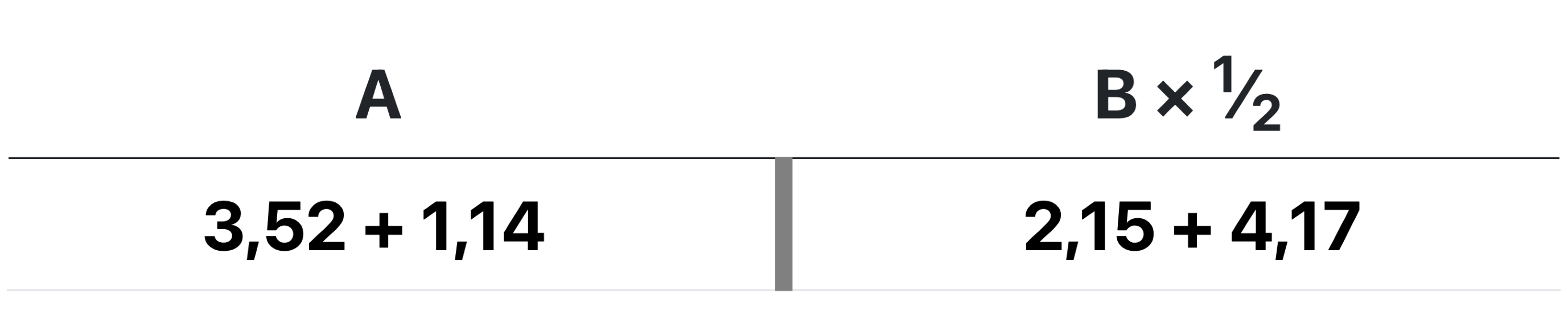}
		\centering\footnotesize
		\textbf{(d)} Number Comparison Treatment
        \end{subfigure}
        \caption[Altruistic Choice and Number Comparison Task]
        {\small Altruistic Choice and Number Comparison Task \quad \textbf{(a)} Decision screen of the Baseline condition featuring a decision between taking a payment $\text{self}$ or giving a payment $\text{other}$. \textbf{(b)} Decision screen of Treatment condition, in which to-be-calculated sums replace monetary values. \textbf{(c)} Baseline condition in the number comparison task. \textbf{(d)} Treatment condition number comparison. Participants choose using the ``a'' (``You''/``A'') and ``l'' key (``Other Person''/``B'') on the (German) keyboard.}\label{fig_experimental_screens}
    \end{figure*}

In line with the theoretical setup, the experiment centers around the following decision: taking a monetary payment $\text{self}$ (and giving nothing) or giving a monetary amount $\text{other}$ to another person (and taking nothing) as depicted in panel (a) of Figure \ref{fig_experimental_screens}. By varying the respective payments of this choice, I can infer a subject's altruistic preference. More specifically, (in the absence of noise) choices should be characterized by a unique switching point, the maximum amount of $\text{self}$ a participant is willing to forego to increase the other person's payment by $\text{other}$. I vary the monetary payments of $\text{self}$ and $\text{other}$ as follows: I choose four distinct values for $\text{other}_{k}$: 6.55 \euro, 9.26 \euro, 13.10 \euro, and 18.52 \euro \footnote{Note that these values follow a series similar to the stakes in \textcite{khawCognitiveImprecisionSmallStakes2021a} as the ratio between each adjacent element in the series is a constant, i.e., $\sqrt{2}$.} and calculate the indifference value $\text{self}_{indiff} \sim \frac{\beta}{1-\beta} \times \text{other}_{k}$ $\forall$ $\beta \in [0,0.05,...,0.55]$ for all four values of $\text{other}_k$. This results in $4 \times 12$ unique combinations of $\text{self}$ and $\text{other}$ (see Figure \ref{fig:trials_overview} for an illustration). Each of these combinations is repeated five times and I call a group of five identical trials a ``game'' throughout. Overall, subjects faced 48 games, i.e., 240 trials, in the altruism choice task of the experiment (with intermediate breaks). Following \textcite{khawCognitiveImprecisionSmallStakes2021a} I use payments including cent values to encourage participants to approach the decisions more approximatively.\footnote{A critique of this approach could be that this leads participants to only focus on the main digit of the payments and simply ignore the cent values. While this would be in line with an extreme form of ``left-digit-bias'', more recent psychological research -- using eye-tracking techniques -- suggests that people often pay as much attention to cents as they do to euros \autocite{laurentHowConsumersRead2023}. Note also that e.g., \textcite[708]{dehaeneCognitiveEuroscienceScalar2002} explicitly avoid round numbers in their stimuli, which are prices of different items.}

At the end of the experiment, one trial is randomly drawn and implemented. Each participant is matched to a person in their session to send their chosen payment of $\text{other}$ and to another person to receive the other person's choice of $\text{other}$. While the matching of the sender to the recipient is randomly determined, no participant can send to and receive from the same person and participants are instructed accordingly. Before making the 240 decisions, participants familiarize themselves with one interactive example of the choice, answer a series of comprehension questions, and encounter 12 practice trials, which are not payoff-relevant and thus remain excluded from the analyses.

\paragraph{Treatment Condition}

In the between-subject treatment condition, \textit{to-be-calculated sums} replace the monetary payments, as shown in panel (b) of Figure \ref{fig_experimental_screens}. Inspired by a variation in \textcite{enkeComplexityHyperbolicDiscounting2023}, the main objective of this condition is to increase the ``cognitive difficulty'' of making altruistic decisions. By \textit{disaggregating} monetary payments into two components, information processing cost increases, which in turn may lead to mis-valuation of true incentives (see \textcite{opreaDecisionsRiskAre2024} for a discussion originally about lotteries). This condition thus aims to reduce the informativeness of the mental signals.\footnote{An analogy to paradigms from cognitive science can also be drawn: For modeling human vision, models of Bayesian observers that integrate noisy visual perceptions with their prior beliefs are very successful in explaining behavior. For example, experiments show that people perceive moving objects as \textit{slower} if the contrast of the visual stimuli is low compared to stimuli with higher contrast, while the actual velocity of the object remains unchanged. This, in turn, is interpreted as evidence that people have a prior belief that things move more slowly \autocite{stockerNoiseCharacteristicsPrior2006, weissMotionIllusionsOptimal2002}.} While one could well expect this variation to impact $\nu_{\frac{\text{self}}{\text{other}}}$ and $\nu_{\frac{\beta}{1-\beta}}$ differently, I leave this as an open empirical question. In terms of the design of the variation, I choose sums as relatively simple mathematical operations to allow participants to still reasonably engage in the repeated trials of the experiment and be able to gauge the values of the monetary payments (i.e., not reducing the informativeness too much). 

The to-be-calculated sums are randomly determined but constructed systematically: I first (uniformly) draw a random number between 0 and the smaller number of the $\text{self}$, $\text{other}$ pair. In the example, I drew $\text{self}_1$ = 3.52 \euro \ from a range between 0 and 4.66 \euro. $\text{self}_1$ then serves as an upper bound for a second random draw, $\text{other}_1$, i.e., 2.15 \euro \ in the example. Both determine $\text{self}_2$ and $\text{other}_2$, the complements of the sums (i.e., 1.14 \euro \ and 4.17 \euro). This specific procedure ensures that no matter the underlying numerical relationship between $\text{self}$ and $\text{other}$, one component of any of the two sums is larger than another component of the other sum and vice versa (e.g., in the example $\text{self}_1 > \text{other}_1$, yet $\text{other} > \text{self}$). Furthermore, the \textit{position} of $\text{self}_1, \text{self}_2$ and $\text{other}_1, \text{other}_2$ is randomly shuffled for each participant individually. This procedure encourages paying attention to all four components in all trials. Further, it hinders the possibility of gauging the underlying value of $\text{self}$ or $\text{other}$ by just focusing on the positions of the components. Table \ref{tab:trials_overview} provides the complete overview of all 240 trials, including the values for $\text{self}_1, \text{self}_2$ and $\text{other}_1, \text{other}_2$, which remain fixed for all participants, yet presented in random order in the experiment.

At the end of Part 1, I gather self-reported data on subjective confidence, how precisely participants calculated during the decisions, and the attention paid to both the values of $\text{self}$ and $\text{other}$ (see Figure \ref{fig_screenshots_meta_cognition} for screenshots).

\subsection{Part 2: Number Comparison}

\paragraph{Number Comparison Task} Part 2 of the experiment features a number comparison task. Participants have to assess which of two columns is numerically larger, either A or B $\times \sfrac{1}{2}$ (see panel (c) in Figure \ref{fig_experimental_screens}). This task features an objectively correct solution (A in the example) while aiming to mirror the ``mechanics'' (or ``mental arithmetics'') of the altruism decisions -- comparing two numbers -- as closely as possible. The term $\sfrac{1}{2}$ replaces the threshold previously determined by each subject's $\beta$ parameter (i.e., their altruistic preference) with an objective and common factor, which in turn is assumed not to give rise to a noisy mental representation, but to remain accurately perceived. This task is inspired by recent work in economics showing a correlation between elementary economic behavior and equivalent number perception \autocite{frydmanEfficientCodingRisky2022, frydmanCoordinationCognitiveNoise2023}.

Importantly, the values of $\text{A}$ ($\text{B}$) are \textit{identical} to those used previously for $\text{self}$ ($\text{other}$). Again, each unique combination of $\text{A}$,$\text{B}$ was repeated five times. To reduce redundancy, I omit the pairs where $\text{A}=0$ and $\text{A}>\text{B}$ in the number comparison task, such that subjects made 200 decisions in total (in 40 unique games). While the Baseline group interacts with the task as depicted in panel (c), the Treatment group again features to-be-calculated sums instead of the numerical values (d).

Similar to the number discrimination task in \textcite{frydmanEfficientCodingRisky2022}, I incentivize this task to reward both speed and accuracy: After the end of Part 2, I calculate the share of correct solutions and the average time in seconds participants took. I then determine their earnings: $10$ \euro \ $\times$ Avg. correct $-$ Avg. time in seconds. Participants thus could earn at most $10$ \euro \ if they solved every task correctly and took 0 seconds on average. Their reward was reduced for each additional second or a lower percentage of correct solutions.\footnote{I chose to implement a time-sensitive incentivization as the task would be much more trivial to solve otherwise. Section \ref{sec:decision_times} shows that the average time participants spent on the number task and the altruism decisions is identical in the baseline and even larger in the Treatment group. I read this as evidence against an argument that participants significantly decided much faster in the number comparison (which might invoke different cognition strategies) than in the altruism decisions. Note also that while participants were effectively put under time pressure, there was no active reminder of their current time usage, which should help prevent high levels of perceived time pressure.}

At the end of Part 2, I elicit beliefs about both participants' number of correct answers and the average amount of seconds they took. One of the belief elicitations was drawn randomly and determined if an additional bonus prize of 1 EUR pays out at the end according to the randomized quadratic scoring rule \autocite{schlagElicitingProbabilitiesMeans2013, hossainBinarizedScoringRule2013}.

\subsection{Part 3: Additional Data Collection}

Finally, Part 3 collects several additional data from participants, which can be grouped into three different categories: (i) cognitive ability, (ii) norms and excuses, and (iii) pro-sociality and demographics.

Participants in the experiment have to answer six questions of the extended Cognitive Reflection Test (CRT) by \textcite{toplakAssessingMiserlyInformation2014}, which entails the original three CRT questions of \textcite{frederickCognitiveReflectionDecision2005a} and adds questions similar in formulation. 
 Figure \ref{screenshot:crt} shows a screenshot of CRT4. One of the six questions is drawn randomly and awards a bonus of 1\euro \ if answered correctly. In addition to the CRT, I conduct the three-question Berlin Numeracy Test \autocite{cokelyMeasuringRiskLiteracy2012} (unincentivized) and gather survey data on the deliberation-intuition scale \autocite{betschPraferenzFurIntuition2004} and the German short-version of the Need for Cognition scale \autocite{cacioppoNeedCognition1982} developed by \textcite{beissertDeutschsprachigeKurzskalaZur2014}. I choose a more extensive set of cognition-related measures to compare the standard CRT questions to alternative measures related to cognitive ability.\footnote{E.g., \textcite{schunkExplainingHeterogeneityUtility2006} show that \textit{self-reported} measures of a preference for deliberative versus intuitive reasoning correlate with individual estimates of utility function parameters.}

The next block of additional data measures private and social norms and two additional survey questions about excuse-taking and (non-)altruistic behavior. I elicit social and private norms regarding behavior in the altruism task in the style of \textcite{krupkaIdentifyingSocialNorms2013}, albeit in a non-incentivized way.\footnote{\textcite{konig-kerstingRobustnessSocialNorm2024} does not identify differences in responses between (non-)incentivizing social norm elicitation in a large-scale experiment.} I show participants from the Baseline and Treatment group an example \textit{both} in the Baseline and Treatment format in randomized order and ask for the subjective appropriateness of the decision (see Figure \ref{screenshot:social_norms} for a screenshot). I also elicit survey questions related to excuse-taking (see Figure \ref{screenshot:survey_excuses}).\footnote{Based on the arguments in \textcite{danaExploitingMoralWiggle2007} and \textcite{exleyMotivatedErrors2024}, the treatment variation could also introduce a ``wiggle-room'' which allows participants to make self-serving miscalculations and thereby justify more selfish behavior.} 


The third block consists of several additional measures. In a simple dictator game, each participant decides how to split 10 \euro \ between themselves and another randomly determined person (see Fig \ref{screenshot:simple_dictator} for a screenshot). I instruct participants that their choice is implemented with a chance of 1\%. Additionally, I obtain answers to the qualitative survey items of the Global Preferences Survey \autocite{falkPreferenceSurveyModule2023}, a visual-analog fatigue scale \autocite{Radbruch2003}, as well as basic demographic information.

\subsection{Implementation}

The experiment ran in January 2023 at the MABELLA lab with 300 student subjects. Each subject was randomly allocated to the Baseline or the Treatment condition within an experimental session (until 150 were in each condition). As stated, subjects could earn rewards from all three parts of the experiment, and the average payment was 16.15\euro. The mean completion time stood at 62 minutes, while the overall session duration averaged 82 minutes, as participants had to wait until everyone in their session was finished. Instructions were presented on-screen and key screens are depicted in Appendix \ref{sec:exp_screenshots} (translated from German). The pre-registration is available at \url{https://aspredicted.org/blind.php?x=5F4_72D}. The joint ethics board of Goethe University Frankfurt and JGU Mainz provided the IRB approval.

\section{Results}\label{sec:results}

This section presents the empirical results, first focusing on differences between the Baseline and Treatment group in altruistic choices and number comparison behavior. Afterward, I discuss the additional hypotheses next to the impact of increasing cognitive noise.

\subsection{Altruistic Choices: Descriptives}

Figure \ref{fig:results_main_treatment_effects} presents the average choice for $\text{self}$ for each unique value of $\frac{\text{self}}{\text{other}}$ featured in the experiment, separately drawn for the Baseline and Treatment group. The Baseline data offers several insights into participants' altruistic preferences: First, unsurprisingly, the larger the payment $\text{self}$ compared to $\text{other}$, the more frequently subjects choose $\text{self}$: If $\text{self} = 0$, only 1,03 \% of choices correspond to $\text{self}$, whereas, if $\text{self} > \text{other}$, 98,4 \% of choices correspond to $\text{self}$. People positively care about the other person's payment, yet more strongly about their own and only very few choices are consistent with spiteful preferences. A local linear interpolation indicates that the Baseline group is indifferent (i.e., the average choice for $\text{self}$ equalling 50\%) if $\frac{\text{self}}{\text{other}} = 0.474$, which implies that participants roughly care twice as much about their payoff compared to the payoff of another person.

\begin{figure}[h]

{\centering

\resizebox{0.9\linewidth}{!}{%
\input{graphs_export/results/main_treatment_effects}}

\mycaption[Altruistic Choices in Baseline and Treatment Group]{\quad This plot shows the association between average choice for $\text{self}$ and distinct values of the ratio $\frac{\text{self}}{\text{other}}$, separately drawn for the Baseline and Treatment group, with 95\% confidence intervals.}\label{fig:results_main_treatment_effects}
} 
\end{figure}

This behavior largely aligns with previous evidence on (structural estimates of) social preferences, primarily that of advantageous inequality/aheadness aversion. There, the aheadness aversion parameter can be similarly interpreted as the $\beta$ parameter in the present framework as the weight a DM places on the well-being of another person (given the DM is better off).\footnote{Note that the framework developed here does not distinguish between being ahead and being behind as, by construction, a subject is ahead if they choose $\text{self}$ and behind if they choose $\text{other}$. Thus, the present setup does not allow for separating these two motivations; instead, it comprises them into one. The fact that subjects overall substantially weigh the other person's payment could also be related to how the decision in Figure \ref{fig_experimental_screens} is displayed, i.e., not including the 0 \euro \ consequence for either person.} Reviewing over 40 articles, \textcite{nunnariMetaAnalysisDistributionalPreferences2024} report a median value for the advantageous inequality aversion of 0.26, indicating that participants often roughly care thrice as much about their payment compared to other people's when ahead, which is in line what, e.g., \textcite{bruhinManyFacesHuman2019a} find. More similar to participants here, \textcite{carpenterMeasuringSociallyAppropriate2024}, \textcite{vonschenkSocialPreferencesHumans2023} and \textcite{klockmannArtificialIntelligenceEthics2022} estimate values that correspond to their subjects caring roughly twice as much about their payment compared to that of another participant. 


In the Treatment condition, these statements about altruistic preferences remain largely true, albeit with subtle differences: First, the association between the average choice for $\text{self}$ and changes in the underlying ratio of $\frac{\text{self}}{\text{other}}$ is \textit{flatter} compared to the Baseline condition. For small values of $\frac{\text{self}}{\text{other}}$, the Treatment group decides more often for $\text{self}$, e.g., 3,2 \% of choices correspond to $\text{self}$ if $\text{self} = 0$, yet less often for larger values of $\frac{\text{self}}{\text{other}}$ as only 95,53 \% of choices correspond to $\text{self}$ if $\text{self} > \text{other}$. Furthermore, over the entire set of trials, the Treatment group behaves less selfishly: While the Baseline group decided in 45,18\% of choices for $\text{self}$, the Treatment group chose $\text{self}$ in 42,93 \% of the cases. This difference is statistically significant, as indicated by a two-sided $t$ and a Fisher exact test (both $p<0.001$). The ratio of $\frac{\text{self}}{\text{other}}$ required for indifference in the Treatment group corresponds to $0.528$, a 5.4 percentage points larger ratio compared to the Baseline. Using a linear probability model, Table \ref{tab:reg_main} confirms that both the overall level of choices for $\text{self}$ is 2.2 percentage points lower and that an increase in $\frac{\text{self}}{\text{other}}$ by 1 has a 6.7 percentage points lower effect on choices for $\text{self}$ in the Treatment group (for a Probit model, see Table \ref{tab:reg_main_probit}). Although their underlying preference should remain the same as in the Baseline, the Treatment group shows a dampened reaction to changes in incentives and chooses $\text{self}$ significantly less frequently, i.e., behaves more altruistically.

\paragraph{Result 1:}\textit{The Treatment group shows both a flatter association between changes in payments and choices and is more altruistic compared to the Baseline group.} \newline 

Both a flatter association between varying payments and choices and a \textit{bias} towards more altruistic choices can be rationalized in light of the theoretical framework.\footnote{Section \ref{sec:appendix_robustness_treatment} discusses if the chosen treatment variation might have invoked behavior other than an increase in cognitive noise. The difference between Treatment and Baseline can not be explained by (i) an exclusive focus on (and comparison of) the first component of the sums (Figure \ref{fig:stakes_and_math}) or (ii) the treatment only working for larger numbers (Figure \ref{fig:stakes_and_math} and Table \ref{tab:reg_stakes}).
} Recall Figure \ref{fig:theory_noise_plot}, which outlines the impact of an increase in noise on the probability of choosing $\text{self}$ (equation \ref{eq:choice_function}). The treatment effect could originate from either an increase in $\nu_{\frac{\text{self}}{\text{other}}}$ ($H_{1b}$) or an increase in $\nu_{\frac{\beta}{1-\beta}}$ ($H_{1c}$), i.e., either through an increase in ``noise in payments'' coupled with an additional adjustment towards an intermediate payment prior or a ``mechanical'' increase in ``noise in preferences''.



\subsection{Altruistic Choices: Probabilistic Model}\label{sec:structural_estimates_altruism}

I now turn to probabilistic modeling to estimate the (posterior) probability of the parameter values given the experimental data and investigate the mechanisms of the treatment effect. Given the choice model formulated in equation \ref{eq:choice_function}, this approach allows to (i) inspect whether (an increase in) noise in payments or preferences dominates the other, (ii) infer the parameters of the prior of monetary payments, and thereby (iii) test the potential mechanisms of the treatment effect. I use Bayesian estimation techniques, which are gaining popularity in experimental economics (see \textcite{blandBayesianModelSelection2023} for an overview and the tutorial by \nptextcite{vieiderBayesianEstimationDecision2024}). The main reason to use Bayesian techniques lies in their practicality: Because they are more flexible than, e.g., maximum likelihood estimation, they can deal more easily with more complex models and still produce meaningful uncertainty estimates of the parameters of the model \autocite[4]{gelmanBayesianDataAnalysis2021}. Here, I estimate a \textit{Bayesian Hierarchical Model} that determines the prior for the individual parameter values from the data. In hierarchical models, individual parameter estimates are partially pooled towards the group mean, which reduces overfitting and thus increases out-of-sample performance \autocite{kruschkeDoingBayesianData2015}. Furthermore, the hierarchical setup allows us to represent potential treatment differences in specific parameters by allowing (some) hyper-parameters to differ between conditions $c$. More specifically, the hierarchical model assumes that the individual parameter vector $\bm{\theta_i} = \left(\nu_{\frac{\text{self}}{\text{other}},i},\nu_{\frac{\beta}{1-\beta},i}, \beta_i, \mu_{\hat{r},i}, \right)$ of individual $i$ is -- on the log-scale -- drawn from a multivariate normal distribution:

\begin{equation}
\bm{\theta_i} \sim \mathcal{N}(\bm{\mu}, \bm{\Sigma})
\end{equation}

where $\bm{\mu} = (\mu^{\nu_{\frac{\text{self}}{\text{other}}}}_c, \mu^{\nu_{\frac{\beta}{1-\beta}}}_c, \mu^{\frac{\beta}{1-\beta}},\mu^{\mu_{\hat{r}}})$ is the vector of the population-means of the parameter distributions. Note that both
$\mu^{\nu_{\frac{\text{self}}{\text{other}}}}_c$ and $\mu^{\nu_{\frac{\beta}{1-\beta}}}_c$ are allowed to differ between Baseline and Treatment group $\mu^{\nu_{\frac{\text{self}}{\text{other}}}}_B$, $\mu^{\nu_{\frac{\text{self}}{\text{other}}}}_T$, $\mu^{\nu_{\frac{\beta}{1-\beta}}}_B$, $\mu^{\nu_{\frac{\beta}{1-\beta}}}_T$ representing potential treatment differences along both types of cognitive noise. All other hyper-parameters remain identical across conditions. $\bm{\Sigma} = \text{diag}(\bm{\tau})\Omega \text{diag}(\bm{\tau})$, where $\Omega$ is the correlation matrix of individual parameters and $\bm{\tau}$ is a vector of standard deviations. Note that, without loss of generality, I set $\sigma_{\hat{r},i} = 1 \, \forall \, i$ (\nptextcite[p.\ 32][]{opreaMindingGapOrigins2024}; \nptextcite{natenzonRandomChoiceLearning2019}). The hierarchical model requires specifying prior distributions for all hyper-parameters and I choose weakly informative priors (see Section \ref{sec:appendix_prob_model_details} for details and prior predictive checks, also see \nptextcite{gelmanStanProbabilisticProgramming2015}). I estimate the model with Numpyro \autocite{bingham2019pyro,phan2019composable}.


%
%
%

\begin{figure}[h]
        \centering
        \begin{subfigure}{0.56\textwidth}  
            \centering 
                  \resizebox{1\linewidth}{!}{%

\begin{tabular}[b]{lllllll}
\toprule
  & mean & median & sd & hdi 2.5\% & hdi 97.5\% & $\hat{R}$\\
\midrule
\textit{Base Parameters:} &  &  &  &  &  & \\
Altr. Preference $\beta$ & 0.316 & 0.316 & 0.008 & 0.3 & 0.332 & 1\\
Prior Mean Monetary Payments $\mu_{\hat{r}}$ & 1.049 & 1.005 & 0.304 & 0.533 & 1.652 & 1.01\\
\addlinespace
\textit{Group Specific:} &  &  &  &  &  & \\
Noise Baseline $\nu_{\frac{\beta}{1-\beta},B}$ & 0.186 & 0.185 & 0.022 & 0.143 & 0.23 & 1\\
Noise Treatment $\nu_{\frac{\beta}{1-\beta},T}$ & 0.181 & 0.18 & 0.032 & 0.121 & 0.243 & 1\\
Noise Baseline $\nu_{\frac{\text{self}}{\text{other}},B}$ & 0.313 & 0.311 & 0.028 & 0.259 & 0.367 & 1\\
Noise Treatment $\nu_{\frac{\text{self}}{\text{other}},T}$ & 0.399 & 0.398 & 0.038 & 0.325 & 0.475 & 1\\
Weight on Payments Baseline $\alpha_B$ & 0.911 & 0.912 & 0.014 & 0.882 & 0.938 & 1\\
Weight on Payments Treatment $\alpha_T$ & 0.862 & 0.863 & 0.023 & 0.817 & 0.905 & 1\\
Prior Threshold Baseline $\delta_B$ & 1 & 1 & 0.025 & 0.951 & 1.05 & 1.01\\
Prior Threshold Treatment $\delta_T$ & 1 & 0.999 & 0.039 & 0.925 & 1.076 & 1.01\\
\bottomrule
\end{tabular}

        }
		\centering\footnotesize
		\textbf{(a)} Posterior parameter summary
        \end{subfigure}
        \hfill
        \begin{subfigure}{0.415\textwidth}   
            \centering 
      \resizebox{1\linewidth}{!}{%
        \input{graphs_export_restricted/structural_estimation/plot_predictions_manual}}
		\centering\footnotesize
		\textbf{(b)} Average and predicted choices
        \end{subfigure}
        \caption[Summary Probabilistic Model Altruistic Choices]
        {\small Summary Probabilistic Model Altruistic Choices \quad \textbf{(a)} Estimated parameter values of equation \ref{eq:choice_function} based on 10000 posterior samples (+ 1000 warmup) per each of four chains. Parameters correspond to the mean of log-normal hyper-distributions. Mean, median and sd refer to the mean, median and standard deviation of the posterior distribution samples. HDI 2.5\% and HDI 97.5\% indicate the borders of the 95\% highest-density interval (HDI). $\hat{R}$ is a diagnostic of convergence of the Markov chains ($\hat{R} = 1 $ indicating convergence). \textbf{(b)} Average (over individuals) and predicted choices, including 95\% HDI.         
         }\label{fig:model_summary}
    \end{figure}

The Table in panel (a) of Figure \ref{fig:model_summary} summarizes the parameters of the model and (b) plots average and predicted choices (including the 95 \% HDI). Given the hierarchical nature of the model, I inspect parameters on the population level, i.e., the mean of the log-normal hyper-distribution of a given parameter (instead of average individual parameters that would assign equal weight to each participant).\footnote{The accompanying online appendix plots the individual choice curves and the individual data for each subject: \url{https://nmwitzig.github.io/noise-app.html}} The table contains the mean, median and standard deviation of the posterior samples of the respective parameter, the 95 \% credible interval, the shortest interval containing 95 \% of probability mass as well as the $\hat{R}$ convergence diagnostic \autocite{vehtariRanknormalizationFoldingLocalization2021} with $\hat{R} < 1.05$ often considered as necessary condition.


I first focus on the ``base parameters'', i.e., parameters that do not differ by treatment group. First, the altruistic preference parameter $\beta = 0.316 \; [0.3 - 0.332]$ aligns with the behavior described previously: on average, participants weigh approximately their payment twice as important as the other person's payment. The probabilistic model further yields an estimate for the mean of the prior distribution of monetary payments $\mu_{\hat{r}} = 1.049 \;[0.533 - 1.652]$, which corresponds -- on average over the posterior distribution -- to an ``ignorance'' intuition as mentioned previously, i.e., that participants intuitively do not distinguish between $\widehat{\text{self}}$ and $\widehat{\text{other}}$. However, note the large degree of uncertainty of this estimate with a 95 \% probability that $\mu_{\hat{r}}$ is between 0.533 and 1.652. This will be important for discussing potential mechanisms of the treatment effect should noise in monetary payment perception be higher in the Treatment group.



This leads to the analysis of group-specific parameters. Recall that the primary goal of the chosen treatment variation was to increase noise levels in the Treatment group, but without pre-specifying if the variation would influence $\nu_{\frac{\beta}{1-\beta}}$ or $\nu_{\frac{\text{self}}{\text{other}}}$ more strongly. Regarding $\nu_{\frac{\beta}{1-\beta}}$, the probabilistic model does not indicate a difference between the Baseline and Treatment group with overall very similar levels of $\nu_{\frac{\beta}{1-\beta},B} = 0.186 \; [0.143 - 0.23]$ and $\nu_{\frac{\beta}{1-\beta},T} = 0.181 \; [0.121 - 0.243]$. Based on the posterior samples, I can also directly calculate probabilistic statements about a potential difference, which indicates that $P(\nu_{\frac{\beta}{1-\beta},B} < \nu_{\frac{\beta}{1-\beta},T}) = 0.434$, confirming the conclusion of no difference. This is also supported by the hyper-parameters of the preference noise distribution (on the log scale) not differing between groups with $\mu^{\nu_{\frac{\beta}{1-\beta}}}_B \text{-} \; \mu^{\nu_{\frac{\beta}{1-\beta}}}_T = \text{-}0.033 \; [\text{-}0.371 - 0.301]$ (see Table \ref{tab:structural_estimation_population}). While there is, therefore no group difference, note that both $P(\nu_{\frac{\beta}{1-\beta},B} > 0) = 1$ and $P(\nu_{\frac{\beta}{1-\beta},T} > 0) = 1$. This implies that altruistic preferences are not perceived ``noiselessly'', yet the chosen treatment variation -- encapsulating monetary payments in to-be-calculated sums -- did not affect (the noise of) this perception.

This is in contrast to noise in perceiving monetary payments: Here, the probabilistic model suggests that the treatment variation did, in fact increase noise levels: $\nu_{\frac{\text{self}}{\text{other}},B} = 0.313 \; [0.259 - 0.367]$, $\nu_{\frac{\text{self}}{\text{other}},T} = 0.399 \; [0.325 - 0.475]$ and $\mu^{\nu_{\frac{\text{self}}{\text{other}}}}_T \text{-} \; \mu^{\nu_{\frac{\text{self}}{\text{other}}}}_B = 0.244 \; [0.045 - 0.445]$ (see Table \ref{tab:structural_estimation_population}). 
Higher noise levels in the Treatment group translate into lower values of $\alpha$, i.e., 
$\alpha_T = 0.862 \; [0.817 - 0.905]$ and $\alpha_B = 0.911 \; [0.882 - 0.938]$. Moreover, with $P(\nu_{\frac{\text{self}}{\text{other}},B} > \nu_{\frac{\beta}{1-\beta},B}) = 0.998$ and $P(\nu_{\frac{\text{self}}{\text{other}},T} > \nu_{\frac{\beta}{1-\beta},T}) = 0.999$,  noise in perceiving monetary payments is also generally higher compared to noise in preferences in addition to the larger group differences.

This leads to a discussion on the origin of the treatment effect, i.e., which hypothesis is most supported by the data. Larger group differences in noise levels in perceiving monetary payments strongly suggest that the treatment variation -- and, in turn, the mechanism of the observed treatment effect -- operates through differences in monetary payment perception. In particular, higher levels of $\nu_{\frac{\text{self}}{\text{other}},T}$ leading to \textit{fewer} choices for $\text{self}$ suggest $H_{1b}$ as a candidate hypothesis and a partial adjustment towards some ``intermediate'' payment perception as the driver of treatment differences. However, the probabilistic model does not conclusively support this conclusion: With the large degree of uncertainty surrounding $\mu_{\hat{r}}$ and the fact that both $\delta_B = 1 \; [0.951 - 1.05]$, $\delta_T = 1 \; [0.925 - 1.076]$ and $P(\delta_T > \delta_B) = 0.496$, i.e., no group difference in the prior-induced threshold despite higher noise levels in the Treatment group, the experimental data does not provide sufficient support for a strong claim in favor of $H_{1b}$.

This becomes more evident following a model comparison in Figure \ref{fig:model_comparison}. There, I compare the predictive power of the ``full'' model (equation \ref{eq:choice_function}) with various simpler variants, each one abstracting from a potential mechanism of the treatment effect.\footnote{More specifically, I formulate models that either set (i) the payment prior mean $\mu_{\hat{r}} = 1$ or (ii) the noise in altruistic preferences $\nu_{\frac{\beta}{1-\beta}} = 0$, or (iii) the noise in monetary payments  $\nu_{\frac{\text{self}}{\text{other}}} = 0$. I also include a (iv) standard random utility benchmark for reference. I inspect $ELPD_{WAIC}$ values which measure the goodness-of-fit minus a model complexity penalty \autocite{watanabeWidelyApplicableBayesian2013} and provide a computationally less demanding approximation to leave-one-out out-of-sample prediction accuracy \autocite{vehtariPracticalBayesianModel2017} while accommodating model uncertainty.} While the ``full'' model provides the highest goodness-of-fit among all models considered, the differences compared to the simpler models are relatively minuscule.\footnote{With all variants of the ``full model'' outperforming the standard random-utility benchmark -- supporting the overall modeling approach.} Importantly, the simpler variants include a model that only incorporates noise in altruistic preferences, which performs nearly as well as the ``full model''. This, in turn, prohibits discarding $H_{1c}$ entirely.\footnote{More concretely, this result suggests that the possibility remains that 
the mechanism of the treatment effect still operates through an increase in noise in preferences, yet that this increase cannot be conclusively identified and separated from an increase in noise in monetary payment perception in the ``full'' model. This also fits the relatively small differences in the \textit{indifference value} between Baseline and Treatment (Figure \ref{fig:results_main_treatment_effects}).}

\paragraph{Result 2:}\textit{The exact origin of the treatment effect in altruistic choices, i.e., its mechanism, cannot be conclusively identified based on the altruism data alone.} \newline 
 
 Aside from the unclear mechanisms of the treatment effect, the probabilistic model nonetheless captures average and individual behavior well. The average predicted choice curve in panel (b) shows a strong overlap between average and predicted choices, with comparably tight HDI. In Figure \ref{fig:scatter_avg_prediction}, I further plot individual average and predicted choices with a rank-correlation of $\rho = 0.99$ between predicted and actual individual average decisions. Overall, these results thus support the general modeling approach.

\subsection{Number Comparison Task: Theory and Descriptives}\label{sec:structural_estimates_number_comp}

As stated, the evidence thus far does not conclusively inform about the origin of the treatment effect in altruistic choices. Therefore, I further investigate the mechanism of the treatment variation with the data from the number comparison task. This data is insightful as the choice in the number comparison task share similar ``mechanics'' with the altruistic choices (comparing two numbers), while abstracting from any subjective altruistic preference. Akin to equation \ref{eq:choice_function}, the choices in the number comparison task can be understood as a result of the following choice function:

\begin{equation}\label{eq:choice_function_number}
{Pr}[(A \succ B \times \sfrac{1}{2}]=\Phi\left(\frac{\alpha^{'} \times \ln \left(\frac{\text{A}}{\text{B}}\right)- \ln (\frac{1}{2}) -\ln (\delta^{'})}{\nu_{\frac{\text{A}}{\text{B}}}^{2} \alpha^{'}}\right)
\end{equation}

where $\frac{\text{A}}{\text{B}}$ is the ratio of numbers $\text{A}$ and $\text{B}$ (previously $\text{self}$ and $\text{other}$), $\alpha^{'} = \frac{1}{1 + \nu_{\frac{\text{A}}{\text{B}}}^{2}}$ and $\delta' = \frac{1}{\mu_{\hat{r'}}^{1-\alpha^{'}}}$. Equation \ref{eq:choice_function_number} assumes that the term $\sfrac{1}{2}$ -- an objectively stated constant -- is perceived without noise (as opposed to the term $\frac{\beta}{1-\beta}$ in equation \ref{eq:choice_function}). Assuming that the treatment variation works similarly across domains (i.e., that the treatment impacts $\nu_{\frac{\text{self}}{\text{other}}}$ and $\nu_{\frac{\text{A}}{\text{B}}}$ similarly) and that this functional form is appropriate, investigating the number comparison data allows to compare between $H_{1b}$ and $H_{1c}$: if $H_{1b}$ is the driver behind the treatment effect, the treatment effect will be qualitatively similar in the number comparison task and an ``intermediate'' perception will lead to fewer choices for $\text{A}$. If, in contrast, $H_{1c}$ is the appropriate hypothesis (and in turn, an increase in noise in preferences dominated noise in monetary payments previously), an increase in noise will \textit{increase} choices for $\text{A}$, given $\mu_{\hat{r}^{'}} \geq 1$.


\begin{figure}[h]

{\centering

\resizebox{0.9\linewidth}{!}{%
\input{graphs_export/results/arithmetics_treatment_effects.tex}}

\mycaption[Number Comparison Baseline and Treatment group]{\quad This plot shows the association between average choice for $\text{A}$ and distinct values of the ratio $\frac{\text{A}}{\text{B}}$, separately drawn for the Baseline and Treatment group with 95\% confidence intervals around mean values.}\label{fig:results_arithmetics_treatment_effects}
} 
\end{figure}

The group differences in behavior in the number comparison task are shown in Figure \ref{fig:results_arithmetics_treatment_effects}, which plots the average choices for $\text{A}$ as a function of $\frac{\text{A}}{\text{B}}$, separately drawn for Baseline and Treatment. This data offers several insights: First, the Baseline group again shows a steeper association between choices and changes in the values of $\frac{\text{A}}{\text{B}}$ compared to the Treatment group. This translates into the Baseline group identifying the correct solution in 96,98\% of trials compared to 92.26 \% in the Treatment group ($p<0.001$).\footnote{To maximize payoffs, subjects should choose $\text{B}$ whenever $\frac{\text{A}}{\text{B}} < 0.5$ and choose $\text{A}$ whenever $\frac{\text{A}}{\text{B}} > 0.5$ (vertical red dashed line at $\frac{\text{A}}{\text{B}} = 0.5$).} The Treatment group also decides less often for $\text{A}$ (i.e., thus errs asymmetrically): $\bar{A}_B = 0.388$, $\bar{A}_T = 0.351 (p<0.001)$. Both observations are confirmed by a linear probability model in Table \ref{tab:reg_number}, which tells that the Treatment group decides 3.7 percentage points less for $\text{A}$ and an increase in $\frac{\text{A}}{\text{B}}$ by 1 has a 9 percentage points lower effect in the Treatment group compared to the Baseline (similar conclusions are drawn based on a Probit model in Table \ref{tab:reg_number_probit}). Another apparent observation is that choices are much more \textit{consistent} in the number comparison task than altruistic choices. This is unsurprising given that the common threshold of $\sfrac{1}{2}$ replaces an individual-specific preference threshold, which eliminates choice differences due to individual heterogeneity in altruistic preferences and noise in its perception.

\paragraph{Result 3:}\textit{In the number comparison task, The Treatment group again shows a flatter association between changes in numerical magnitudes and choices and decides less often for $\text{A}$.} \newline 

Transporting these findings to the previous results in altruism choices -- and assuming the treatment variation works similarly across tasks -- a common explanation for the treatment effect in both groups would be the mechanism underlying $H_{1b}$: Participants rely on an intuition that $0<\frac{\widehat{\text{A}}}{\widehat{\text{B}}} < 1$ (and $0<\frac{\widehat{\text{self}}}{\widehat{\text{other}}} < 1$), which in particular biases the perception of larger ratios downwards and turns the Treatment group towards fewer choices for $\text{A}$ and $\text{self}$. I now turn to a closer inspection of the mechanism, again using a probabilistic model.

\subsection{Number Comparison Task: Probabilistic Model}\label{sec:structural_estimates_number_comp2}

Equivalent to Section \ref{sec:structural_estimates_altruism}, I can estimate a probabilistic model based on the number comparison data and investigate the probability of the parameter values of equation \ref{eq:choice_function_number}. The estimated parameters are shown in panel (a) of Figure \ref{fig:model_summary_arithmetics}.

\begin{figure}[h	]
        \begin{subfigure}{0.615\textwidth}  
            \centering 
                  \resizebox{1\linewidth}{!}{%

\begin{tabular}[b]{lllllll}
\toprule
  & mean & median & sd & hdi 2.5\% & hdi 97.5\% & $\hat{R}$\\
\midrule
\textit{Base Parameters:} &  &  &  &  &  & \\
Prior Mean Num. Magnitudes $\mu_{\hat{r}'}$ & 0.515 & 0.51 & 0.056 & 0.411 & 0.625 & 1\\
\addlinespace
\textit{Group Specific:} &  &  &  &  &  & \\
Noise Baseline $\nu_{\frac{\text{A}}{\text{B}},B}$ & 0.198 & 0.197 & 0.009 & 0.18 & 0.216 & 1\\
Noise Treatment $\nu_{\frac{\text{A}}{\text{B}},T}$ & 0.286 & 0.286 & 0.014 & 0.26 & 0.315 & 1\\
Weight on Payments Baseline $\alpha_{B}^{'}$ & 0.962 & 0.962 & 0.003 & 0.955 & 0.969 & 1\\
Weight on Payments Treatment $\alpha_{T}^{'}$ & 0.924 & 0.924 & 0.007 & 0.91 & 0.937 & 1\\
Prior Threshold Baseline $\delta_{B}^{'}$ & 1.026 & 1.026 & 0.004 & 1.017 & 1.034 & 1\\
Prior Threshold Treatment $\delta_{T}^{'}$ & 1.052 & 1.052 & 0.01 & 1.032 & 1.073 & 1\\
\bottomrule
\end{tabular}

        }
		\centering\footnotesize
		\textbf{(a)} Parameters posterior summary
        \end{subfigure}
\hfill
        \begin{subfigure}{0.375\textwidth}   
            \centering 
      \resizebox{1\linewidth}{!}{%
        \input{graphs_export_restricted/structural_estimation/plot_predictions_arithmetics.tex}}
		\centering\footnotesize
		\textbf{(b)} Average and predicted choices
        \end{subfigure}
        \caption[Summary Probabilistic Model Number Comparison]
        {\small \quad Model Summary Number Comparison \textbf{(a)} Estimated parameter values of equation \ref{eq:choice_function} based on 10000 posterior samples (+ 1000 warmup) per each of four chains. Parameters correspond to the mean of log-normal hyper-distributions. Mean, median and sd refer to the mean, median and standard deviation of the posterior distribution draws. HDI 2.5\% and HDI 97.5\% indicate the borders of the 95\% highest-density interval (HDI). $\hat{R}$ is a diagnostic of convergence of the Markov chains ($\hat{R} = 1 $ indicating convergence).
        \textbf{(b)} Average and predicted choices, including 95\% HDI.         
         }\label{fig:model_summary_arithmetics}
    \end{figure}
    
Due to the absence of altruistic preferences in the choice rule, the treatment-group invariant parameters now consist of only the mean of the prior numerical magnitude. The probabilistic model indicates that $\mu_{\hat{r}^{'}} = 0.483 \; [0.393 - 0.603]$, which corresponds to an ``intermediate'' intuitive perception of numerical magnitudes, that intuitively $\frac{\widehat{\text{A}}}{\widehat{\text{B}}} < 1$. Before interpreting the impact of this prior on choices in more detail, it is useful to inspect the differences in noise(-related) parameters across groups first: Again, noise (now in perceiving numerical magnitudes) is larger in the Treatment versus the Baseline group with $\nu_{\frac{\text{A}}{\text{B}},T} = 0.286 \; [0.26 - 0.315]$ and $\nu_{\frac{\text{A}}{\text{B}},B} = 0.198 \; [0.18 - 0.216]$ and $P(\nu_{\frac{\text{A}}{\text{B}},T} >\nu_{\frac{\text{A}}{\text{B}},B}) = 1$. Note also that $\mu_T^{\nu_{\frac{\text{A}}{\text{B}}}} - \mu_B^{\nu_{\frac{\text{A}}{\text{B}}}} = 0.370$ $[0.236 - 0.505]$ (c.f. Table \ref{tab:structural_estimation_population_number}). Due to larger noise levels in the Treatment group, the weight on payments is consequently smaller with $\alpha'_T = 0.924 \; [0.91 - 0.937]$, $\alpha'_B = 0.962 \; [0.955 - 0.969]$.

Comparing these parameter estimates to the values in Figure \ref{fig:model_summary}, several differences emerge: First, the values of $\nu_{\frac{\text{A}}{\text{B}}}$ are smaller compared to the values of $\nu_{\frac{\text{self}}{\text{other}}}$, yet the \textit{treatment effect}, i.e., $\mu_T^{\nu_{\frac{\text{A}}{\text{B}}}} - \mu_B^{\nu_{\frac{\text{A}}{\text{B}}}}$ is larger compared to $\mu_T^{\nu_{\frac{\text{self}}{\text{other}}}} - \mu_B^{\nu_{\frac{\text{self}}{\text{other}}}}$. One possible explanation could be that noise in preferences previously counteracted some of the effects of the noise in number increase, leading to a smaller treatment effect in parameters and behavior. 

The most striking difference, however, is that the mean of the numerical magnitude prior, i.e., $\mu_{\hat{r}^{'}}$ is now much better identified (tighter HDI) and its value allows for a clear statement on the origin of the treatment effect: Based on the posterior samples, I calculate that $P(\mu_{\hat{r}'} < 1) = 1$, lending strong support to an ``intermediate'' intuitive numerical magnitude perception. Importantly, this translates into $P(\delta'_B > 1) = 1, P(\delta'_T > 1) = 1$, too, which implies strong evidence for a (noise-induced) bias towards \textit{fewer} choices for $\text{A}$ in both groups (matching the direction of the treatment effect). This bias is also \textit{larger} in the Treatment compared to the Baseline group given $P(\delta'_T > \delta'_B) = 1$.

Overall, the probabilistic model based on the number comparison data thus yields a much clearer indication of how the treatment effect operates, namely through a biased perception of numerical magnitudes. In particular, larger values of $\frac{\text{A}}{\text{B}}$ are being perceived as smaller under noise, which leads to a bias towards $\text{B}$ more generally. This more precise interpretation contrasts the more ambiguous explanations for the treatment effect discussed in Figure \ref{fig:model_summary}.

\paragraph{Result 4:}\textit{The probabilistic model of the number comparison behavior indicates a high probability of an ``intermediate'' perception of numerical magnitudes as the driver of the treatment effect.} \newline 
    
Furthermore, the individual choice curves depicted in panel (b) of Figure \ref{fig:model_summary_arithmetics} show that the average choices are close to the HDI areas, indicating that the structural estimates reasonably recover average behavior. Figure \ref{fig:scatter_avg_prediction} supports this with a rank-correlation of $\rho = 0.94$ between average and predicted individual choices for $\text{A}$. However, especially in the Baseline group, the intervals sometimes do not include the average behavior, which indicates that the chosen functional form can not fully explain these data points. However, in comparison with a ``pure noise'' model that abstracts from any influence of the numerical magnitude prior, the model specification in equation \ref{eq:choice_function_number} is superior as indicated by the $ELPD_{WAIC}$ values (see Figure \ref{fig:model_comparison_number}). This strongly suggests that a single parameter $\nu_{\frac{\text{A}}{\text{B}}}$ is unable to capture the behavior of participants in the number comparison task and suggests that an additional driver, here a numerical magnitude prior with an ``intermediate'' mean, is at play.



Applying the findings from the number comparison data to the altruistic choices provides much stronger support for $H_{1b}$: Under the treatment variation, participants relied relatively more strongly on an intermediate intuitive perception that $\widehat{\text{self}} < \widehat{\text{other}}$, i.e., that the payment prior mean $0 < \mu_{\hat{r}} = \frac{\widehat{\text{self}}}{\widehat{\text{other}}} < 1$. This interpretation also fits to the nature of the treatment variation: Encasing monetary payments in to-be-calculated sums instead of showing plain values predominantly biases the perception of monetary payments instead of altruistic preferences. This also matches the previous finding in Figure \ref{fig:model_summary} that the treatment effect mainly increased noise in perceiving monetary payments.

\subsection{Identification, Noise in Altruism and Nature of the Treatment Effect}\label{sec:noise_altruism}

However, a caveat to this interpretation is that it requires an explicit ``logical transfer'' from the number comparison to the altruism domain, as the altruism data alone did not allow to uncover the origin of the treatment effect. This is related to the \textit{identification} of the model, which is -- despite the simplifying assumption of $\sigma_{\hat{b}}\rightarrow \infty$ (i.e., no influence of a prior about altruistic preferences) not entirely given and thus may characterize a weakness of the approach. While identifiability has a different connotation in Bayesian models compared to a more classical understanding, one way to think about identifiability is the difference between the prior and posterior (parameter) distribution, i.e., how informative the data is (see, e.g., \nptextcite{xieMeasuresBayesianLearning2006}). Prior predictive checks in Figure \ref{fig:model_summary_prior} show that the parameter values and HDI (and corresponding behavior) based on the chosen priors differ from the posterior parameters in Figure \ref{fig:model_summary}. A notable exception to this is the mean of the monetary payment prior, where the posterior distribution is \textit{wider}, again questioning the full identifiability of the model.

Related is a critique that, in turn, calls into doubt whether (modeling) noise in altruistic preferences is necessary to explain altruistic behavior in the present setting and if -- instead -- assuming only noise in perceiving numerical magnitudes is perhaps the more appropriate (and parsimonious) assumption. This seems especially pertinent given that the treatment effect likely operates through biased numerical magnitude perception as discussed above and that a model that excludes noise in altruistic preferences performs almost as well as the full model in the model comparison in Figure \ref{fig:model_comparison}.

To address both critiques—that of a ``mere logical transfer'' and the possibility that modeling noise in altruistic preferences may be unnecessary, I now combine the datasets from both tasks. This approach allows to \textit{jointly} estimate the parameters of equations \ref{eq:choice_function} and \ref{eq:choice_function_number}.
 More specifically, I define a joint monetary payment and numerical magnitude noise term $\nu_{\frac{\text{self,A}}{\text{other,B}}}$ that, alongside joint $\mu_{\hat{r}}$, is estimated from both number comparison \textit{and} altruistic choices, whereas $\nu_{\frac{\beta}{1-\beta}}$ and $\beta$ are estimated from only the altruistic choice data. 
For the model estimation, this simply requires to include two likelihoods, again demonstrating the high flexibility of Bayesian methods. 

The resulting parameter values of the combined estimation are shown in the appended Table \ref{tab:parameters_combined}. The combined model confirms the conclusions drawn previously: With $P(\mu_{\hat{r}} < 1) = 1$, $P(\delta_T > 1) = 1$ and $P(\delta_B > 1) = 1$, the combined model provides strong support in favor of a (noise-induced) biased ''intermediate'' numerical perception in favor of fewer choices for $\text{self}$ and $\text{A}$. Compared to Figure \ref{fig:model_summary}, these more assertive probabilistic statements are due to much tighter posterior distributions, particularly around the prior mean, and speak in favor of increased identifiability of the jointly estimated model. The combined model further indicates -- similar to above -- a larger treatment effect for noise in monetary payments and numerical magnitudes than altruistic preferences (see Table \ref{tab:parameters_combined}). 

One notable difference between the model based on the combined dataset and the original model is that, with $P(\nu_{\frac{\beta}{1-\beta},B} > \nu_{\frac{\text{self,A}}{\text{other,B}},B}) = 1$ and $P(\nu_{\frac{\beta}{1-\beta},T} > \nu_{\frac{\text{self,A}}{\text{other,B}},T}) = 0.997$ the combined model now indicates a \textit{higher} level of noise in altruistic preferences. This is first evidence against the argument that altruistic preferences are perceived without noise. Further evidence against this argument is provided by the model comparison in Figure \ref{fig:model_comparison_combined}. There, I formulate a model that assumes $\nu_{\frac{\beta}{1-\beta}} = 0$, but now use the combined dataset for model estimation. In contrast to Figure \ref{fig:model_comparison}, such a simpler model now performs considerably worse in explaining altruistic choices. Accounting only for noise in numerical magnitude and monetary payment perception is thus insufficient for explaining altruistic choices. Finally, a simple linear regression in Table \ref{tab:inconsistencies_compared_tasks} shows that, given participant and game fixed effects (which account for individual differences in altruism and cognitive noise), the inconsistency in a given trial is significantly higher in the altruism compared to the number comparison task, underscoring the previous line of argument.

\subsection{Alternative Explanations for the Treatment Effect}\label{sec:mechanism_nature_treatment}

So far, the discussion has centered around the mechanism of the treatment effect operating through an increase in cognitive noise. On a more critical note, one might argue that some other (unintended) effect of the chosen variation is responsible for the differences between treatment groups beyond the mechanisms proposed by the theoretical model.

I discuss several alternative explanations in Section \ref{sec:appendix_robustness_treatment}. For example, one argument could be that the treatment effect (i.e., the aforementioned intuition) is ``learned'' over the repeated trials of the experiment, which in turn could limit the external validity of the results. However, treatment differences towards more altruism already materialize in the initial 10 (hypothetical) practice trials. Similarly, I find to impact of the round variable, i.e., in which trial a decision was made, on altruistic choices. I also do not find evidence for a (growing) difference in fatigue as the driver of the treatment effect and provide arguments against a purely ``mechanical'' increase in altruism due to how I constructed the sums of the Treatment group. Furthermore, in Section \ref{sec:het_treatment}, I estimate heterogeneous treatment effects and show that the treatment variation did not work systematically differently for participants, who, e.g., expect or hold different norms between the baseline and treatment variants of the altruism task. Most personal characteristics do not meaningfully contribute to heterogeneity in the treatment effect; if anything, the treatment effect is slightly weaker for participants scoring high on self-reported altruism and ``Need for Cognition''.

Overall, the best guess on the treatment effect's origin is that participants quickly understand how the task works: ``less-for-me'' vs.\ ``more-for-other'', which is reflected in their intuitive perception of the respective monetary payments and numerical magnitudes. I will return to this point and its potential implications in more detail in Section \ref{sec:discussion}.

\subsection{Altruism, Number Comparison and Cognitive Ability}

I now explore $H_2$ and $H_3$, i.e., whether behavior in altruistic choices and number comparison is correlated and if measures of cognitive ability correlate with individual measures of cognitive noise.

\subsubsection{Altruistic Choices and Number Comparison}\label{sec:altruism_number_comp}

As stated earlier, if similar cognitive processes (partly) guide altruistic choices and number comparison, I expect to see some association between behavior in both tasks. Given that the numbers featured in the trials of the altruism choices and the number comparison task are \textit{identical}, I can closely examine possible relationships. Table \ref{tab:choices_tasks} contains the results of a linear probability model that explores if choices for $\text{self}$ in the altruism choices correlate with future choices in the number comparison task \textit{in the exact same trial}.\footnote{As I have five repetitions of each unique trial in both tasks, I need to match the data between tasks on an occurrence variable, which tracks in which order a participant encountered a given trial in a given game (i.e., a group of identical trials). I thus match the choices of the first altruism trial of a given game to the first number comparison choice in the same game, and so on.} With multiple iterations per group of trials, I can include both participant- and game- (one game consisting of the five repetitions of a given trial) fixed effects, with the former including the treatment effect.

\begin{table}[H]
\caption{Correlation of choices between tasks}\label{tab:choices_tasks}  	
\centering
\resizebox{0.85\textwidth}{!}{%
\begin{tabular}[!hptb]{lccc}
\toprule
  & (1) & (2) & (3)\\
\midrule
$A$ chosen & 0.048*** &  & 0.045***\\
 & (0.008) &  & (0.010)\\
Correct Number Comparison &  & 0.029*** & 0.006\\
 &  & (0.009) & (0.010)\\
\midrule
$N$ & 60000 & 60000 & 60000\\
Participant Fixed Effects & Yes & Yes & Yes\\
Game Fixed Effects & Yes & Yes & Yes\\
Clustered Standard Erorrs & Yes & Yes & Yes\\
Unique Obs & 300 & 300 & 300\\
$R^2$ & 0.001 & 0.000 & 0.001\\
\bottomrule
\multicolumn{4}{p{0.675\textwidth}}{\footnotesize \textit{Note:} Linear Probability Model. Dependent variable is the choice for $\text{self}$. Clustered standard errors (participant-level) in parentheses. * p $<$ 0.1, ** p $<$ 0.05, *** p $<$ 0.01.}
\end{tabular}

}
\end{table}

Column 1 shows that if a person will choose $\text{A}$ in a given trial, they are 4.8 percentage points more likely to choose $\text{self}$. Moreover, by Column 2, more \textit{correct} choices in a given trial also positively correlate with choices for $\text{self}$: A person is 2.9 percentage points more likely to choose $\text{self}$ if they will identify the correct solution in the number comparison in that trial. This implies that factual errors in the number comparison correspond to more altruistic choices, which is consistent with the Treatment group being both more altruistic and making more errors in the number comparison task (but recall that I control for treatment differences with participant fixed effects). However, the correlation between correct and selfish choices vanishes once I include the choice for $\text{A}$ in column 3. This can be explained by the fact that both the Baseline and Treatment group errs on the side of $\text{A}$, i.e., chooses $\text{A}$ not often enough (see Figure \ref{fig:results_arithmetics_treatment_effects}).

In addition to choices, \textit{inconsistencies} across tasks are (moderately) correlated, too: The average standard deviations in the altruism task and the number comparison are positively correlated in both the Baseline ($\rho = 0.256$, $p=0.0015$) and the Treatment ($\rho = 0.139$, $p=0.089$) group. Participants who are more inconsistent in their altruistic choices are thus also slightly more inconsistent in the number comparison task (yet to a smaller extent in the Treatment group). 

Leveraging the multivariate normal setup of both the base (Section \ref{sec:structural_estimates_altruism}) and combined probabilistic model (Section \ref{sec:noise_altruism}), I can further inspect individual correlations between the noise in monetary payments (and numerical magnitudes) and altruistic preferences independent of the treatment effects. The base model yields a high positive correlation with $\rho(\nu_{\frac{\text{self}}{\text{other}}},\nu_{\frac{\beta}{1-\beta}}) = 0.576 \; [0.295 - 0.845]$, yet the combined model a very small and even slightly negative correlation of $\rho(\nu_{\frac{\text{self,A}}{\text{other,B}}},\nu_{\frac{\beta}{1-\beta}}) = \text{-}0.096 \; [\text{-}0.218 - 0.024]$. Overall, this, therefore, only yields mixed evidence in favor of a positive association between noise across domains.

Alternatively, I can also correlate individual values (i.e., means of posterior distributions) of $\nu_{\frac{\text{self}}{\text{other}}}$ from the base model and $\nu_{\frac{\text{A}}{\text{B}}}$ from the number comparison model (Section \ref{sec:structural_estimates_number_comp2}). The overall correlation between these values is $\rho = 0.155$, $p=0.007$, which indicates a small positive association between the two. If I separate the data by treatment group, I obtain $\rho = 0.245$, $p=0.002$ for the Treatment and $\rho = 0.059$, $p=0.470$ for the Baseline group, which suggests that noise across tasks is positively related only within the Treatment variant of the task. Notably, the Treatment correlation coefficient of 0.245 is very similar in magnitude to the reported rank correlation coefficient of 0.26 in \textcite{frydmanEfficientCodingRisky2022}. There, a parameter $n$ (that indicates the precision of the mental representation of monetary payoffs in their model) correlates between a risky lottery and a ``perceptual'' choice task in which participants had to identify if a given number shown is larger or smaller compared to some reference number. The fact that the correlation is smaller in the Baseline group here could be related to the fact that the number comparison task is relatively easy given sufficient time which in turn leads to a high choice consistency that somewhat mutes the impact of individual noise.\footnote{Note also that, in the number comparison task, participants know there exists a \textit{correct solution} and even though taking longer reduces the eventual payoff, they often invest ample time to find the correct answer. This is a marked difference compared to the altruism choices, where no objectively correct solution exists.} As Section \ref{sec:decision_times} shows, thinking times -- a common measure of decision difficulty -- are positively correlated across tasks in \textit{both} the Baseline and Treatment group. 

Overall, I evaluate the presented evidence as tentative support for $H_2$ and that processes which guide imprecisions in number comparison also partly guide imprecisions in altruism choices if gathered in a similar way, although the link is not as straightforward as in previous work.

\paragraph{Result 5:}\textit{Behavior and choice inconsistency, as well as individual noise measures are moderately positively correlated between altruistic choices and number comparison.} \newline 

\subsubsection{(Self-reported) Cognitive Ability and Individual Measures of Noise}

I now investigate $H_3$, i.e., test for a negative relationship between individual measures of cognitive ability and cognitive noise. I compute correlations between the CRT, BNT, and NFC scale as well as the preferences for intuition and deliberation scale and self-reported math abilities with individual structural measures of noise, i.e., both $\nu_{\frac{\text{self}}{\text{other}},i}$ and $\nu_{\frac{\beta}{1-\beta},i}$ from the base model (Section \ref{sec:structural_estimates_altruism}). For further reference, I also include the altruistic preference parameter $\beta_i$ as well as more altruism-related measures. Table \ref{tab:params_correlations} contains the rank correlation coefficients between the various measures and structural parameters:

\begin{table}[h]
\centering
\caption{Correlation structural parameters and individual characteristics: Cognition and altruism}\label{tab:params_correlations}
\resizebox{1\textwidth}{!}{%
	\begin{tabular}[!hptb]{lccc}
\toprule
  & \makecell{Noise Altr. Preference \\ $\nu_{\frac{\beta}{1-\beta},i}$} & \makecell{Noise Monetary Payments \\ $\nu_{\frac{\text{self}}{\text{other}},i}$} & \makecell{Altr. Preference \\ $\beta_i$}\\
\midrule
\textit{Cognition-related:} &  &  & \\
No. Correct CRT & -0.327*** & -0.274*** & 0.126*\\
Berlin Numeracy Test & -0.247*** & -0.276*** & 0.173**\\
'I am good at math' & -0.197*** & -0.149** & 0.059\\
Avg. Need for Cognition & -0.150** & -0.163** & 0.125*\\
Avg. Deliberation & 0.098 & 0.150** & -0.142*\\
Avg. Intuition & 0.066 & 0.119* & -0.097\\
\addlinespace
\textit{Altruism-related:} &  &  & \\
Dictator Game Other & -0.120* & -0.357*** & 0.495***\\
GPS Donation & 0.063 & -0.049 & 0.146*\\
GPS Value Gift & -0.048 & -0.107 & 0.137*\\
\bottomrule
\multicolumn{4}{p{0.925\textwidth}}{\footnotesize \textit{Note:} Need for Cognition, Deliberation and Intuition averaged values. Self-reported math abilities are elicited on a 0-10 scale. Individual parameter estimates taken from model in Section \ref{sec:structural_estimates_altruism}. $p$-values from pairwise rank-correlation tests ($n=300$). * p $<$ 0.1, ** p $<$ 0.05, *** p $<$ 0.01.}
\end{tabular}

}
\end{table}

Focusing on the first column, I observe a negative correlation between the number of correct items in the Cognitive Reflection Test and individual measures of both $\nu_{\frac{\beta}{1-\beta},i}$ and $\nu_{\frac{\text{self}}{\text{other}},i}$. Similarly, the higher the score on the Berlin Numeracy Test, the higher self-reported math capabilities and ``Need for Cognition'', the lower the individual estimate of both noise terms. These associations -- though only correlational -- underscore an important point: The proposed theoretical model and the $\nu_{\frac{\beta}{1-\beta},i}$ and $\nu_{\frac{\text{self}}{\text{other}},i}$ parameters, in particular, indeed appear to relate to a \textit{cognitive} component of the process of making altruistic choices, which provides validating evidence for the overall approach. This is in line with the above-mentioned work that shows how CRT performance correlates with biases and mistakes in choices \autocite{augenblickOverinferenceWeakSignals2022, opreaDecisionsRiskAre2024, assenzaPerceivedWealthCognitive2019, chewMultipleswitchingBehaviorChoicelist2022}. In contrast, self-reported preferences for deliberation and intuition do not meaningfully correlate with $\nu_{\frac{\beta}{1-\beta},i}$ and only to a slight extent with $\nu_{\frac{\text{self}}{\text{other}},i}$.

\paragraph{Result 6:}\textit{Individual measures of cognitive noise negatively correlate with cognitive ability as measured by performance in the Cognitive Reflection Test and Berlin Numeracy Test.} \newline 



 Table \ref{tab:params_correlations} further contains correlations with the values of the altruistic preference parameter $\beta_i$. These values correlate positively with the amount a participant gave to another person in the simple dictator game and also, albeit to a much lesser extent, with the hypothetical donation and gift-giving decision from the GPS. Furthermore, $\beta_i$ positively correlates with the CRT and BNT performance\footnote{But note that there is no correlation between CRT performance and the average choice for $\text{self}$: $\rho = -0.089, p = 0.123$.}, whereas $\nu_{\frac{\beta}{1-\beta},i}$ and even more so $\nu_{\frac{\text{self}}{\text{other}},i}$ negatively correlates with the amount given in the simple dictator game. 

 While I abstain from hypothesizing on the origins of this nexus, it could be related to the particular structure of the hierarchical model: Both the correlation between noise in monetary payments and altruistic preferences, $\rho(\nu_{\frac{\text{self}}{\text{other}}},\frac{\beta}{1-\beta}) = \text{-}0.277 \; [\text{-}0.519 - \text{-}0.034]$, as well as noise in altruistic preferences and altruistic preferences $\rho(\nu_{\frac{\beta}{1-\beta}},\frac{\beta}{1-\beta}) = \text{-}0.768 \; [\text{-}0.868 - \text{-}0.661]$ are negatively correlated, which in turn could explain the above-mentioned patterns.


\subsection{Response Times and Metacognition}

I now turn to study two core components insightful for choice processes: response times and measures of ``metacognition,'' which -- as understood here -- comprise several measures of participants' subjective thinking about their choices.

\subsubsection{Response Time}\label{sec:decision_times}
I begin by investigating response times (RT), i.e., the amount of time a participant took to decide in both tasks. RT is a highly informative variable of the choice process in psychology and cognitive science (see e.g., \nptextcite{luceResponseTimesTheir1991}) with the following ``standard results``: for discriminating between stimuli, RT is higher the more similar the stimuli which is often attributed to a higher trial difficulty. This is true both for physical stimuli, such as the brightness of two lights (see, e.g., \nptextcite{pinsRelationStimulusIntensity1996}, ``Pierons Law''), as well as for numerical stimuli, such as two Arabic numerals (see, e.g., \nptextcite{moyerTimeRequiredJudgements1967}).


For economic research, arguably the most important insight from RT stems from its close relationship to the strength of preference. Similar to the perceptual difficulty described above, the closer a subject is to indifference in an economic choice task, the longer their RT (see, e.g., \nptextcite{alos-ferrerStrengthPreferenceDecisions2022}). 
RT has also been used to investigate social preferences, especially under the umbrella of dual-process models with fast (slow) decisions usually attributed to intuitive (deliberate) reasoning. Time pressure studies concluded that people intuitively tend towards cooperation \autocite{randSpontaneousGivingCalculated2012, randReflectionDoesNot2014} (``Social Heuristics Hypothesis''), and that ``fairness is intuitive'' \autocite{cappelenFairnessIntuitive2016}.\footnote{
This conclusion, however, has been challenged by subsequent work: \textcite{krajbichRethinkingFastSlow2015} show how such claims are often unwarranted once discriminability of choice options is accounted for. Similar findings are obtained by \textcite{merkelFairnessIntuitiveExperiment2019}.} 
 
In the present setting, I analyze RT and its correlation with behavior from various angles. First, a straightforward test is to investigate if the treatment variation, aimed at increasing the cognitive difficulty, actually leads to higher RT in the Treatment group. This is the case: On average, participants in the Baseline group took 1.33 seconds to decide on the altruism task, whereas participants in the Treatment condition took 2.02 seconds ($p < 0.001$). The difference is more pronounced in the number comparison task, with a mean RT of 1.36 seconds in the Baseline and 2.6 seconds in the Treatment group ($p < 0.001$). This supports the treatment intention that the to-be-calculated sums increased the difficulty of deciding in both tasks. Within individuals, RT between the two tasks is (moderately positively) correlated both in the Baseline ($\rho = 0.25, p<0.01$,) and Treatment ($\rho = 0.238, p<0.01$) group. Participants for which the altruism task was more difficult thus also had a higher difficulty in identifying the solution in the number comparison task (equating longer RT with choice difficulty).

I also investigate the distribution of RT and its relationship with the strength of preference. For this, the individual structural estimates outlined in Section \ref{sec:structural_estimates_altruism} and \ref{sec:structural_estimates_number_comp} can be utilized: These estimates allow to infer the (mean) indifference values, i.e., at which value a subject is indifferent between $\text{self}$ and $\text{other}$ ($\widetilde{\frac{\text{self}}{\text{other}}_i}$), resp. $\text{A}$ and $\text{B}$ ($\widetilde{\frac{\text{A}}{\text{B}}_i}$). I can then calculate the difference of the ratio of a current trial $j$ to that indifference value ($\Delta \widetilde{\frac{\text{self}}{\text{other}}_{ij}} = \frac{\text{self}}{\text{other}}_{j} - \widetilde{\frac{\text{self}}{\text{other}}_i}$) and investigate how the RT of participant $i$ in trial $j$ relates to that difference.
 
 \begin{figure}[h]
{\centering

\resizebox{0.8\textwidth}{!}{%
\input{graphs_export_restricted/decision_times/distance_indifference_ratio_manual}}

\mycaption[Distribution of RT and Distance to Predicted Indifference Ratio]{\quad $\Delta{\widetilde{\frac{\text{self}}{\text{other}}}}_{ij} = \frac{\text{self}}{\text{other}}_j - \widetilde{\frac{\text{self}}{\text{other}}}_i$, where $\widetilde{\frac{\text{self}}{\text{other}}}_i = \frac{\widetilde{self}}{\widetilde{other}}_i: Pr(self_i, \widetilde{\frac{\text{self}}{\text{other}}}_i) = 0.5$ and $\widetilde{\frac{\text{A}}{\text{B}}}$ and $\Delta{\widetilde{\frac{\text{A}}{\text{B}}}}$ are constructed accordingly. The fit is from a local polynomial regression (with 95\% confidence intervals). In addition, average data points are depicted with the size of the point proportional to its relative frequency.
\label{fig:time_indifference}}}
\footnotesize\justifying \noindent

\end{figure}
 
 Figure \ref{fig:time_indifference} plots the average RT (in a given trial) as a function of the difference to the individually predicted indifference value. For the data points, I aggregate over individuals according to the value of $\Delta{\widetilde{\frac{\text{self}}{\text{other}}}}$ and $\Delta{\widetilde{\frac{\text{A}}{\text{B}}}}$ and scale the size of the data points proportional to their relative frequency. The polynomials are fitted to the non-aggregated data. From this Figure, it becomes apparent that RT follows the usual pattern with its peak around the indifference value, i.e., that RT is largest at those ratios where the model predicts indifference.\footnote{\textcite{vieiderDecisionsUncertaintyBayesian2024} goes a step further and shows that the distribution of individual predictions of indifference exhibits a more pronounced pattern with RT compared to the expected value of lotteries, but in the present setting, such a direct benchmark is not available (at least not for the altruism data).} I consider this as validating evidence that the probabilistic model and the proposed decision rule in Section \ref{sec:theory} (and Section \ref{sec:structural_estimates_number_comp}) are useful in conceptualizing how subjects made their choices in both tasks and in understanding the respective decision difficulty.
 
 \paragraph{Result 7:}\textit{Response Times are larger in the Treatment Group and largest where the probabilistic model predicts indifference. } \newline


\paragraph{RT and Choices}

I also investigate correlations of RT with choices. Table \ref{tab:decision_times} contains 4 Probit models that regress choices for $\text{self}$ respectively $\text{A}$ on the amount of RT. I log-transform the RT variable to reduce the impact of outliers (see e.g., \nptextcite{alos-ferrerCognitiveReflectionDecision2016}). I add participant fixed effects, which contain the treatment effect (columns 1 and 3) as well as game fixed effects (columns 2 and 4). In the first two columns, I observe a small positive and insignificant coefficient of the RT variable on the probability of choosing $\text{self}$. I thus do not observe a strong correlation between the amount of time a person took to decide and the level of altruism (even though the general treatment effect would also be consistent with a ``fairness is intuitive'', i.e., quick, narrative). In contrast, I observe a pronounced positive relationship between RT and choices for $\text{A}$ in columns 3 and 4. Longer RT is thus associated with a higher probability of choosing $\text{A}$, which could be interpreted that fast, intuitive answers lead participants to choose $\text{B}$, while more careful deliberation leads to $\text{A}$ more often. 

\begin{table}[h]
\centering
\caption{Correlation of RT with behavior}\label{tab:decision_times}
\resizebox{0.9\textwidth}{!}{%
\begin{tabular}{ @{} l cccc @{}}

\toprule
  & (1) & (2) & (3) & (4)\\
\midrule
$\frac{\text{self}}{\text{other}}$,$\frac{\text{A}}{\text{B}}$ & 5.160*** &  & 7.824*** & \\
 & (0.086) &  & (0.107) & \\
RT (log.) & 0.034 & 0.020 & 0.482*** & 0.320***\\
 & (0.021) & (0.022) & (0.018) & (0.019)\\
\midrule
Data & Altruism & Altruism & Number Comp. & Number Comp.\\
Participant FE & Yes & Yes & Yes & Yes\\
Game FE & No & Yes & No & Yes\\
$N$ & 72000 & 72000 & 60000 & 60000\\
Clustered Standard Errors & Yes & Yes & Yes & Yes\\
Unique Obs & 300 & 300 & 300 & 300\\
\bottomrule
\multicolumn{5}{p{0.775\textwidth}}{\footnotesize \textit{Note:} Probit Model. Columns (1) and (2) use data from the altruism choices and the dependent variable is the choice for $\text{self}$, columns (3) and (4) from the number comparison with choice for $\text{A}$ as dependent variable. Clustered standard errors (participant-level, ``bias-reduced linearization'' \citealt{pustejovskySmallSampleMethodsClusterRobust2018}) in parentheses. * p $<$ 0.1, ** p $<$ 0.05, *** p $<$ 0.01.}
\end{tabular}
}
\end{table}

 \paragraph{Result 8:}\textit{In the number comparison task, higher RT (i.e., more ``deliberate'' choices) corresponds to more choices for $\text{A}$.} \newline


\subsubsection{``Metacognition''}

In addition to RT, I investigate the relationship between behavior and measures of ``metacognition'', i.e., measures on how participants think about their decisions. Recent literature shows how metacognition can play an important role in explaining (biases in) economic choices. \textcite{enkeCognitiveUncertainty2023} show how self-reported cognitive uncertainty, i.e., ``people's subjective uncertainty over which decision maximizes their expected utility'' 
\textcite[p.\ 2021]{enkeCognitiveUncertainty2023} is predictive of a compression effect in various domains from risky choice to belief updating. \textcite{olschewskiWhatSampleEpistemic2024} illustrate how information on metacognitive awareness of one's cognitive imprecisions improves Bayesian decision models in sample estimation tasks. Further, \textcite{opreaDecisionsRiskAre2024} documents that self-reported measures of attention and noise correlate with prospect-theoretic behavior. 

I can, too, explore the links between (noise in) altruistic choices and number comparison and metacognitive self-reports. These comprise of self-reported measures of confidence (similar to the inverse of cognitive uncertainty), attention, and the precision of comparison as well as additional belief-based measures from the number comparison task. I (i) first test for treatment differences in these metacognitive measures, and (ii) investigate their correlation (on a subject level) with the main choice data.

\begin{table}[H]
\centering
\caption{Treatment effects metacognition}\label{tab:meta_cognition_treatment}
\resizebox{0.85\textwidth}{!}{%
\centering
\centering\resizebox{0.9\textwidth}{!}{
\begin{tabular}[!hptb]{lccc}
\toprule
  & Baseline (avg.) & Treatment (avg.) & $p$\\
\midrule
\textit{Altruism:} &  &  & \\
Negative Confidence & 0.304 & 0.318 & 0.643\\
Avg. Attention & 0.731 & 0.698 & 0.127\\
Precision & 0.364 & 0.382 & 0.573\\
 &  &  & \\

\textit{Number Comparison:} &  &  & \\
Avg. Attention & 0.761 & 0.705 & 0.007***\\
Precision & 0.568 & 0.447 & $<$0.001***\\
$|\Delta$ Belief Correct$|$ & 0.084 & 0.163 & $<$0.001***\\
Belief Correct Confidence & 0.785 & 0.651 & $<$0.001***\\
$|\Delta$ Belief Time Spent$|$ & 0.674 & 0.976 & 0.01***\\
Belief Time Spent Confidence & 0.638 & 0.593 & 0.093*\\
\bottomrule
\multicolumn{4}{p{0.70\textwidth}}{\footnotesize \textit{Note:} 7 participants are omitted, where $|\Delta$ Belief Time Spent$| > 10$ in all tests (i.e., $n = 293$). $p$-values from two-sided $t$-test. * p $<$ 0.1, ** p $<$ 0.05, *** p $<$ 0.01.}
\end{tabular}}

}
\end{table}

Table \ref{tab:meta_cognition_treatment} displays the average values of various measures of metacognition, both from the domain of altruistic choices as well as number comparison, separately for the Baseline and Treatment group, alongside the p-value of a two-sided t-test. The first set of measures in the table contains the negatively-recoded self-reported confidence (the inverse of how confident subjects are they made the for-them correct decision), the average attention (a subject paid to the values of $\text{self}$ and $\text{other}$), and precision, (i.e., if participants compared the payments more approximatively or did a precise comparison). The negative confidence measure is very similar between both groups with an average of 0.304 in the Baseline and 0.318 in the Treatment ($p=0.643$).\footnote{This, in turn, is similar in magnitude to average cognitive uncertainty measures from typical lottery or balls-and-urns tasks (see \nptextcite{enkeCognitiveUncertainty2023} and \nptextcite{amelioCognitiveUncertaintyOverconfidence2022}).} For both the self-reported average attention (Baseline: 0.731, Treatment 0.698, $p=0.127$) and precision (Baseline: 0.364, Treatment 0.382, $p=0.573$) there is also no group difference. Overall, there is no evidence of a treatment effect in the metacognitive measures.
 
This is markedly different in the number comparison domain: here, participants in the Treatment group report lower average attention (Baseline: 0.761, Treatment 0.705, $p<0.01$) and precision (Baseline: 0.568, Treatment 0.447, $p<0.01$). In addition to these self-reports, the number comparison task offers several additional measures, which all show clear group differences: in the Treatment group, participants deviate more strongly in their beliefs from their true performance\footnote{See also Figure \ref{fig:beliefs_number_comparison}, which plots participants' beliefs of the average of correct answers (time spent) and their actual share of correct answers (time spent). Participants consistently underestimate the amount of correctly solved tasks \textit{and} overestimate the amount of time spent, which results in a strong pessimistic bias in the beliefs of the number comparison task.}, i.e., have a larger $|\Delta$ Belief Correct$|$ (Baseline: 0.084, Treatment 0.163, $p<0.01$), report lower confidence in these belief statements (Baseline: 0.785, Treatment 0.651, $p<0.01$), deviated more strongly in their belief how much time they think they needed in the number comparison (Baseline: 0.674 sec., Treatment 0.976 sec., $p=0.01$) and again report lower confidence in these estimates (Baseline: 0.638, Treatment 0.593, $p=0.093$). Similarly, correlations between choices, RT, metacognitive measures and choice inconsistencies are overall more pronounced in the number comparison domain compared to the altruism choices (see Table \ref{tab:meta_cognition_correlations} in the appendix).

 \paragraph{Result 9:}\textit{Measures of ``metacognition'' exhibit a strong treatment effect only in the number comparison task and not in the altruism task. Further, the correlation between behavior and metacognition is more pronounced for number comparison than for altruism.} \newline 






\section{Discussion of Results and Next Steps}\label{sec:discussion}

In this paper, I established the following main results: (i) Encasing monetary payments in to-be-calculated sums causes more altruistic choices in the a simple give vs. take task. This effect most likely operates through the perception of monetary payment values, as the effect manifests comparably in the number comparison task. (ii) I observe correlations in behavior between altruism choices and number comparison, (iii) a positive association between individual measures of cognitive noise and cognitive ability, and finally, (iv) a link between RT, metacognition and choices, which however both reacts more strongly to the treatment variation and is more pronounced in the number comparison compared to the altruism domain. I discuss each result in more detail, outlining potential avenues for future research in turn.

\paragraph{(i) Implications of the Treatment Effect} 

A similar treatment effect in the altruism and number comparison task demonstrates how an intermediate intuition of $\widehat{\text{A}} < \widehat{\text{B}}$ (and $\widehat{\text{self}} < \widehat{\text{other}}$) is a candidate driver of the group differences. The exact origins of this intuition are less clear, but a possible explanation could be an instinctive understanding of the ``rules'' of the task, i.e., ``less-for-me'' vs.\ ``more-for-other''. Not only does the altruism task carry such simple (and easy to grasp) ''rules'', it consequently also fits to the statistical environment of the tasks of the experiment as the empirical average ratio in the average trial amounts to $\bar{\frac{\text{self}}{\text{other}}} = 0.466$ and $\bar{\frac{\text{A}}{\text{B}}} = 0.437$. While it remains possible that an \textit{adaptation} to the statistics of both tasks over the course of the experiments explains the origin of the treatment effect, the early emergence of treatment differences (in the practice trials) and the absence of strong evidence for learning effects (see Section \ref{sec:mechanism_nature_treatment}) make a quick, intuitive grasp of the task a more likely explanation.

Either way, this challenges a common assumption in the noisy cognition literature that perceptions of monetary payments of different choice options are intuitively perceived to be the same (see e.g., \nptextcite{khawCognitiveImprecisionSmallStakes2021a}). As in the present tasks, other tasks inherently imply certain statistical proportions and also allow for an instinctive understanding of the numerical relationship of stakes, e.g., in intertemporal decision-making (``smaller-and-sooner'' vs.\ ``larger-and-later''), or in lotteries, where risky and safe payoffs are necessarily different.\footnote{However, note that e.g., in \textcite{vieiderDecisionsUncertaintyBayesian2024} the objects of perception are ``benefits'' and ``costs'' of risky and safe payoffs, where an intuitive understanding of them being equal is more convincing.} This relates to a point in \textcite[p.\ 33]{opreaMindingGapOrigins2024} who explicitly discuss differences between ``naive'' versus more sophisticated decision-makers when specifying a prior mean parameter. Here, I provide strong support for the presence of such non-naive intuitions. This has important implications: If people quickly grasp the ``rules'' or statistics of typical tasks, this potentially alters the direction of an increase in cognitive noise.\footnote{But also note that an ``overfitting'' of prior intuitions to a given statistical environment is not necessarily a given and not a good strategy across tasks.} While \textcite{khawCognitiveImprecisionSmallStakes2021a} attribute risk-aversion to higher levels of cognitive noise (and \textcite{barretto-garciaIndividualRiskAttitudes2023} show the neurological underpinnings of the model), a \textit{causal} test of this direction is still to be performed that identifies how behavior actually reacts to an increase in noise, which in turn depends on the prior (mean). The to-be-calculated sums proposed here are a candidate for such a causal test.

However, I fully acknowledge that the interpretation of an intuition reminiscent of the ``rules'' of the task is so far purely speculative and not the result of an empirical test. Future work could therefore investigate the drivers of a potential adaptation and e.g., exogenously manipulate choice environments that induce differences in intuitions (akin to efficient coding studies such as \textcite{frydmanEfficientCodingRisky2022, polaniaEfficientCodingSubjective2019} or \nptextcite{prat-carrabinBayesVsWeber2024}) or explicitly model the noisy learning process (see \textcite{poggiLearningDynamicsOptimal2021} for a start). 

In addition, while I \textit{theoretically} demonstrate how noise in perceiving altruistic preferences can affect choices, the implemented to-be-calculated sums likely operate through monetary payment and numerical magnitude perception. An important next step would thus be to either design and implement a variation that exclusively affects the noise in the perception of altruistic preferences without affecting numerical perception. Another possible next step is to compare the effects of the present treatment to more standard time pressure or cognitive load treatments, which potentially could also aid in better-identifying parameters of a more extensive theoretical model (that, e.g., includes prior beliefs over preferences).




\paragraph{(ii) Correlation of Altruism and Number Comparison}

The second set of results shows a (moderate) correlation between behavior and measures of noise in altruistic choices and number comparison: I both observe an increase in choices for $\text{self}$ if a person chooses $\text{A}$ in a future ``twin''-trial and also if this person identifies the correct solution later on. Participants who are more inconsistent in choosing between $\text{self}$ and $\text{other}$ are also more inconsistent in the number comparison task. Both facts point towards some common driver between both domains, for which the noisy representation of monetary payments is a potential candidate. This is similar to conclusions in \textcite{frydmanEfficientCodingRisky2022} and \textcite{barretto-garciaIndividualRiskAttitudes2023}, although the relationship between economic choice and numerical perception is weaker in the present setting.\footnote{Note that, in \textcite{frydmanEfficientCodingRisky2022} and \textcite{khawCognitiveImprecisionSmallStakes2021a}, the probability of the lottery payoff is not only an objective quantity but remains fixed over all trials. Only differences in the payoffs are thus important for choices, which could render their lottery choice task into a number perception task (see \autocite[313]{alos-ferrerStrengthPreferenceDecisions2022}.} Nonetheless, common to both domains is the necessity to \textit{compare}, which in turn could be related to common cognitive processes. Note that I specifically designed both tasks to be similar to each other. In turn, if such relationships between economic choice and numerical perception across domains manifest in other settings or what characteristics of a chosen setting determine this relationship, remains a largely open question and seems worthy of future investigation.\footnote{For risk elicitation methods, \textcite{holzmeisterRiskElicitationPuzzle2021} show that the within-person inconsistency in risk elicitations across different methods is related to the subjectively perceived elicitation complexity, suggesting that the complexity of a setting could potentially guide the impact of overarching concepts.}
 
\paragraph{(iii) Cognition and Altruistic Preferences}

The third result shows a correlation between measures of cognitive noise and cognitive ability. 
Cognitive processes are thus likely to play a role at \textit{expressing} one's (subjective) preferences, yet a directional association with different levels of altruism is less clear (recall that I did not observe any correlation between measures of cognitive noise and altruism per se). Similar to the present treatment effect, this implies that e.g., across contexts of varying complexity, differences in cognitive ability could nonetheless lead to systematically different behavior. Recall also that the association between changes in payments in the Treatment group was flatter compared to the Baseline group. This ``flatness'' (or insensitivity) is at the center of discussions in \textcite{enkeCognitiveUncertainty2023}, \textcite{enkeComplexityHyperbolicDiscounting2023} and especially \textcite{enkeBehavioralAttenuation2024}, who establish cognitive uncertainty as the common driver of such inattentive behavior across over 30 experiments in various decision domains. Transporting this argument to the present setting, it speaks in favor of a dampened expression of selfish preferences in the Treatment group. A similar point is raised by \textcite{enkeCognitiveTurnBehavioral2024} -- in light of discussions on the link between confusion and public goods contribution -- in that information-processing constraints impact the translation of social preferences into behavior. Similarly, \textcite{baoCognitiveUncertaintyGPT2024} interpret cognitive uncertainty as a complementary driver to social preferences in public goods contributions (see also the public goods game results in \textcite{enkeBehavioralAttenuation2024}). What this paper adds to this discussion is that sheds closer light on the \textit{mechanism} of the dampened expression, namely through numerical magnitude perception in the present setting, and uncovering precise estimates on the location of the prior -- in the absence of a clear default option -- that dictates how higher noise leads to differences in behavior. At the same time, self-reported ``meta-cognitive'' measures seem less relevant for explaining altruistic choice in the present setting (see iv below).

Ultimately, a dampened expression as the mechanism also provides a more nuanced angle on the discussion that investigates associations between cognitive ability and economic preferences more generally \autocite{burksCognitiveSkillsAffect2009,chapmanEconographics2023, falkGlobalEvidenceEconomic2018,stangoWeAreAll2023}, especially for associations between social preferences and cognitive ability \autocite{haugeAreSocialPreferences2009, chenTooSmartBe2013,pontiSocialPreferencesCognitive2015}, which mostly focus on associations between the level of preferences and cognitive ability thus far. In line with the interpretation put forward here, \textcite{olschewskiTaxingCognitiveCapacities2018} find that in ultimatum game choices, cognitive load mostly increases choice variability only and does not impact preferences per se.\footnote{See also an ongoing discussion regarding risk and time preferences and whether cognitive ability is related to choice mistakes ``only'' or preferences \autocite{amador-hidalgoCognitiveAbilitiesRisktaking2021, olschewskiLinkCognitiveAbilities2023}.}

\paragraph{(iv) RT and Metacognition}

The fourth set of results is related to the link between behavior, RT and metacognition. The main result is the presence of a treatment effect in ``metacognitive'' measures in the number comparison, yet an absence of such an effect in the altruism domain. Correlational analyses further show that the link between metacognition, RT and choices is \textit{weaker} in the altruism domain compared to the number comparison domain. One possible explanation for this could be that in domains where an objectively correct solution exists, RT and metacognition (which could be formed from a recollection of the latter, see \nptextcite{kianiChoiceCertaintyInformed2014}) are \textit{better calibrated} because a more direct notion of a ``correct'' solution is available. In turn, the treatment variation, aimed at increasing cognitive difficulty, could have only an effect on metacognitive reasoning with a clear indication of what a ``correct'' choice is. This does not imply that metacognitive judgments are detached from internal processes (see the discussion in \textcite{flemingMetacognitionConfidenceReview2024} for value-based decisions), yet their determinants and consequences are possibly different and generally remain less well understood \autocite{brusSourcesConfidenceValuebased2021}. Economic tasks often contain a strong subjective component of what is ``correct'' and in turn, could imply that the overall link between metacognition and subjective preferences ``plays out'' differently compared to settings with more clear notions of choice correctness. This points towards a difference between lottery choices (which also remain dependent on subjective preferences) and altruism choices: In the former, a benchmark choice, i.e., the one that maximizes expected value is available. Such a ``virtually objective'' benchmark is lacking when making altruistic choices.




\section{Conclusion}\label{sec:conclusion}

In this paper, I study altruistic choices through the lens of a cognitively noisy decision-maker. I ran an experiment that elicited altruistic choices, i.e., choosing between taking an amount $\text{self}$ or giving an amount $\text{other}$. Crucially, the experiment featured a between-subject manipulation of the cognitive difficulty of choosing in the Treatment group, which was shown to-be-calculated sums instead of plain monetary values. I observe both a flatter association between changes in payments and choices as well as overall more altruistic choices in the Treatment group. After the altruistic choices, I repeated the trials of the experiment in a comparable number comparison task, where participants had to judge which of two numbers was larger. In this task, I observe a similar treatment effect, which suggests that the perception of numerical magnitudes -- in particular an intuitive ``intermediate'' perception of numerical values -- is responsible for the observed group differences in both tasks. In addition to these treatment differences, I observe (correlational) associations between number comparison, cognitive ability and altruistic choice.

The expression of altruistic preferences -- and social preferences more generally -- is thus not immune to the cognitive difficulty of their implementation. This further implies that at least part of observed pro-social choices are due to (individual differences in) cognitive noise, which in turn may be related to cognitive ability. This also suggests that the expression of social preferences is likely to be context-dependent if different contexts invoke differences in the ``noisiness of perception'' or have different complexity. Ultimately, this is an important implication if social preferences are used as the basis for welfare calculations.

A caveat of this paper remains in that the treatment effect in the experiment remained relatively small and altruistic behavior did not react that much to increasing noise. However, this could be related both to the fact that altruism choices as operationalized here remained relatively simple and that the chosen treatment variation represents a relatively mild increase in cognitive noise. Both, in turn, imply that the observed group differences are likely a \textit{lower bound} on the influence of cognitive noise on social preferences more generally. Other decisions involving social preferences are often much more complex to carry out: Both in more involved laboratory environments, e.g., in choosing between payoff allocations as in the popular binary dictator game and in real-world scenarios featuring social preferences that often involve multiple trade-offs, decisions are likely \textit{more} prone to be affected by cognitive noise. Exploring these effects is a promising avenue for future research.

\clearpage

\FloatBarrier
\appendix
\hypertarget{appendix}{%
	\section{Appendix}\label{appendix}}

\setcounter{table}{0}
\setcounter{figure}{0}
\renewcommand{\thetable}{\Alph{section}\arabic{table}}
\renewcommand{\thefigure}{\Alph{section}\arabic{figure}}

\subsection{Theory}

\begin{figure}[H]
\hspace{-5cm}
{\centering
\resizebox{1.1\linewidth}{!}{%
\input{graphs_export_restricted/theory/number_theory_plot_manual.tex}}
\vspace{-6.5cm}
\mycaption[Impact of Cognitive Noise on Number Comparison]{\quad This shows the impact of changes in noise $\nu_{\frac{\text{A}}{\text{B}}}$ on the choice function \ref{eq:choice_function_number} depending on different values of $\mu_{\hat{r'}}$. Note that $\sigma_{\hat{r'}} = 1$.}\label{fig:theory_noise_number}}
\footnotesize\justifying \noindent
\end{figure}

\newpage
\subsection{Graphs and Figures Experiment}

\begin{figure}[H]
{\centering
\resizebox{0.85\textwidth}{!}{%
\tikzstyle{box} = [rectangle, draw, text centered, rounded corners, minimum height=1cm, minimum width=16cm, line width=0.5mm]
\tikzstyle{line} = [draw, -latex', thick]

\begin{tikzpicture}[auto, node distance=15mm, scale=0.7, every node/.style={scale=0.9}]
	
	\node[align=center, font=\bfseries, below=2pt] (title1) {\textbf{Part 1: Altruistic Choices}};
	\node [box, below=2pt of title1, align=left, text width=15cm] (stage1) {
		\begin{itemize}
			\setlength{\itemsep}{3pt}
			\item Decide between taking payment $\text{self}$ or giving $\text{other}$. 240 total trials with intermediate breaks possible.
			\item One decision randomly implemented at the end of the experimental session.
		\end{itemize}
	};
	
	\node[align=center, font=\bfseries, below=10pt of stage1] (title2) {\textbf{Part 2: Number Comparison}};
	\node [box, below=2pt of title2, align=left, text width=15cm] (stage2) {
		\begin{itemize}
			\setlength{\itemsep}{3pt}
			\item Judge which of two numbers $\text{A}$ (previously $\text{self}$) or $\text{B}$ (previously $\text{other}$) $\times \sfrac{1}{2}$ is larger. 200 trials total.
			\item Final payment = $10$ \EUR $\times$ avg. correct - avg. time in seconds.
		\end{itemize}
	};
	
	\node[align=center, font=\bfseries, below=10pt of stage2] (title3) {\textbf{Part 3: Additional Data}};
	\node [box, below=2pt of title3, align=left, text width=15cm] (stage3) {
		\begin{itemize}
			\setlength{\itemsep}{3pt}
			\item Dictator game featuring 10 \EUR, implemented with 1\% probability.
			\item Social and private norm elicitation for altruistic choices, both for Baseline and Treatment variant; ``Excuses'' survey.
			\item Cognitive Reflection Test, Berlin Numeracy Test.
			\item Surveys (Preferences for Deliberation and Intuition, GPS Altruism, Need for Cognition, Machiavellianism).
			\item Demographics.
		\end{itemize}
	};
	
\end{tikzpicture}}
\caption{Graphical Outline of an Experimental Session}\label{fig:session_outline}}
\end{figure}

\begin{figure}[H]
\centering
\resizebox{1\textwidth}{!}{%
\input{graphs_export/experiment/trials}}

\mycaption[Payment Combinations in Altruistic Choice Task]{\quad This graph shows the 48 unique combinations of $\text{self}$ and $\text{other}$ in the experimental trials. Each combination is repeated five times, totaling 240 decisions, with one decision randomly implemented. Note that I instructed participants precisely like this, but not each trial had the same chance of being drawn: Instead of drawing from a uniform distribution across trials, I overweighted trials of smaller stakes (i.e., where the sum of $\text{self}$ and $\text{other}$ is small) to be more likely to be drawn. Details and implementation are available upon request. The indifference threshold for a noiseless decision maker with a $\beta = 0.2$ is drawn for illustration purposes. This DM always decides for $\text{other}$ in the trials below and for $\text{self}$ above this line. Payments are in Eurocents.}\label{fig:trials_overview}
\end{figure}

\newpage
	\begin{longtable}[c]{p{1.25cm}p{1.25cm}p{1.25cm}p{1.25cm}p{1.25cm}p{1.25cm}p{1.25cm}p{1.25cm}}
\caption{Overview of 240 trials of the altruism task }\label{tab:trials_overview}
	\setlength{\tabcolsep}{0.4em}
\renewcommand{\arraystretch}{7}\\
\toprule
\multicolumn{2}{c}{Game Identifiers} & \multicolumn{2}{c}{Payments} & \multicolumn{4}{c}{Components of Sums (Treatment)} \\
\cmidrule(l{3pt}r{3pt}){1-2} \cmidrule(l{3pt}r{3pt}){3-4} \cmidrule(l{3pt}r{3pt}){5-8}
Game ID & Game Group & Other & Self & Self 1 & Self 2 & Other 1 & Other 2\\
\midrule
\endfirsthead
	\multicolumn{7}{c}{{\bfseries \tablename\ \thetable{} -- continued from previous page \vspace{0.25cm}}} \\
\toprule
\multicolumn{2}{c}{Game Identifiers} & \multicolumn{2}{c}{Payments} & \multicolumn{4}{c}{Components of Sums (Treatment)} \\
\cmidrule(l{3pt}r{3pt}){1-2} \cmidrule(l{3pt}r{3pt}){3-4} \cmidrule(l{3pt}r{3pt}){5-8}
Game ID & Game Group & Other & Self & Self 1 & Self 2 & Other 1 & Other 2\\
\midrule
\endhead

\midrule
	\multicolumn{7}{r}{{Continued on next page}} \\
	\midrule
\endfoot
\bottomrule
\endlastfoot
0 & 1 & 655 & 0 & 0 & 0 & 543 & 112\\
0 & 1 & 655 & 0 & 0 & 0 & 629 & 26\\
0 & 1 & 655 & 0 & 0 & 0 & 490 & 165\\
0 & 1 & 655 & 0 & 0 & 0 & 32 & 623\\
0 & 1 & 655 & 0 & 0 & 0 & 540 & 115\\
\addlinespace
1 & 1 & 655 & 34 & 14 & 20 & 638 & 17\\
1 & 1 & 655 & 34 & 23 & 11 & 20 & 635\\
1 & 1 & 655 & 34 & 22 & 12 & 643 & 12\\
1 & 1 & 655 & 34 & 15 & 19 & 14 & 641\\
1 & 1 & 655 & 34 & 22 & 12 & 10 & 645\\
\addlinespace
2 & 1 & 655 & 72 & 24 & 48 & 627 & 28\\
2 & 1 & 655 & 72 & 35 & 37 & 34 & 621\\
2 & 1 & 655 & 72 & 58 & 14 & 26 & 629\\
2 & 1 & 655 & 72 & 61 & 11 & 40 & 615\\
2 & 1 & 655 & 72 & 37 & 35 & 35 & 620\\
\addlinespace
3 & 1 & 655 & 115 & 31 & 84 & 33 & 622\\
3 & 1 & 655 & 115 & 36 & 79 & 621 & 34\\
3 & 1 & 655 & 115 & 39 & 76 & 617 & 38\\
3 & 1 & 655 & 115 & 28 & 87 & 645 & 10\\
3 & 1 & 655 & 115 & 34 & 81 & 585 & 70\\
\addlinespace
4 & 1 & 655 & 163 & 62 & 101 & 11 & 644\\
4 & 1 & 655 & 163 & 12 & 151 & 10 & 645\\
4 & 1 & 655 & 163 & 15 & 148 & 643 & 12\\
4 & 1 & 655 & 163 & 13 & 150 & 643 & 12\\
4 & 1 & 655 & 163 & 140 & 23 & 554 & 101\\
\addlinespace
5 & 1 & 655 & 218 & 167 & 51 & 47 & 608\\
5 & 1 & 655 & 218 & 164 & 54 & 509 & 146\\
5 & 1 & 655 & 218 & 172 & 46 & 622 & 33\\
5 & 1 & 655 & 218 & 81 & 137 & 41 & 614\\
5 & 1 & 655 & 218 & 18 & 200 & 122 & 533\\
\addlinespace
6 & 1 & 655 & 280 & 170 & 110 & 588 & 67\\
6 & 1 & 655 & 280 & 234 & 46 & 60 & 595\\
6 & 1 & 655 & 280 & 90 & 190 & 104 & 551\\
6 & 1 & 655 & 280 & 161 & 119 & 506 & 149\\
6 & 1 & 655 & 280 & 126 & 154 & 15 & 640\\
\addlinespace
7 & 1 & 655 & 352 & 107 & 245 & 557 & 98\\
7 & 1 & 655 & 352 & 161 & 191 & 68 & 587\\
7 & 1 & 655 & 352 & 227 & 125 & 583 & 72\\
7 & 1 & 655 & 352 & 310 & 42 & 645 & 10\\
7 & 1 & 655 & 352 & 170 & 182 & 486 & 169\\
\addlinespace
8 & 1 & 655 & 436 & 68 & 368 & 51 & 604\\
8 & 1 & 655 & 436 & 331 & 105 & 97 & 558\\
8 & 1 & 655 & 436 & 425 & 11 & 485 & 170\\
8 & 1 & 655 & 436 & 326 & 110 & 634 & 21\\
8 & 1 & 655 & 436 & 312 & 124 & 158 & 497\\
\addlinespace
9 & 1 & 655 & 535 & 199 & 336 & 471 & 184\\
9 & 1 & 655 & 535 & 413 & 122 & 71 & 584\\
9 & 1 & 655 & 535 & 398 & 137 & 27 & 628\\
9 & 1 & 655 & 535 & 426 & 109 & 478 & 177\\
9 & 1 & 655 & 535 & 222 & 313 & 443 & 212\\
\addlinespace
10 & 1 & 655 & 655 & 253 & 402 & 79 & 576\\
10 & 1 & 655 & 655 & 277 & 378 & 82 & 573\\
10 & 1 & 655 & 655 & 332 & 323 & 575 & 80\\
10 & 1 & 655 & 655 & 361 & 294 & 565 & 90\\
10 & 1 & 655 & 655 & 156 & 499 & 419 & 236\\
\addlinespace
11 & 1 & 655 & 800 & 740 & 60 & 515 & 140\\
11 & 1 & 655 & 800 & 678 & 122 & 260 & 395\\
11 & 1 & 655 & 800 & 503 & 297 & 635 & 20\\
11 & 1 & 655 & 800 & 311 & 489 & 39 & 616\\
11 & 1 & 655 & 800 & 744 & 56 & 244 & 411\\
\addlinespace
12 & 2 & 926 & 0 & 0 & 0 & 850 & 76\\
12 & 2 & 926 & 0 & 0 & 0 & 836 & 90\\
12 & 2 & 926 & 0 & 0 & 0 & 254 & 672\\
12 & 2 & 926 & 0 & 0 & 0 & 418 & 508\\
12 & 2 & 926 & 0 & 0 & 0 & 391 & 535\\
\addlinespace
13 & 2 & 926 & 48 & 13 & 35 & 10 & 916\\
13 & 2 & 926 & 48 & 11 & 37 & 10 & 916\\
13 & 2 & 926 & 48 & 36 & 12 & 19 & 907\\
13 & 2 & 926 & 48 & 32 & 16 & 26 & 900\\
13 & 2 & 926 & 48 & 30 & 18 & 914 & 12\\
\addlinespace
14 & 2 & 926 & 102 & 84 & 18 & 11 & 915\\
14 & 2 & 926 & 102 & 39 & 63 & 32 & 894\\
14 & 2 & 926 & 102 & 76 & 26 & 909 & 17\\
14 & 2 & 926 & 102 & 35 & 67 & 19 & 907\\
14 & 2 & 926 & 102 & 66 & 36 & 20 & 906\\
\addlinespace
15 & 2 & 926 & 163 & 70 & 93 & 69 & 857\\
15 & 2 & 926 & 163 & 138 & 25 & 28 & 898\\
15 & 2 & 926 & 163 & 111 & 52 & 48 & 878\\
15 & 2 & 926 & 163 & 39 & 124 & 114 & 812\\
15 & 2 & 926 & 163 & 25 & 138 & 17 & 909\\
\addlinespace
16 & 2 & 926 & 231 & 214 & 17 & 16 & 910\\
16 & 2 & 926 & 231 & 62 & 169 & 53 & 873\\
16 & 2 & 926 & 231 & 35 & 196 & 914 & 12\\
16 & 2 & 926 & 231 & 67 & 164 & 864 & 62\\
16 & 2 & 926 & 231 & 161 & 70 & 126 & 800\\
\addlinespace
17 & 2 & 926 & 308 & 90 & 218 & 45 & 881\\
17 & 2 & 926 & 308 & 101 & 207 & 169 & 757\\
17 & 2 & 926 & 308 & 22 & 286 & 905 & 21\\
17 & 2 & 926 & 308 & 84 & 224 & 38 & 888\\
17 & 2 & 926 & 308 & 173 & 135 & 99 & 827\\
\addlinespace
18 & 2 & 926 & 396 & 11 & 385 & 10 & 916\\
18 & 2 & 926 & 396 & 184 & 212 & 733 & 193\\
18 & 2 & 926 & 396 & 74 & 322 & 45 & 881\\
18 & 2 & 926 & 396 & 170 & 226 & 896 & 30\\
18 & 2 & 926 & 396 & 325 & 71 & 187 & 739\\
\addlinespace
19 & 2 & 926 & 498 & 63 & 435 & 51 & 875\\
19 & 2 & 926 & 498 & 264 & 234 & 879 & 47\\
19 & 2 & 926 & 498 & 330 & 168 & 273 & 653\\
19 & 2 & 926 & 498 & 325 & 173 & 779 & 147\\
19 & 2 & 926 & 498 & 366 & 132 & 50 & 876\\
\addlinespace
20 & 2 & 926 & 617 & 486 & 131 & 200 & 726\\
20 & 2 & 926 & 617 & 410 & 207 & 148 & 778\\
20 & 2 & 926 & 617 & 409 & 208 & 775 & 151\\
20 & 2 & 926 & 617 & 416 & 201 & 186 & 740\\
20 & 2 & 926 & 617 & 106 & 511 & 470 & 456\\
\addlinespace
21 & 2 & 926 & 757 & 565 & 192 & 171 & 755\\
21 & 2 & 926 & 757 & 604 & 153 & 106 & 820\\
21 & 2 & 926 & 757 & 480 & 277 & 152 & 774\\
21 & 2 & 926 & 757 & 557 & 200 & 821 & 105\\
21 & 2 & 926 & 757 & 733 & 24 & 470 & 456\\
\addlinespace
22 & 2 & 926 & 926 & 224 & 702 & 647 & 279\\
22 & 2 & 926 & 926 & 79 & 847 & 730 & 196\\
22 & 2 & 926 & 926 & 117 & 809 & 83 & 843\\
22 & 2 & 926 & 926 & 46 & 880 & 727 & 199\\
22 & 2 & 926 & 926 & 370 & 556 & 567 & 359\\
\addlinespace
23 & 2 & 926 & 1132 & 827 & 305 & 711 & 215\\
23 & 2 & 926 & 1132 & 669 & 463 & 146 & 780\\
23 & 2 & 926 & 1132 & 1072 & 60 & 863 & 63\\
23 & 2 & 926 & 1132 & 598 & 534 & 222 & 704\\
23 & 2 & 926 & 1132 & 867 & 265 & 323 & 603\\
\addlinespace
24 & 3 & 1310 & 0 & 0 & 0 & 963 & 347\\
24 & 3 & 1310 & 0 & 0 & 0 & 898 & 412\\
24 & 3 & 1310 & 0 & 0 & 0 & 726 & 584\\
24 & 3 & 1310 & 0 & 0 & 0 & 876 & 434\\
24 & 3 & 1310 & 0 & 0 & 0 & 459 & 851\\
\addlinespace
25 & 3 & 1310 & 68 & 35 & 33 & 1288 & 22\\
25 & 3 & 1310 & 68 & 28 & 40 & 1294 & 16\\
25 & 3 & 1310 & 68 & 31 & 37 & 26 & 1284\\
25 & 3 & 1310 & 68 & 26 & 42 & 10 & 1300\\
25 & 3 & 1310 & 68 & 29 & 39 & 1283 & 27\\
\addlinespace
26 & 3 & 1310 & 145 & 121 & 24 & 1288 & 22\\
26 & 3 & 1310 & 145 & 42 & 103 & 1293 & 17\\
26 & 3 & 1310 & 145 & 79 & 66 & 1286 & 24\\
26 & 3 & 1310 & 145 & 121 & 24 & 1212 & 98\\
26 & 3 & 1310 & 145 & 14 & 131 & 1298 & 12\\
\addlinespace
27 & 3 & 1310 & 231 & 80 & 151 & 1292 & 18\\
27 & 3 & 1310 & 231 & 140 & 91 & 10 & 1300\\
27 & 3 & 1310 & 231 & 34 & 197 & 110 & 1200\\
27 & 3 & 1310 & 231 & 35 & 196 & 28 & 1282\\
27 & 3 & 1310 & 231 & 127 & 104 & 74 & 1236\\
\addlinespace
28 & 3 & 1310 & 327 & 182 & 145 & 1238 & 72\\
28 & 3 & 1310 & 327 & 167 & 160 & 1259 & 51\\
28 & 3 & 1310 & 327 & 62 & 265 & 1257 & 53\\
28 & 3 & 1310 & 327 & 242 & 85 & 1154 & 156\\
28 & 3 & 1310 & 327 & 79 & 248 & 1124 & 186\\
\addlinespace
29 & 3 & 1310 & 436 & 346 & 90 & 1293 & 17\\
29 & 3 & 1310 & 436 & 261 & 175 & 153 & 1157\\
29 & 3 & 1310 & 436 & 112 & 324 & 1233 & 77\\
29 & 3 & 1310 & 436 & 143 & 293 & 215 & 1095\\
29 & 3 & 1310 & 436 & 145 & 291 & 72 & 1238\\
\addlinespace
30 & 3 & 1310 & 561 & 298 & 263 & 1040 & 270\\
30 & 3 & 1310 & 561 & 550 & 11 & 10 & 1300\\
30 & 3 & 1310 & 561 & 202 & 359 & 1254 & 56\\
30 & 3 & 1310 & 561 & 515 & 46 & 934 & 376\\
30 & 3 & 1310 & 561 & 470 & 91 & 884 & 426\\
\addlinespace
31 & 3 & 1310 & 705 & 655 & 50 & 1273 & 37\\
31 & 3 & 1310 & 705 & 442 & 263 & 1161 & 149\\
31 & 3 & 1310 & 705 & 665 & 40 & 278 & 1032\\
31 & 3 & 1310 & 705 & 627 & 78 & 21 & 1289\\
31 & 3 & 1310 & 705 & 597 & 108 & 10 & 1300\\
\addlinespace
32 & 3 & 1310 & 873 & 849 & 24 & 521 & 789\\
32 & 3 & 1310 & 873 & 704 & 169 & 1292 & 18\\
32 & 3 & 1310 & 873 & 758 & 115 & 1271 & 39\\
32 & 3 & 1310 & 873 & 395 & 478 & 62 & 1248\\
32 & 3 & 1310 & 873 & 832 & 41 & 783 & 527\\
\addlinespace
33 & 3 & 1310 & 1071 & 512 & 559 & 416 & 894\\
33 & 3 & 1310 & 1071 & 246 & 825 & 217 & 1093\\
33 & 3 & 1310 & 1071 & 942 & 129 & 72 & 1238\\
33 & 3 & 1310 & 1071 & 493 & 578 & 1262 & 48\\
33 & 3 & 1310 & 1071 & 718 & 353 & 656 & 654\\
\addlinespace
34 & 3 & 1310 & 1310 & 108 & 1202 & 814 & 496\\
34 & 3 & 1310 & 1310 & 63 & 1247 & 1149 & 161\\
34 & 3 & 1310 & 1310 & 750 & 560 & 219 & 1091\\
34 & 3 & 1310 & 1310 & 944 & 366 & 235 & 1075\\
34 & 3 & 1310 & 1310 & 304 & 1006 & 254 & 1056\\
\addlinespace
35 & 3 & 1310 & 1601 & 1045 & 556 & 575 & 735\\
35 & 3 & 1310 & 1601 & 1005 & 596 & 637 & 673\\
35 & 3 & 1310 & 1601 & 1465 & 136 & 800 & 510\\
35 & 3 & 1310 & 1601 & 1459 & 142 & 667 & 643\\
35 & 3 & 1310 & 1601 & 1134 & 467 & 587 & 723\\
\addlinespace
36 & 4 & 1852 & 0 & 0 & 0 & 622 & 1230\\
36 & 4 & 1852 & 0 & 0 & 0 & 1453 & 399\\
36 & 4 & 1852 & 0 & 0 & 0 & 1022 & 830\\
36 & 4 & 1852 & 0 & 0 & 0 & 1110 & 742\\
36 & 4 & 1852 & 0 & 0 & 0 & 734 & 1118\\
\addlinespace
37 & 4 & 1852 & 97 & 85 & 12 & 10 & 1842\\
37 & 4 & 1852 & 97 & 34 & 63 & 27 & 1825\\
37 & 4 & 1852 & 97 & 43 & 54 & 41 & 1811\\
37 & 4 & 1852 & 97 & 27 & 70 & 1828 & 24\\
37 & 4 & 1852 & 97 & 37 & 60 & 54 & 1798\\
\addlinespace
38 & 4 & 1852 & 205 & 54 & 151 & 1740 & 112\\
38 & 4 & 1852 & 205 & 91 & 114 & 1777 & 75\\
38 & 4 & 1852 & 205 & 57 & 148 & 1818 & 34\\
38 & 4 & 1852 & 205 & 83 & 122 & 1785 & 67\\
38 & 4 & 1852 & 205 & 122 & 83 & 113 & 1739\\
\addlinespace
39 & 4 & 1852 & 326 & 216 & 110 & 24 & 1828\\
39 & 4 & 1852 & 326 & 109 & 217 & 198 & 1654\\
39 & 4 & 1852 & 326 & 45 & 281 & 198 & 1654\\
39 & 4 & 1852 & 326 & 64 & 262 & 239 & 1613\\
39 & 4 & 1852 & 326 & 222 & 104 & 81 & 1771\\
\addlinespace
40 & 4 & 1852 & 463 & 74 & 389 & 188 & 1664\\
40 & 4 & 1852 & 463 & 293 & 170 & 77 & 1775\\
40 & 4 & 1852 & 463 & 385 & 78 & 1821 & 31\\
40 & 4 & 1852 & 463 & 150 & 313 & 1706 & 146\\
40 & 4 & 1852 & 463 & 259 & 204 & 171 & 1681\\
\addlinespace
41 & 4 & 1852 & 617 & 367 & 250 & 61 & 1791\\
41 & 4 & 1852 & 617 & 331 & 286 & 1708 & 144\\
41 & 4 & 1852 & 617 & 85 & 532 & 1817 & 35\\
41 & 4 & 1852 & 617 & 414 & 203 & 1838 & 14\\
41 & 4 & 1852 & 617 & 302 & 315 & 277 & 1575\\
\addlinespace
42 & 4 & 1852 & 793 & 50 & 743 & 1358 & 494\\
42 & 4 & 1852 & 793 & 649 & 144 & 42 & 1810\\
42 & 4 & 1852 & 793 & 698 & 95 & 66 & 1786\\
42 & 4 & 1852 & 793 & 431 & 362 & 348 & 1504\\
42 & 4 & 1852 & 793 & 245 & 548 & 142 & 1710\\
\addlinespace
43 & 4 & 1852 & 997 & 239 & 758 & 1188 & 664\\
43 & 4 & 1852 & 997 & 964 & 33 & 960 & 892\\
43 & 4 & 1852 & 997 & 768 & 229 & 1774 & 78\\
43 & 4 & 1852 & 997 & 374 & 623 & 330 & 1522\\
43 & 4 & 1852 & 997 & 772 & 225 & 1442 & 410\\
\addlinespace
44 & 4 & 1852 & 1235 & 1069 & 166 & 1318 & 534\\
44 & 4 & 1852 & 1235 & 1055 & 180 & 759 & 1093\\
44 & 4 & 1852 & 1235 & 103 & 1132 & 1781 & 71\\
44 & 4 & 1852 & 1235 & 715 & 520 & 1516 & 336\\
44 & 4 & 1852 & 1235 & 307 & 928 & 1636 & 216\\
\addlinespace
45 & 4 & 1852 & 1515 & 1276 & 239 & 1119 & 733\\
45 & 4 & 1852 & 1515 & 1276 & 239 & 118 & 1734\\
45 & 4 & 1852 & 1515 & 1237 & 278 & 921 & 931\\
45 & 4 & 1852 & 1515 & 1165 & 350 & 749 & 1103\\
45 & 4 & 1852 & 1515 & 928 & 587 & 1768 & 84\\
\addlinespace
46 & 4 & 1852 & 1852 & 566 & 1286 & 1161 & 691\\
46 & 4 & 1852 & 1852 & 536 & 1316 & 1809 & 43\\
46 & 4 & 1852 & 1852 & 1502 & 350 & 441 & 1411\\
46 & 4 & 1852 & 1852 & 454 & 1398 & 1820 & 32\\
46 & 4 & 1852 & 1852 & 773 & 1079 & 1455 & 397\\
\addlinespace
47 & 4 & 1852 & 2264 & 655 & 1609 & 1022 & 830\\
47 & 4 & 1852 & 2264 & 1457 & 807 & 1310 & 542\\
47 & 4 & 1852 & 2264 & 1287 & 977 & 530 & 1322\\
47 & 4 & 1852 & 2264 & 100 & 2164 & 176 & 1676\\
47 & 4 & 1852 & 2264 & 107 & 2157 & 1339 & 513\\*
\end{longtable}

\newpage
\subsection{Additional Results}\label{sec:appendix_results}
\paragraph{Regressions}\indent

\begin{table}[H]
\centering
\caption{Altruistic choice treatment effect regression}\label{tab:reg_main}
\resizebox{0.8\textwidth}{!}{%
\centering\centering
\fontsize{10}{12}\selectfont
\begin{tabular}[b]{lcccc}
\toprule
  & (1) & (2) & (3) & (4)\\
\midrule
Treatment Group & -0.022*** & -0.022 & 0.009 & -0.014\\
 & (0.004) & (0.024) & (0.030) & (0.023)\\
Ratio $\frac{\text{self}}{\text{other}}$ &  &  & 0.900*** & 0.867***\\
 &  &  & (0.018) & (0.013)\\
Treatment Group * Ratio $\frac{\text{self}}{\text{other}}$ &  &  & -0.067** & \\
 &  &  & (0.026) & \\
Intercept & 0.452*** & 0.452*** & 0.032 & 0.052***\\
 & (0.003) & (0.016) & (0.021) & (0.019)\\
\midrule
Random Effects & No & No & No & Yes\\
Clustered Standard Errors & No & Yes & Yes & Yes\\
$N$ & 72000 & 72000 & 72000 & 72000\\
Unique Obs & 300 & 300 & 300 & 300\\
$R^2$ & 0.001 & 0.001 & 0.431 & 0.517\\
\bottomrule
\multicolumn{5}{p{0.7\textwidth}}{\textit{Note:} Linear probability model with choice for $\text{self}$ as dependent variable. Clustered standard errors (participant-level, ``bias-reduced linearization'' \citep{pustejovskySmallSampleMethodsClusterRobust2018}) in parentheses. * p $<$ 0.1, ** p $<$ 0.05, *** p $<$ 0.01.}

\end{tabular}

}
\end{table}

\begin{table}[H]
\centering
\caption{Altruistic choice treatment effect regression probit model}\label{tab:reg_main_probit}
\resizebox{0.8\textwidth}{!}{%
\centering\centering
\fontsize{10}{12}\selectfont
\begin{tabular}[b]{lcccc}
\toprule
  & (1) & (2) & (3) & (4)\\
\midrule
Treatment Group & -0.057*** & -0.057 & 0.154 & 0.663***\\
 & (0.009) & (0.060) & (0.137) & (0.171)\\
Ratio $\frac{\text{self}}{\text{other}}$ &  &  & 3.380*** & 6.225***\\
 &  &  & (0.153) & (0.065)\\
Treatment Group $\times$ Ratio $\frac{\text{self}}{\text{other}}$ &  &  & -0.566*** & -1.757***\\
 &  &  & (0.197) & (0.078)\\
Intercept & -0.121*** & -0.121*** & -1.654*** & -3.056***\\
 & (0.007) & (0.041) & (0.100) & (0.122)\\
\midrule
Random Effects & No & No & No & Yes\\
Clustered Standard Errors & No & Yes & Yes & No\\
Unique Obs & 300 & 300 & 300 & 300\\
pseudo $R^2$ & 0 & 0 & 0.375 & -\\
$N$ & 72000 & 72000 & 72000 & 72000\\
\bottomrule
\multicolumn{5}{p{0.7\textwidth}}{\textit{Note:} Probit Model with choice for $\text{self}$ as dependent variable. Clustered standard errors (participant-level, ``bias-reduced linearization'' \citep{pustejovskySmallSampleMethodsClusterRobust2018}) in parentheses. * p $<$ 0.1, ** p $<$ 0.05, *** p $<$ 0.01.}
\end{tabular}

}
\end{table}

\begin{table}[H]
\centering
\caption{Number comparison treatment effect regression}\label{tab:reg_number}
\resizebox{0.7\textwidth}{!}{%
\centering
\fontsize{10}{12}\selectfont
\begin{tabular}[b]{lccc}
\toprule
  & (1) & (2) & (3)\\
\midrule
Treatment Group & -0.037*** & -0.220*** & -0.220***\\
 & (0.005) & (0.022) & (0.022)\\
Ratio $\frac{\text{A}}{\text{B}}$ &  & -0.932*** & -0.932***\\
 &  & (0.005) & (0.005)\\
Treatment Group $\times$ Ratio $\frac{\text{A}}{\text{B}}$ &  & 0.114*** & 0.114***\\
 &  & (0.011) & (0.011)\\
Intercept & 0.388*** & 1.880*** & 1.880***\\
 & (0.002) & (0.009) & (0.009)\\
\midrule
Random Effects & No & No & Yes\\
Clustered Standard Erorrs & Yes & Yes & Yes\\
$N$ & 60000 & 60000 & 60000\\
Unique Obs & 300 & 300 & 300\\
$R^2$ & 0.001 & 0.794 & 0.798\\
\bottomrule
\multicolumn{4}{p{0.6\textwidth}}{\textit{Note:} Linear probability model with choice for $\text{A}$ as dependent variable. Clustered standard errors (participant-level, ``bias-reduced linearization'' \citep{pustejovskySmallSampleMethodsClusterRobust2018}) in parentheses. * p $<$ 0.1, ** p $<$ 0.05, *** p $<$ 0.01.}
\end{tabular}

}
\end{table}

\begin{table}[H]
\centering
\caption{Number comparison treatment effect probit model}\label{tab:reg_number_probit}
\resizebox{0.8\textwidth}{!}{%
\centering\centering
\fontsize{10}{12}\selectfont
\begin{tabular}[b]{lcccc}
\toprule
  & (1) & (2) & (3) & (4)\\
\midrule
Treatment Group & -0.099*** & -0.099*** & 0.839*** & 0.800***\\
 & (0.010) & (0.012) & (0.286) & (0.077)\\
Ratio $\frac{\text{A}}{\text{B}}$ &  &  & 8.693*** & 9.131***\\
 &  &  & (0.576) & (0.104)\\
Treatment Group $\times$ Ratio $\frac{\text{A}}{\text{B}}$ &  &  & -2.298*** & -2.259***\\
 &  &  & (0.639) & (0.126)\\
Intercept & -0.284*** & -0.284*** & -4.431*** & -4.643***\\
 & (0.007) & (0.005) & (0.261) & (0.060)\\
\midrule
Random Effects & No & No & No & Yes\\
Clustered Standard Errors & No & Yes & Yes & No\\
$N$ & 300 & 300 & 300 & 300\\
pseudo $R^2$ & 0.001 & 0.001 & 0.696 & -\\
Num.Obs. & 60000 & 60000 & 60000 & 60000\\
\bottomrule
\multicolumn{5}{p{0.7\textwidth}}{\textit{Note:} Probit model with choice for $\text{A}$ as dependent variable. Clustered standard errors (participant-level, ``bias-reduced linearization'' \citep{pustejovskySmallSampleMethodsClusterRobust2018}) in parentheses. * p $<$ 0.1, ** p $<$ 0.05, *** p $<$ 0.01.}
\end{tabular}

}
\end{table}

\begin{table}[H]
\centering
\caption{Inconsistencies across tasks regression}\label{tab:inconsistencies_compared_tasks}
\resizebox{0.7\textwidth}{!}{%
\centering\centering
\fontsize{10}{12}\selectfont
\begin{tabular}[t]{lccc}
		\toprule
		& (1) & (2) & (3)\\
		\midrule
		Altruism Task & 0.023*** & 0.023*** & 0.037***\\
		& (0.002) & (0.002) & (0.002)\\
		Intercept & 0.079*** & 0.073*** & -0.015\\
		& (0.002) & (0.020) & (0.021)\\
		\midrule
		$N$ & 26400 & 26400 & 26400\\
		$R^2$ & 0.004 & 0.070 & 0.191\\
		Participant FE & No & Yes & Yes \\
		Game FE & No & No & Yes \\		
		Clustered Standard Errors & No & Yes & Yes \\ \bottomrule
		\multicolumn{4}{p{0.6\textwidth}}{\textit{Note:} Linear Probability Model. Dependent variable is the standard deviation in a particular game. Clustered standard errors (participant-level) in parentheses. * p $<$ 0.1, ** p $<$ 0.05, *** p $<$ 0.01.}\\
	\end{tabular}
}
\end{table}

\newpage
\subsubsection{Probabilistic Model}\label{sec:appendix_prob_model_details}


\paragraph{Priors}
\indent 

\begin{equation*}
  \begin{split}
  \mu^{\nu_{\frac{\text{self}}{\text{other}}}} &\sim \mathcal{N}(-0.5, 0.25)\\  
\mu^{\nu_{\frac{\beta}{1-\beta}}} &\sim \mathcal{N}(-0.5, 0.25)\\
\mu^{\frac{\beta}{1-\beta}} &\sim \mathcal{N}(-0.5, 0.25)\\
\mu^{\mu_{\hat{r}}} &\sim \mathcal{N}(-0.5, 0.25)\\
	 \mu^{\nu_{\frac{\text{self}}{\text{other}}}}_B - \mu^{\nu_{\frac{\text{self}}{\text{other}}}}_T & \sim \mathcal{N}(0, 0.25) \\
  \end{split}
\quad\quad
  \begin{split}
\sigma^{\nu_{\frac{\text{self}}{\text{other}}}} &\sim \mathcal{N}^+(0,0.25)\\ 
\sigma^{\nu_{\frac{\beta}{1-\beta}}} &\sim \mathcal{N}^+(0,0.25)\\ 
\sigma^{\frac{\beta}{1-\beta}} &\sim \mathcal{N}^+(0, 0.25)\\  
\sigma^{\mu_{\hat{r}}} &\sim \mathcal{N}^+(0,0.25)\\ 
	 \mu^{\nu_{\frac{\beta}{1-\beta}}}_B - \mu^{\nu_{\frac{\beta}{1-\beta}}}_T & \sim \mathcal{N}(0, 0.25) \\
  \end{split}
\end{equation*}
\vspace{-0.5cm}
\begin{equation*}
	 \Omega \sim \text{LKJ}(2)
\end{equation*}

\paragraph{Prior Summary and Predictive Checks}
\indent 

\begin{figure}[H]
        \centering
        \begin{subfigure}{0.54\textwidth}  
            \centering 
                  \resizebox{1\linewidth}{!}{%
        
\begin{tabular}[b]{llllll}
\toprule
  & mean & median & sd & hdi 2.5\% & hdi 97.5\%\\
\midrule
\textit{Base Parameters:} &  &  &  &  & \\
Altr. Preference $\beta$ & 0.395 & 0.392 & 0.111 & 0.197 & 0.619\\
Prior Mean Outcomes $\mu^{\hat{r}}$ & 0.65 & 0.619 & 0.175 & 0.34 & 1.004\\
\addlinespace
\textit{Group Specific:} &  &  &  &  & \\
Noise Baseline $\nu_{\frac{\beta}{1-\beta},B}$ & 0.634 & 0.616 & 0.159 & 0.34 & 0.911\\
Noise Treatment $\nu_{\frac{\beta}{1-\beta},T}$ & 0.653 & 0.617 & 0.229 & 0.279 & 1.104\\
Noise Baseline $\nu_{\frac{\text{self}}{\text{other}},B}$ & 0.638 & 0.613 & 0.169 & 0.361 & 1.012\\
Noise Treatment $\nu_{\frac{\text{self}}{\text{other}},T}$ & 0.653 & 0.612 & 0.24 & 0.289 & 1.116\\
Weight on Payments Baseline $\alpha_B$ & 0.713 & 0.727 & 0.101 & 0.482 & 0.872\\
Weight on Payments Treatment $\alpha_T$ & 0.708 & 0.728 & 0.135 & 0.445 & 0.923\\
Prior Threshold Baseline $\delta_B$ & 1.148 & 1.133 & 0.111 & 0.969 & 1.38\\
Prior Threshold Treatment $\delta_T$ & 1.153 & 1.127 & 0.131 & 0.969 & 1.443\\
\bottomrule
\end{tabular}
}
		\centering\footnotesize
		\textbf{(a)} Prior parameter summary
        \end{subfigure}
        \hfill
        \begin{subfigure}{0.425\textwidth}   
            \centering 
      \resizebox{1\linewidth}{!}{%
        \input{graphs_export_restricted/structural_estimation/plot_predictions_prior}}
		\centering\footnotesize
		\textbf{(b)} Average and prior-predicted choices
        \end{subfigure}
        \caption[Summary Prior Probabilistic Model Altruistic Choices]
        {\small Summary Prior Probabilistic Model Altruistic Choices \quad \textbf{(a)} Prior parameter values of equation \ref{eq:choice_function} based on 10000 prior samples (i.e., before providing experimental data). Parameters correspond to the mean of log-normal hyper-distributions. Mean, median and sd mean, median and standard deviation of the prior distribution samples. HDI 2.5\% and HDI 97.5\% indicate the borders of the 95\% highest-density interval (HDI). \textbf{(b)} Average and prior-predicted choices, including 95\% HDI.         
         }\label{fig:model_summary_prior}
    \end{figure}

\newpage
\paragraph{Posterior Summaries}
\indent 
\begin{table}[H]
\caption{Posterior parameter summary average individual parameters altruistic choice}\label{tab:structural_estimation_population}  
\centering

\begin{tabular}[b]{lrrrrr}
\toprule
  & mean & median & hdi 2.5\% & hdi 97.5\% & $\hat{R}$\\
\midrule
$\mu^{\nu_{\frac{\beta}{1-\beta}}}_B - \mu^{\nu_{\frac{\beta}{1-\beta}}}_T$ & -0.033 & -0.028 & -0.371 & 0.301 & 1.00\\
$\mu^{\nu_{\frac{\text{self}}{\text{other}}}}_T - \mu^{\nu_{\frac{\text{self}}{\text{other}}}}_B$ & 0.244 & 0.245 & 0.045 & 0.445 & 1.00\\
$\mu^{\nu_{\frac{\beta}{1-\beta}}}$ & -1.783 & -1.778 & -2.019 & -1.554 & 1.01\\
$\mu^{\nu_{\frac{\text{self}}{\text{other}}}}$ & -1.342 & -1.339 & -1.523 & -1.165 & 1.00\\
$\mu^{\mu^{^{\hat{r}}}}$ & -0.316 & -0.316 & -0.779 & 0.131 & 1.01\\
\addlinespace
$\mu^{\frac{\beta}{1-\beta}}$ & -0.795 & -0.795 & -0.875 & -0.715 & 1.00\\
$\sigma^{\nu_{\frac{\beta}{1-\beta}}}$ & 0.415 & 0.405 & 0.230 & 0.621 & 1.00\\
$\sigma^{\nu_{\frac{\text{self}}{\text{other}}}}$ & 0.586 & 0.580 & 0.434 & 0.748 & 1.00\\
$\sigma^{\frac{\beta}{1-\beta}}$ & 0.206 & 0.204 & 0.151 & 0.265 & 1.00\\
$\sigma^{\mu_{\hat{r}}}$ & 0.786 & 0.781 & 0.447 & 1.145 & 1.00\\
\bottomrule
\end{tabular}

\end{table}


\begin{table}[H]
\caption{Posterior parameter summary hyper-parameters number comparison}\label{tab:structural_estimation_population_number}
\centering

\begin{tabular}[b]{lrrrrr}
\toprule
  & mean & median & hdi 2.5\% & hdi 97.5\% & $\hat{R}$\\
\midrule
$\mu^{\nu_{\frac{\text{A}}{\text{B}}}}$ & -1.754 & -1.754 & -1.844 & -1.664 & 1\\
$\mu^{\mu_{\hat{r}}}$ & -1.101 & -1.102 & -1.269 & -0.938 & 1\\
$\mu^{\nu_{\frac{\text{A}}{\text{B}}}}_T - \mu^{\nu_{\frac{\text{A}}{\text{B}}}}_B$ & 0.370 & 0.371 & 0.236 & 0.505 & 1\\
$\sigma^{\nu_{\frac{\text{A}}{\text{B}}}}$ & 0.513 & 0.512 & 0.465 & 0.562 & 1\\
$\sigma^{\mu_{\hat{r}}}$ & 0.927 & 0.925 & 0.800 & 1.062 & 1\\
\bottomrule
\end{tabular}

\end{table}


\begin{figure}[H]
        \centering
        \begin{subfigure}{0.49\textwidth}
        \setcounter{subfigure}{0}
            \centering
                \resizebox{1\linewidth}{!}{%
	\input{graphs_export_restricted/structural_estimation/scatter_average_maintask_manual.tex}}
		\centering\footnotesize
		\textbf{(a)} Altruism Average and Individually Predicted 
        \end{subfigure}
        \hfill
        \begin{subfigure}{0.49 \textwidth}  
            \centering 
                  \resizebox{1\linewidth}{!}{%
	\input{graphs_export_restricted/structural_estimation/scatter_average_number_manual.tex}}		\centering\footnotesize
		\textbf{(b)} Number Comparison Average and Individually Predicted
        \end{subfigure}
        \mycaption[Individual Average and Predicted Behavior: Altruism and Number Comparison]{\quad Correlation between average choice for $\text{self}$ (a) and $\text{A}$ (b) and predicted choice at $\frac{\text{self}}{other}$ and $\frac{\text{A}}{B}$ implemented in the experiment. Rank-correlations are $\rho = 0.999$ (a) and $0.939$ (b).}\label{fig:scatter_avg_prediction}
\end{figure}

\paragraph{Model Comparisons}\indent

\begin{figure}[H]
        \begin{subfigure}{0.615\textwidth}  
            \centering 
                  \resizebox{1\linewidth}{!}{%
     \begin{tabular}{@{}lll@{}}
\toprule
Model & Choice Function & $ELPD_{WAIC}$ \\ \midrule
Full Model (equation \ref{eq:choice_function})           &          ${Pr}([\text{self} \succ \text{other}])=\Phi(\frac{\alpha \times \ln \left(\frac{\text{self}}{\text{other}}\right)- \ln \left(\frac{\beta}{1-\beta}\right) -\ln (\delta)}{\sqrt{\nu_{\frac{\text{self}}{\text{other}}}^2\alpha^2 + \nu_{\frac{\beta}{1-\beta}}^2}})$ & -14,927.47 \\
Payment Prior Mean $\mu_{\hat{r}}$ = 1           &    ${Pr}([\text{self} \succ \text{other}])=\Phi(\frac{\alpha \times \ln \left(\frac{\text{self}}{\text{other}}\right)- \ln \left(\frac{\beta}{1-\beta}\right)}{\sqrt{\nu_{\frac{\text{self}}{\text{other}}}^2\alpha^2 + \nu_{\frac{\beta}{1-\beta}}^2}})$       & -14,931.52 \\
Preference Noise $\nu_{\frac{\beta}{1-\beta}} = 0$           &   ${Pr}([\text{self} \succ \text{other}])=\Phi(\frac{\alpha \times \ln \left(\frac{\text{self}}{\text{other}}\right)- \ln \left(\frac{\beta}{1-\beta}\right)}{\alpha \times \nu_{\frac{\text{self}}{\text{other}}}})$ & -14,930.33 \\
Monetary Payment Noise $\nu_{\frac{\text{self}}{\text{other}}} = 0$           &   ${Pr}([\text{self} \succ \text{other}])=\Phi(\frac{\ln \left(\frac{\text{self}}{\text{other}}\right)- \ln \left(\frac{\beta}{1-\beta}\right)}{\nu_{\frac{\beta}{1-\beta}}})$ & -14,935.94 \\
Random Utility           &   ${Pr}([\text{self} \succ \text{other}])= \frac{e^{\sigma (1-\beta)\text{self}}}{e^{\sigma (1-\beta)\text{self}} + e^{\sigma \beta \text{other}}}$       & -15,656.30 \\
 \end{tabular}}
		\centering\footnotesize
		\textbf{(a)} Models Altruistic Choice
        \end{subfigure}
\hfill
        \begin{subfigure}{0.375\textwidth}   
            \centering 
      \resizebox{1\linewidth}{!}{%
\begin{tikzpicture}[x=1pt,y=1pt]
\definecolor{fillColor}{RGB}{255,255,255}
\path[use as bounding box,fill=fillColor,fill opacity=0.00] (0,0) rectangle (578.16,361.35);
\begin{scope}
\path[clip] (273.86, 44.29) rectangle (570.16,353.35);
\definecolor{drawColor}{gray}{0.92}

\path[draw=drawColor,line width= 1.7pt,line join=round] (273.86, 79.95) --
	(570.16, 79.95);

\path[draw=drawColor,line width= 1.7pt,line join=round] (273.86,139.39) --
	(570.16,139.39);

\path[draw=drawColor,line width= 1.7pt,line join=round] (273.86,198.82) --
	(570.16,198.82);

\path[draw=drawColor,line width= 1.7pt,line join=round] (273.86,258.26) --
	(570.16,258.26);

\path[draw=drawColor,line width= 1.7pt,line join=round] (273.86,317.69) --
	(570.16,317.69);
\definecolor{fillColor}{RGB}{0,0,0}

\path[fill=fillColor] (514.53,317.69) --
	(519.17,322.33) --
	(523.81,317.69) --
	(519.17,313.05) --
	cycle;

\path[fill=fillColor] (513.76,258.26) --
	(518.40,262.90) --
	(523.04,258.26) --
	(518.40,253.62) --
	cycle;

\path[fill=fillColor] (513.44,198.82) --
	(518.08,203.46) --
	(522.72,198.82) --
	(518.08,194.18) --
	cycle;

\path[fill=fillColor] (512.26,139.39) --
	(516.90,144.03) --
	(521.54,139.39) --
	(516.90,134.75) --
	cycle;

\path[fill=fillColor] (319.17, 79.95) --
	(323.81, 84.59) --
	(328.45, 79.95) --
	(323.81, 75.31) --
	cycle;
\definecolor{drawColor}{RGB}{0,0,0}

\path[draw=drawColor,line width= 1.4pt,dash pattern=on 4pt off 4pt ,line join=round] (519.17, 44.29) -- (519.17,353.35);

\path[draw=drawColor,line width= 1.4pt,line join=round] (556.69,311.75) --
	(556.69,323.63);

\path[draw=drawColor,line width= 1.4pt,line join=round] (556.69,317.69) --
	(481.64,317.69);

\path[draw=drawColor,line width= 1.4pt,line join=round] (481.64,311.75) --
	(481.64,323.63);

\path[draw=drawColor,line width= 1.4pt,line join=round] (556.00,252.31) --
	(556.00,264.20);

\path[draw=drawColor,line width= 1.4pt,line join=round] (556.00,258.26) --
	(480.80,258.26);

\path[draw=drawColor,line width= 1.4pt,line join=round] (480.80,252.31) --
	(480.80,264.20);

\path[draw=drawColor,line width= 1.4pt,line join=round] (555.64,192.88) --
	(555.64,204.77);

\path[draw=drawColor,line width= 1.4pt,line join=round] (555.64,198.82) --
	(480.53,198.82);

\path[draw=drawColor,line width= 1.4pt,line join=round] (480.53,192.88) --
	(480.53,204.77);

\path[draw=drawColor,line width= 1.4pt,line join=round] (554.50,133.44) --
	(554.50,145.33);

\path[draw=drawColor,line width= 1.4pt,line join=round] (554.50,139.39) --
	(479.30,139.39);

\path[draw=drawColor,line width= 1.4pt,line join=round] (479.30,133.44) --
	(479.30,145.33);

\path[draw=drawColor,line width= 1.4pt,line join=round] (360.29, 74.01) --
	(360.29, 85.90);

\path[draw=drawColor,line width= 1.4pt,line join=round] (360.29, 79.95) --
	(287.33, 79.95);

\path[draw=drawColor,line width= 1.4pt,line join=round] (287.33, 74.01) --
	(287.33, 85.90);
\definecolor{drawColor}{RGB}{0,0,0}

\path[draw=drawColor,draw opacity=0.50,line width= 1.4pt,line join=round] (519.17,325.12) --
	(519.17,328.09);

\path[draw=drawColor,draw opacity=0.50,line width= 1.4pt,line join=round] (519.17,326.60) --
	(519.17,326.60);

\path[draw=drawColor,draw opacity=0.50,line width= 1.4pt,line join=round] (519.17,325.12) --
	(519.17,328.09);

\path[draw=drawColor,draw opacity=0.50,line width= 1.4pt,line join=round] (518.96,265.68) --
	(518.96,268.66);

\path[draw=drawColor,draw opacity=0.50,line width= 1.4pt,line join=round] (518.96,267.17) --
	(517.84,267.17);

\path[draw=drawColor,draw opacity=0.50,line width= 1.4pt,line join=round] (517.84,265.68) --
	(517.84,268.66);

\path[draw=drawColor,draw opacity=0.50,line width= 1.4pt,line join=round] (519.07,206.25) --
	(519.07,209.22);

\path[draw=drawColor,draw opacity=0.50,line width= 1.4pt,line join=round] (519.07,207.74) --
	(517.10,207.74);

\path[draw=drawColor,draw opacity=0.50,line width= 1.4pt,line join=round] (517.10,206.25) --
	(517.10,209.22);

\path[draw=drawColor,draw opacity=0.50,line width= 1.4pt,line join=round] (518.58,146.82) --
	(518.58,149.79);

\path[draw=drawColor,draw opacity=0.50,line width= 1.4pt,line join=round] (518.58,148.30) --
	(515.22,148.30);

\path[draw=drawColor,draw opacity=0.50,line width= 1.4pt,line join=round] (515.22,146.82) --
	(515.22,149.79);

\path[draw=drawColor,draw opacity=0.50,line width= 1.4pt,line join=round] (341.05, 87.38) --
	(341.05, 90.35);

\path[draw=drawColor,draw opacity=0.50,line width= 1.4pt,line join=round] (341.05, 88.87) --
	(306.57, 88.87);

\path[draw=drawColor,draw opacity=0.50,line width= 1.4pt,line join=round] (306.57, 87.38) --
	(306.57, 90.35);
\end{scope}
\begin{scope}
\path[clip] (  0.00,  0.00) rectangle (578.16,361.35);
\definecolor{drawColor}{RGB}{0,0,0}

\node[text=drawColor,anchor=base east,inner sep=0pt, outer sep=0pt, scale=  1.33] at (267.56, 74.45) {Random Utility Model};

\node[text=drawColor,anchor=base east,inner sep=0pt, outer sep=0pt, scale=  1.33] at (267.56,133.88) {Monetary Payment Noise $\nu_{\frac{self}{other}} = 0$};

\node[text=drawColor,anchor=base east,inner sep=0pt, outer sep=0pt, scale=  1.33] at (267.56,193.31) {Payment Prior Mean $\mu_{\hat{r}} = 1$};

\node[text=drawColor,anchor=base east,inner sep=0pt, outer sep=0pt, scale=  1.33] at (267.56,252.75) {Preferences Noise $\nu_{\frac{\beta}{1-\beta}} = 0$};

\node[text=drawColor,anchor=base east,inner sep=0pt, outer sep=0pt, scale=  1.33] at (267.56,312.18) {Full Model (equation \ref{eq:choice_function})};
\end{scope}
\begin{scope}
\path[clip] (  0.00,  0.00) rectangle (578.16,361.35);
\definecolor{drawColor}{gray}{0.30}

\node[text=drawColor,anchor=base,inner sep=0pt, outer sep=0pt, scale=  1.33] at (298.69, 26.98) {-15750};

\node[text=drawColor,anchor=base,inner sep=0pt, outer sep=0pt, scale=  1.33] at (365.70, 26.98) {-15500};

\node[text=drawColor,anchor=base,inner sep=0pt, outer sep=0pt, scale=  1.33] at (432.72, 26.98) {-15250};

\node[text=drawColor,anchor=base,inner sep=0pt, outer sep=0pt, scale=  1.33] at (499.73, 26.98) {-15000};

\node[text=drawColor,anchor=base,inner sep=0pt, outer sep=0pt, scale=  1.33] at (566.74, 26.98) {-14750};
\end{scope}
\begin{scope}
\path[clip] (  0.00,  0.00) rectangle (578.16,361.35);
\definecolor{drawColor}{RGB}{0,0,0}

\node[text=drawColor,anchor=base,inner sep=0pt, outer sep=0pt, scale=  1.17] at (422.01, 10.72) {$ELPD\_{WAIC}$};
\end{scope}
\begin{scope}
\path[clip] (  0.00,  0.00) rectangle (578.16,361.35);
\definecolor{drawColor}{RGB}{0,0,0}

\node[text=drawColor,rotate= 90.00,anchor=base,inner sep=0pt, outer sep=0pt, scale=  1.17] at ( 17.64,198.82) { };
\end{scope}
\end{tikzpicture}}
		\centering\footnotesize
		\textbf{(b)} $ELPD_{WAIC}$ values
        \end{subfigure}
\mycaption[Model Comparison Altruistic Choices]{\quad $ELPD_{WAIC}$ refers to the expected log predictive density as based on the widely-applicable information criterion (WAIC); A larger $ELPD_{WAIC}$ indicates a better model fit. Error bars show the standard error of the respective $ELPD_{WAIC}$ value and the standard error of the $\Delta ELPD_{WAIC}$ value, the $ELPD_{WAIC}$ difference to the best model. Model comparison done via the \texttt{arviz}-package \autocite{kumarArviZUnifiedLibrary2019}.} 
        \label{fig:model_comparison}

    \end{figure}

\begin{figure}[H]
        \begin{subfigure}{0.615\textwidth}  
            \centering 
                  \resizebox{1\linewidth}{!}{%
\begin{tabular}{@{}llc@{}}
\toprule
Model & Choice Function & $ELPD_{WAIC}$ \\ \midrule
Main Model (equation \ref{eq:choice_function_number})           &          ${Pr}([\text{A} \succ \text{B} \times \sfrac{1}{2}]) = \Phi \frac{\alpha^{'} \log \frac{\text{A}}{\text{B}} - \log \frac{1}{2} - \log \delta^{'}}{\nu^{'} \times \alpha^{'}}$ & 10,470.85 \\
Magnitude Prior Mean  $\mu_{\hat{r}'} = 1$ &                   $({Pr}[(\text{A} \succ \text{B} \times \sfrac{1}{2}]) = \Phi \frac{\alpha^{'} \log \frac{\text{A}}{\text{B}} - \log \frac{1}{2}}{\nu^{'} \times \alpha^{'}}$ & -11,523.44	 \\
 \end{tabular}}

		\centering\footnotesize
		\vspace{1.5cm}
		\textbf{(a)} Models Number Comparison
        \end{subfigure}
\hfill
        \begin{subfigure}{0.375\textwidth}   
            \centering 
      \resizebox{1\linewidth}{!}{%
\begin{tikzpicture}[x=1pt,y=1pt]
\definecolor{fillColor}{RGB}{255,255,255}
\path[use as bounding box,fill=fillColor,fill opacity=0.00] (0,0) rectangle (578.16,361.35);
\begin{scope}
\path[clip] (240.19, 44.29) rectangle (570.16,353.35);
\definecolor{drawColor}{gray}{0.92}

\path[draw=drawColor,line width= 1.7pt,line join=round] (240.19,128.58) --
	(570.16,128.58);

\path[draw=drawColor,line width= 1.7pt,line join=round] (240.19,269.06) --
	(570.16,269.06);
\definecolor{fillColor}{RGB}{0,0,0}

\path[fill=fillColor] (515.94,269.06) --
	(520.58,273.70) --
	(525.22,269.06) --
	(520.58,264.42) --
	cycle;

\path[fill=fillColor] (287.43,128.58) --
	(292.07,133.22) --
	(296.71,128.58) --
	(292.07,123.94) --
	cycle;
\definecolor{drawColor}{RGB}{0,0,0}

\path[draw=drawColor,line width= 1.4pt,dash pattern=on 4pt off 4pt ,line join=round] (520.58, 44.29) -- (520.58,353.35);

\path[draw=drawColor,line width= 1.4pt,line join=round] (555.16,255.01) --
	(555.16,283.11);

\path[draw=drawColor,line width= 1.4pt,line join=round] (555.16,269.06) --
	(485.99,269.06);

\path[draw=drawColor,line width= 1.4pt,line join=round] (485.99,255.01) --
	(485.99,283.11);

\path[draw=drawColor,line width= 1.4pt,line join=round] (328.96,114.53) --
	(328.96,142.63);

\path[draw=drawColor,line width= 1.4pt,line join=round] (328.96,128.58) --
	(255.19,128.58);

\path[draw=drawColor,line width= 1.4pt,line join=round] (255.19,114.53) --
	(255.19,142.63);
\definecolor{drawColor}{RGB}{0,0,0}

\path[draw=drawColor,draw opacity=0.50,line width= 1.4pt,line join=round] (520.58,286.62) --
	(520.58,293.65);

\path[draw=drawColor,draw opacity=0.50,line width= 1.4pt,line join=round] (520.58,290.13) --
	(520.58,290.13);

\path[draw=drawColor,draw opacity=0.50,line width= 1.4pt,line join=round] (520.58,286.62) --
	(520.58,293.65);

\path[draw=drawColor,draw opacity=0.50,line width= 1.4pt,line join=round] (302.36,146.14) --
	(302.36,153.17);

\path[draw=drawColor,draw opacity=0.50,line width= 1.4pt,line join=round] (302.36,149.65) --
	(281.79,149.65);

\path[draw=drawColor,draw opacity=0.50,line width= 1.4pt,line join=round] (281.79,146.14) --
	(281.79,153.17);
\end{scope}
\begin{scope}
\path[clip] (  0.00,  0.00) rectangle (578.16,361.35);
\definecolor{drawColor}{RGB}{0,0,0}

\node[text=drawColor,anchor=base east,inner sep=0pt, outer sep=0pt, scale=  1.33] at (233.89,123.07) {Magnitude Prior Mean $\mu_{\hat{r}'} = 1$};

\node[text=drawColor,anchor=base east,inner sep=0pt, outer sep=0pt, scale=  1.33] at (233.89,263.55) {Main Model};
\end{scope}
\begin{scope}
\path[clip] (  0.00,  0.00) rectangle (578.16,361.35);
\definecolor{drawColor}{gray}{0.30}

\node[text=drawColor,anchor=base,inner sep=0pt, outer sep=0pt, scale=  1.33] at (318.87, 26.98) {-11400};

\node[text=drawColor,anchor=base,inner sep=0pt, outer sep=0pt, scale=  1.33] at (405.71, 26.98) {-11000};

\node[text=drawColor,anchor=base,inner sep=0pt, outer sep=0pt, scale=  1.33] at (492.54, 26.98) {-10600};
\end{scope}
\begin{scope}
\path[clip] (  0.00,  0.00) rectangle (578.16,361.35);
\definecolor{drawColor}{RGB}{0,0,0}

\node[text=drawColor,anchor=base,inner sep=0pt, outer sep=0pt, scale=  1.17] at (405.17, 10.72) {$ELPD\_{WAIC}$};
\end{scope}
\begin{scope}
\path[clip] (  0.00,  0.00) rectangle (578.16,361.35);
\definecolor{drawColor}{RGB}{0,0,0}

\node[text=drawColor,rotate= 90.00,anchor=base,inner sep=0pt, outer sep=0pt, scale=  1.17] at ( 17.64,198.82) { };
\end{scope}
\end{tikzpicture}}
		\centering\footnotesize
		\textbf{(b)} $ELPD_{WAIC}$ values
        \end{subfigure}
\mycaption[Model Comparison Number Comparison]{\quad $ELPD_{WAIC}$ refers to the expected log predictive density as based on the widely-applicable information criterion (WAIC); A larger $ELPD_{WAIC}$ indicates a better model fit. Error bars show the standard error of the respective $ELPD_{WAIC}$ value and the standard error of the $\Delta ELPD_{WAIC}$ value, the $ELPD_{WAIC}$ difference to the best model. Model comparison done via the \texttt{arviz}-package \autocite{kumarArviZUnifiedLibrary2019}.} 
        \label{fig:model_comparison_number}

    \end{figure}

\subsubsection{Combined Estimation}\label{sec:appendix_combined_estimation}

\begin{table}[H]
\caption{Parameter summary combined estimation}\label{tab:parameters_combined}
\centering
\resizebox{0.9\textwidth}{!}{%

\begin{tabular}[b]{lllllll}
\toprule
  & mean & median & sd & hdi 2.5\% & hdi 97.5\% & $\hat{R}$\\
\midrule
\textit{Base Parameters:} &  &  &  &  &  & \\
Altr. Preference $\beta$ & 0.315 & 0.315 & 0.012 & 0.292 & 0.339 & 1\\
Prior Mean Outcomes $\mu^{\hat{r}}$ & 0.474 & 0.467 & 0.062 & 0.366 & 0.6 & 1\\
\addlinespace
\textit{Group Specific:} &  &  &  &  &  & \\
Noise Baseline $\nu_{\frac{\beta}{1-\beta},B}$ & 0.339 & 0.335 & 0.038 & 0.27 & 0.415 & 1\\
Noise Treatment $\nu_{\frac{\beta}{1-\beta},T}$ & 0.323 & 0.32 & 0.035 & 0.262 & 0.394 & 1\\
Noise Baseline $\nu_{\frac{\text{self,A}}{\text{other,B}},B}$ & 0.172 & 0.172 & 0.008 & 0.158 & 0.188 & 1\\
Noise Treatment $\nu_{\frac{\text{self,A}}{\text{other,B}},T}$ & 0.255 & 0.255 & 0.012 & 0.233 & 0.279 & 1\\
Weight on Payments Baseline $\alpha_B$ & 0.971 & 0.971 & 0.003 & 0.966 & 0.976 & 1\\
Weight on Payments Treatment $\alpha_T$ & 0.939 & 0.939 & 0.005 & 0.928 & 0.949 & 1\\
Prior Threshold Baseline $\delta_B$ & 1.022 & 1.022 & 0.004 & 1.014 & 1.03 & 1\\
Prior Threshold Treatment $\delta_T$ & 1.047 & 1.048 & 0.009 & 1.029 & 1.066 & 1\\
\bottomrule
\end{tabular}
}
\end{table}

\begin{table}[h]
\caption{Posterior parameter summary hyper-parameters combined estimation}\label{tab:structural_estimation_population_combined}
\centering

\begin{tabular}[b]{lrrrrr}
\toprule
  & mean & median & hdi 2.5\% & hdi 97.5\% & $\hat{R}$\\
\midrule
$\mu^{\nu_{\frac{\beta}{1-\beta}}}_B - \mu^{\nu_{\frac{\beta}{1-\beta}}}_T$ & -0.047 & -0.046 & -0.251 & 0.153 & 1.00\\
$\mu^{\nu_{\frac{\text{self,A}}{\text{other,B}}}}_T - \mu^{\nu_{\frac{\text{self,A}}{\text{other,B}}}}_B$ & 0.394 & 0.396 & 0.260 & 0.520 & 1.01\\
$\mu^{\nu_{\frac{\beta}{1-\beta}}}$ & -1.519 & -1.519 & -1.667 & -1.366 & 1.00\\
$\mu^{\nu_{\frac{\text{self,A}}{\text{other,B}}}}$ & -1.792 & -1.793 & -1.878 & -1.703 & 1.00\\
$\mu^{\mu_{\hat{r}}}$ & -1.132 & -1.133 & -1.297 & -0.967 & 1.00\\
\addlinespace
$\mu^{\frac{\beta}{1-\beta}}$ & -0.980 & -0.980 & -1.072 & -0.894 & 1.00\\
$\sigma^{\nu_{\frac{\beta}{1-\beta}}}$ & 0.924 & 0.920 & 0.752 & 1.109 & 1.00\\
$\sigma^{\nu_{\frac{\text{self,A}}{\text{other,B}}}}$ & 0.254 & 0.252 & 0.206 & 0.302 & 1.00\\
$\sigma^{\frac{\beta}{1-\beta}}$ & 0.638 & 0.635 & 0.543 & 0.745 & 1.00\\
$\sigma^{\mu_{\hat{r}}}$ & 0.863 & 0.859 & 0.659 & 1.079 & 1.00\\
\bottomrule
\end{tabular}

\end{table}

\begin{figure}[H]
        \begin{subfigure}{1\textwidth}  
            \centering 
                  \resizebox{1\linewidth}{!}{%
\begin{tabular}{@{}lccc@{}}
\toprule
Model & Combined Choice Functions & $ELPD_{WAIC,A}$ & $ELPD_{WAIC,NC}$ \\ \midrule
Full Model Combined           &          ${Pr}([\text{self} \succ \text{other}])=\Phi\left(\frac{\alpha \times \ln \left(\frac{\text{self}}{\text{other}}\right)- \ln \left(\frac{\beta}{1-\beta}\right) -\ln (\delta)}{\sqrt{\alpha^2 \times \nu_{\frac{\text{self,A}}{\text{other,B}}}^2 + \nu_{\frac{\beta}{1-\beta}}^2}}\right); {Pr}([\text{A} \succ \text{B} \times \sfrac{1}{2}])=\Phi\left(\frac{\alpha \times \ln \left(\frac{\text{A}}{\text{B}}\right)- \ln (\frac{1}{2}) -\ln (\frac{1}{\mu_{\hat{r}^{1-\alpha}}})}{\alpha \times \nu_{\frac{\text{self,A}}{\text{other,B}}} }\right)$   & -15,103.79	 & -10,519.21 \\
Preferences Noise $\nu_{\frac{\beta}{1-\beta}} = 0$ Combined           &          ${Pr}([\text{self} \succ \text{other}])=\Phi\left(\frac{\alpha \times \ln \left(\frac{\text{self}}{\text{other}}\right)- \ln \left(\frac{\beta}{1-\beta}\right) -\ln (\frac{1}{\mu_{\hat{r}^{1-\alpha}}})}{\alpha \times \nu_{\frac{\text{self,A}}{\text{other,B}}} }\right); {Pr}([\text{A} \succ \text{B} \times \sfrac{1}{2}])=\Phi\left(\frac{\alpha \times \ln \left(\frac{\text{A}}{\text{B}}\right)- \ln (\frac{1}{2}) -\ln (\frac{1}{\mu_{\hat{r}^{1-\alpha}}})}{\alpha \times \nu_{\frac{\text{self,A}}{\text{other,B}}} }\right)$  & -15,640.32 & -11,853.78 \\
 \end{tabular}}
 \vspace{0.1cm}

		\centering\footnotesize
		\textbf{(a)} Models Altruism and Number Comparison
        \end{subfigure}
\hfill
        \begin{subfigure}{0.45\textwidth}   
            \centering 
      \resizebox{1\linewidth}{!}{%
\begin{tikzpicture}[x=1pt,y=1pt]
\definecolor{fillColor}{RGB}{255,255,255}
\path[use as bounding box,fill=fillColor,fill opacity=0.00] (0,0) rectangle (578.16,361.35);
\begin{scope}
\path[clip] (288.42, 44.29) rectangle (570.16,353.35);
\definecolor{drawColor}{gray}{0.92}

\path[draw=drawColor,line width= 1.7pt,line join=round] (288.42,128.58) --
	(570.16,128.58);

\path[draw=drawColor,line width= 1.7pt,line join=round] (288.42,269.06) --
	(570.16,269.06);
\definecolor{fillColor}{RGB}{0,0,0}

\path[fill=fillColor] (510.88,269.06) --
	(515.52,273.70) --
	(520.16,269.06) --
	(515.52,264.42) --
	cycle;

\path[fill=fillColor] (349.71,128.58) --
	(354.35,133.22) --
	(358.99,128.58) --
	(354.35,123.94) --
	cycle;
\definecolor{drawColor}{RGB}{0,0,0}

\path[draw=drawColor,line width= 1.4pt,dash pattern=on 4pt off 4pt ,line join=round] (515.52, 44.29) -- (515.52,353.35);

\path[draw=drawColor,line width= 1.4pt,line join=round] (557.35,255.01) --
	(557.35,283.11);

\path[draw=drawColor,line width= 1.4pt,line join=round] (557.35,269.06) --
	(473.69,269.06);

\path[draw=drawColor,line width= 1.4pt,line join=round] (473.69,255.01) --
	(473.69,283.11);

\path[draw=drawColor,line width= 1.4pt,line join=round] (407.47,114.53) --
	(407.47,142.63);

\path[draw=drawColor,line width= 1.4pt,line join=round] (407.47,128.58) --
	(301.22,128.58);

\path[draw=drawColor,line width= 1.4pt,line join=round] (301.22,114.53) --
	(301.22,142.63);
\definecolor{drawColor}{RGB}{0,0,0}

\path[draw=drawColor,draw opacity=0.50,line width= 1.4pt,line join=round] (515.52,286.62) --
	(515.52,293.65);

\path[draw=drawColor,draw opacity=0.50,line width= 1.4pt,line join=round] (515.52,290.13) --
	(515.52,290.13);

\path[draw=drawColor,draw opacity=0.50,line width= 1.4pt,line join=round] (515.52,286.62) --
	(515.52,293.65);

\path[draw=drawColor,draw opacity=0.50,line width= 1.4pt,line join=round] (369.06,146.14) --
	(369.06,153.17);

\path[draw=drawColor,draw opacity=0.50,line width= 1.4pt,line join=round] (369.06,149.65) --
	(339.63,149.65);

\path[draw=drawColor,draw opacity=0.50,line width= 1.4pt,line join=round] (339.63,146.14) --
	(339.63,153.17);
\end{scope}
\begin{scope}
\path[clip] (  0.00,  0.00) rectangle (578.16,361.35);
\definecolor{drawColor}{RGB}{0,0,0}

\node[text=drawColor,anchor=base east,inner sep=0pt, outer sep=0pt, scale=  1.33] at (282.12,123.07) {Preferences Noise $\nu_{\frac{\beta}{1-\beta}} = 0$ Combined};

\node[text=drawColor,anchor=base east,inner sep=0pt, outer sep=0pt, scale=  1.33] at (282.12,263.55) {Full Model Combined};
\end{scope}
\begin{scope}
\path[clip] (  0.00,  0.00) rectangle (578.16,361.35);
\definecolor{drawColor}{gray}{0.30}

\node[text=drawColor,anchor=base,inner sep=0pt, outer sep=0pt, scale=  1.33] at (306.38, 26.98) {-15800};

\node[text=drawColor,anchor=base,inner sep=0pt, outer sep=0pt, scale=  1.33] at (366.46, 26.98) {-15600};

\node[text=drawColor,anchor=base,inner sep=0pt, outer sep=0pt, scale=  1.33] at (426.54, 26.98) {-15400};

\node[text=drawColor,anchor=base,inner sep=0pt, outer sep=0pt, scale=  1.33] at (486.62, 26.98) {-15200};

\node[text=drawColor,anchor=base,inner sep=0pt, outer sep=0pt, scale=  1.33] at (546.70, 26.98) {-15000};
\end{scope}
\begin{scope}
\path[clip] (  0.00,  0.00) rectangle (578.16,361.35);
\definecolor{drawColor}{RGB}{0,0,0}

\node[text=drawColor,anchor=base,inner sep=0pt, outer sep=0pt, scale=  1.17] at (429.29, 10.72) {$ELPD\_{WAIC}$};
\end{scope}
\begin{scope}
\path[clip] (  0.00,  0.00) rectangle (578.16,361.35);
\definecolor{drawColor}{RGB}{0,0,0}

\node[text=drawColor,rotate= 90.00,anchor=base,inner sep=0pt, outer sep=0pt, scale=  1.17] at ( 17.64,198.82) { };
\end{scope}
\end{tikzpicture}}
		\centering\footnotesize
		\textbf{(b)} $ELPD_{WAIC,A}$ Altruism
        \end{subfigure}
                \begin{subfigure}{0.45\textwidth}   
            \centering 
      \resizebox{1\linewidth}{!}{%
\begin{tikzpicture}[x=1pt,y=1pt]
\definecolor{fillColor}{RGB}{255,255,255}
\path[use as bounding box,fill=fillColor,fill opacity=0.00] (0,0) rectangle (578.16,361.35);
\begin{scope}
\path[clip] (288.42, 44.29) rectangle (570.16,353.35);
\definecolor{drawColor}{gray}{0.92}

\path[draw=drawColor,line width= 1.7pt,line join=round] (288.42,128.58) --
	(570.16,128.58);

\path[draw=drawColor,line width= 1.7pt,line join=round] (288.42,269.06) --
	(570.16,269.06);
\definecolor{fillColor}{RGB}{0,0,0}

\path[fill=fillColor] (526.15,269.06) --
	(530.79,273.70) --
	(535.43,269.06) --
	(530.79,264.42) --
	cycle;

\path[fill=fillColor] (318.97,128.58) --
	(323.61,133.22) --
	(328.25,128.58) --
	(323.61,123.94) --
	cycle;
\definecolor{drawColor}{RGB}{0,0,0}

\path[draw=drawColor,line width= 1.4pt,dash pattern=on 4pt off 4pt ,line join=round] (530.79, 44.29) -- (530.79,353.35);

\path[draw=drawColor,line width= 1.4pt,line join=round] (557.35,255.01) --
	(557.35,283.11);

\path[draw=drawColor,line width= 1.4pt,line join=round] (557.35,269.06) --
	(504.23,269.06);

\path[draw=drawColor,line width= 1.4pt,line join=round] (504.23,255.01) --
	(504.23,283.11);

\path[draw=drawColor,line width= 1.4pt,line join=round] (345.99,114.53) --
	(345.99,142.63);

\path[draw=drawColor,line width= 1.4pt,line join=round] (345.99,128.58) --
	(301.22,128.58);

\path[draw=drawColor,line width= 1.4pt,line join=round] (301.22,114.53) --
	(301.22,142.63);
\definecolor{drawColor}{RGB}{0,0,0}

\path[draw=drawColor,draw opacity=0.50,line width= 1.4pt,line join=round] (530.79,286.62) --
	(530.79,293.65);

\path[draw=drawColor,draw opacity=0.50,line width= 1.4pt,line join=round] (530.79,290.13) --
	(530.79,290.13);

\path[draw=drawColor,draw opacity=0.50,line width= 1.4pt,line join=round] (530.79,286.62) --
	(530.79,293.65);

\path[draw=drawColor,draw opacity=0.50,line width= 1.4pt,line join=round] (332.10,146.14) --
	(332.10,153.17);

\path[draw=drawColor,draw opacity=0.50,line width= 1.4pt,line join=round] (332.10,149.65) --
	(315.11,149.65);

\path[draw=drawColor,draw opacity=0.50,line width= 1.4pt,line join=round] (315.11,146.14) --
	(315.11,153.17);
\end{scope}
\begin{scope}
\path[clip] (  0.00,  0.00) rectangle (578.16,361.35);
\definecolor{drawColor}{RGB}{0,0,0}

\node[text=drawColor,anchor=base east,inner sep=0pt, outer sep=0pt, scale=  1.33] at (282.12,123.07) {Preferences Noise $\nu_{\frac{\beta}{1-\beta}} = 0$ Combined};

\node[text=drawColor,anchor=base east,inner sep=0pt, outer sep=0pt, scale=  1.33] at (282.12,263.55) {Full Model Combined};
\end{scope}
\begin{scope}
\path[clip] (  0.00,  0.00) rectangle (578.16,361.35);
\definecolor{drawColor}{gray}{0.30}

\node[text=drawColor,anchor=base,inner sep=0pt, outer sep=0pt, scale=  1.33] at (300.91, 26.98) {-12000};

\node[text=drawColor,anchor=base,inner sep=0pt, outer sep=0pt, scale=  1.33] at (378.53, 26.98) {-11500};

\node[text=drawColor,anchor=base,inner sep=0pt, outer sep=0pt, scale=  1.33] at (456.15, 26.98) {-11000};

\node[text=drawColor,anchor=base,inner sep=0pt, outer sep=0pt, scale=  1.33] at (533.77, 26.98) {-10500};
\end{scope}
\begin{scope}
\path[clip] (  0.00,  0.00) rectangle (578.16,361.35);
\definecolor{drawColor}{RGB}{0,0,0}

\node[text=drawColor,anchor=base,inner sep=0pt, outer sep=0pt, scale=  1.17] at (429.29, 10.72) {$ELPD\_{WAIC}$};
\end{scope}
\begin{scope}
\path[clip] (  0.00,  0.00) rectangle (578.16,361.35);
\definecolor{drawColor}{RGB}{0,0,0}

\node[text=drawColor,rotate= 90.00,anchor=base,inner sep=0pt, outer sep=0pt, scale=  1.17] at ( 17.64,198.82) { };
\end{scope}
\end{tikzpicture}}
		\centering\footnotesize
		\textbf{(c)} $ELPD_{WAIC,NC}$ Number Comparison
        \end{subfigure}
        
\mycaption[Model Comparison Combined Models]{\quad (a) Altruism Choices ($ELPD_{WAIC,A}$) and (b) Number Comparison ($ELPD_{WAIC,NC}$).} 
        \label{fig:model_comparison_combined}

    \end{figure}

\subsubsection{Robustness of Treatment Variation}\label{sec:appendix_robustness_treatment}

\paragraph{Learning Effects and Fatigue} First, I investigate the role of learning effects on altruistic and number comparison behavior. A straightforward way to do so is to augment the linear probability models of Tables \ref{tab:reg_main} and \ref{tab:reg_number} by a $Round$ variable, which indicates in which of the 300 (200) rounds a decision was made. If the treatment effect is ``learned,'' I expect a negative coefficient of the interaction effect between the treatment dummy and the round variable, i.e., a treatment effect that grows over time. The result of the corresponding linear probability model is depicted in Table \ref{tab:reg_learning}, where the first two columns refer to the altruism data and the last to the number comparison data. In the first two specifications, the coefficient of the interaction effect is indeed negative ($-0.00006$) and comparing round 0 to round 300 implies a 1.8 percentage point difference in selfish choices, which is sizable compared to the overall treatment effect. However, the coefficient is statistically insignificant ($p > 0.1$) in both specifications. In addition, if I take the results of column 1 at face value, already in round 0 the Treatment group decides 1.43 percentage points less often for $\text{self}$, which speaks against the responsibility of learning effects for the treatment difference. I arrive at a similar conclusion, albeit with different evidence, for the number comparison data: In columns 3 and 4, I include the mentioned interaction effect. I observe a statistically significant \textit{positive} coefficient of the interaction effect of ($0.00013$), which implies an \textit{increase} in 2.6 percentage points to choose $\text{A}$ between round 0 and round 200. Instead of growing over time, this implies that the treatment effect shrinks. Supporting this argument is that in round 0, the treatment effect is sizable and statistically significant and the Treatment group decides 4.916, respectively 5.84 percentage points less for $\text{A}$. The columns thus do not provide evidence for a learned treatment effect and instead, point towards some attenuation over time in the number comparison data.

\begin{table}[h]
\caption{Treatment effect and learning regression}
\centering
\fontsize{10}{12}\selectfont
\resizebox{1\textwidth}{!}{%
\begin{tabular}[!htpb]{lcccc}
\toprule
  & (1) & (2) & (3) & (4)\\
\midrule
Treatment Group & -0.01430** & -0.01430 & -0.04916*** & -0.05084***\\
 & (0.00603) & (0.02342) & (0.00362) & (0.00690)\\
Ratio $\frac{\text{self}}{\text{other}}$ / $\frac{\text{A}}{\text{B}}$ & 0.86707*** & 0.86707*** & -0.87519*** & 1.32264***\\
 & (0.00372) & (0.01323) & (0.00184) & (0.00736)\\
Treatment Group * Round No. & -0.00006 & -0.00006 & 0.00012*** & 0.00013***\\
 & (0.00004) & (0.00006) & (0.00003) & (0.00005)\\
Intercept & 0.05165*** & 0.05165*** & 1.78986*** & -0.18654***\\
 & (0.00459) & (0.01862) & (0.00390) & (0.00444)\\
\midrule
$N$ & 72000 & 72000 & 60000 & 60000\\
Data & Altruism & Altruism & Number Comp. & Number Comp.\\
Clustered Standard Errors & No & Yes & No & Yes\\
Random Effects & No & Yes & No & Yes\\
Unique Obs & 300 & 300 & 300 & 300\\
$R^2$ & 0.430 & 0.517 & 0.791 & 0.679\\
\bottomrule
\multicolumn{5}{p{0.8\textwidth}}{\textit{Note:} Linear Probability Model. Clustered standard errors (participant-level, ``bias-reduced linearization'' \citep{pustejovskySmallSampleMethodsClusterRobust2018}) in parentheses. * p $<$ 0.1, ** p $<$ 0.05, *** p $<$ 0.01.}
\end{tabular}}

\label{tab:reg_learning}
\end{table}

Two other facts that are insightful for learning over time come from the decision in the \textit{very first} illustrative example as well as the 12 consecutive practice trials. At the very beginning and as part of explaining the study, participants had to make a non-consequential decision whether to take $2.31$ \EUR (= 1.72 \EUR + 0.59 \EUR) for themselves or give $4.66$ \EUR (= 1.14 \EUR + 3.52 \EUR) to another person. In this decision, there is no treatment difference as $\bar{\text{self}}_T = 0.393$, $\bar{\text{self}}_B = 0.367$ ($p = 0.6356$). However, in the 12 practice trials\footnote{
Recall that in the practice trials, I fixed $\text{other}$ = 10.00 \EUR and varied $\text{self} \in [0, 0.52, 1.11, 1.76, 2.50, 3.33, 4.28, 5.38, 6.66, 8.18, 10.00, 12.22]$ \EUR, i.e., these trials in principle already allow to infer something about $\beta$.} the Treatment group decides significantly more often for the other person with an average of $\bar{\text{self}}_T = 0.364$, $\bar{\text{self}}_B = 0.414$ ($p < 0.01$). Thus, during the practice trials, the Treatment group is much more pro-social. While I acknowledge limits for drawing conclusions from this data -- given it is non-incentivized, only for practice purposes and contains only 10 decisions -- this behavior would be consistent with the following explanation: Participants quickly understand how the task works, i.e., ``less-for-me'' vs ``more-for-other''. This, in turn, could translate into the intuition that $\mu_{\hat{r}} < 1$ and thus a higher pro-sociality of the Treatment group (throughout the experiment). All in all, this data suggests that the treatment effect is not learned over the repeated trials of the experiment but that participants formed a quick intuition about the rules of the task.

The treatment variation could also introduce differences in \textit{fatigue levels}, which are then responsible for the difference in choices between both groups. If differences in fatigue are not already present in the very first choices, the above analysis already provides some evidence against this argument. In addition, I can test both for group differences in (i) revealed (effects of) fatigue and (ii) subjectively reported fatigue levels. Regarding the first, often-discussed consequences of fatigue are more errors in choices and higher levels \textit{choice inconsistency}, an argument especially relevant for survey design (see e.g., \nptextcite{bechDoesNumberChoice2011, schwappachQuickDirtyNumbers2006, ozdemirWhoPaysAttention2010}). My data offers a unique way of analyzing the determinants of choice inconsistency: Recall that each trial of the altruism and number comparison task was repeated five times (which is one game), yet the order of trials was randomly determined. This implies that some participants encountered the fifth iteration of a given game earlier in the experiment compared to other participants, which induces exogenous differences in the completion rounds of a given trial group. If fatigue (differences) increase throughout the experiment, \textit{later completions} should be associated with a higher choice inconsistency.

\begin{table}[h]
\caption{Inconsistency regression on trial level}\label{tab:reg_inconsistency}
\centering
\resizebox{1\textwidth}{!}{%
\begin{tabular}[!hptb]{lcccc}
\toprule
  & (1) & (2) & (3) & (4)\\
\midrule
Trial Final Round No. & -0.00005 & -0.00006 & -0.00005 & -0.00005\\
 & (0.00007) & (0.00006) & (0.00008) & (0.00007)\\
Trial Final Round No. * Treatment Group & 0.00000 & 0.00003 & 0.00001 & 0.00006\\
 & (0.00009) & (0.00009) & (0.00011) & (0.00010)\\
Intercept & 0.11693*** & 0.04352 & 0.10210*** & 0.03006\\
 & (0.03063) & (0.03111) & (0.03079) & (0.02785)\\
\midrule
Data & Altruism & Altruism & Number Comp. & Number Comp.\\
Trial Group Fixed Effects & No & Yes & No & Yes\\
Participant Fixed Effects & Yes & Yes & Yes & Yes\\
$N$ & 14400 & 14400 & 12000 & 12000\\
$R^2$ & 0.131 & 0.206 & 0.066 & 0.309\\
\bottomrule
\multicolumn{5}{p{0.9\textwidth}}{\textit{Note:} Clustered standard errors (participant-level, ``bias-reduced linearization'' \citep{pustejovskySmallSampleMethodsClusterRobust2018}) in parentheses. * p $<$ 0.1, ** p $<$ 0.05, *** p $<$ 0.01.}
\end{tabular}

}

\end{table}

\begin{figure}[H]
\centering
	\includegraphics[width=1\linewidth]{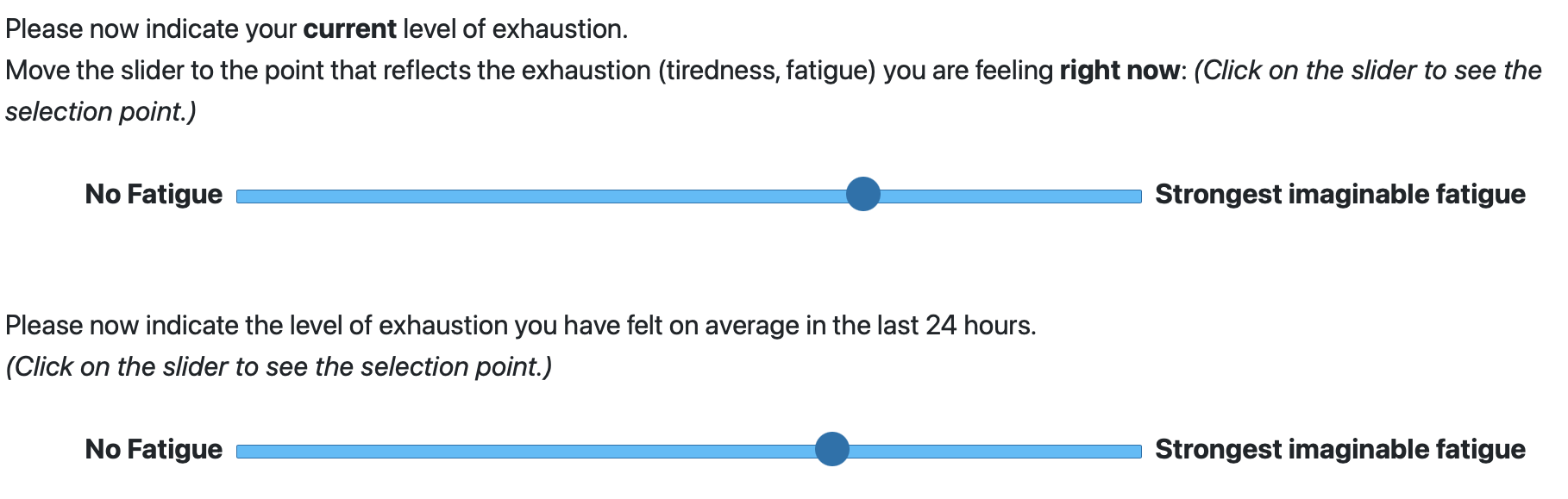}

\caption{Screenshot: Fatigue Visual Analog Scales}\label{screenshot:fatigue}
\end{figure}

Table \ref{tab:reg_inconsistency} performs a linear regression on a dataset with 300 (subject) $\times$ 48 (trial groups, see Figure \ref{fig:trials_overview}) = 14,400 respectively 200 $\times$ 40 = 12,000 observations based on the Altruism and Number Comparison data. Each row of this dataset contains the standard deviation of a given trial group by a given participant alongside its completion round, i.e., in which round a participant encountered the fifth and last iteration of that trial group. Crucially, this dataset allows for the inclusion both of participant and trial group fixed effects. Every specification in Table \ref{tab:reg_inconsistency}, regardless of the data source, does neither provide evidence for a growing inconsistency, nor a growing difference in inconsistency between both groups. Thus, participants' choices do not get more inconsistent over time nor does the choice inconsistency develop differently between the treatment and control group. In addition to implied fatigue effects, I collected self-reported measures of fatigue levels using visual analog scales \autocite{Radbruch2003}. I asked participants both about their current level of fatigue as well as the average during the past 24 hours on a scale of 0-10 using a slider (see Figure \ref{screenshot:fatigue}). The Treatment group indeed does report slightly higher levels of current fatigue ($\bar{fatigue}_T = 4.775$,$\bar{fatigue}_B = 4.310$, $p=0.1005$). However, it is not obvious that higher levels of self-reported fatigue necessarily translate into different choices. I will pick up this point in more detail in Section \ref{sec:het_treatment} during the estimation of heterogeneous treatment effects.

\paragraph{Mechanical Difference in Choices and Increase in Inconsistency}

Alternative to the above-mentioned points, an alternative explanation for the treatment effect could be a purely ``mechanical'' one: If participants only focus on the \textit{first components} of the sums in the Treatment group and simply pick the larger they would behave both more variable due to the random placement of the position of the components and behave less selfish if the first component of the $\text{other}$ variable ($\text{other}_1$) is larger more often. If this argument holds, I should observe a higher level of selfishness (compared to Baseline) if $\text{self}_1 > \text{other}_1$ and a lower level once $\text{self}_1 < \text{other}_1$.

\begin{figure}[h]
        \centering
        \begin{subfigure}{0.49\textwidth}
        \setcounter{subfigure}{0}
            \centering
                \resizebox{1\linewidth}{!}{%
	\input{graphs_export/robustness/math_plot.tex}}
		\centering\footnotesize
		\textbf{(a)} Math components: Altruism
        \end{subfigure}
        \hfill
        \begin{subfigure}{0.49 \textwidth}  
            \centering 
                  \resizebox{1\linewidth}{!}{%
	\input{graphs_export/robustness/treatment_stakes.tex}}  
		\centering\footnotesize
		\textbf{(b)} Stake sizes: Altruism
        \end{subfigure}
        \begin{subfigure}{0.49\textwidth}
        \setcounter{subfigure}{0}
            \centering
                \resizebox{1\linewidth}{!}{%
	\input{graphs_export/robustness/math_plot_numbercomp.tex}}
		\centering\footnotesize
		\textbf{(c)} Math components: Number Comparison
        \end{subfigure}
        \hfill
        \begin{subfigure}{0.49 \textwidth}  
            \centering 
                  \resizebox{1\linewidth}{!}{%
	\input{graphs_export/robustness/treatment_stakes_numbercomp.tex}}  
		\centering\footnotesize
		\textbf{(d)} Stake sizes: Number comparison
        \end{subfigure}
        \mycaption[Components of to-be-calculated Sums and Stake Sizes] {\quad Panels (a) and (c) plot the average choice for $\text{self}$, respectively $\text{A}$ as a function of $\frac{\text{self}}{\text{other}}$, $\frac{\text{A}}{\text{B}}$, for Baseline and Treatment whereas the latter is divided into cases where the first math component $\text{self}_1 > \text{other}_1$ and $\text{self}_1 < \text{other}_1$, with $\text{self} = \text{self}_1 + \text{self}_2$ and $\text{other} = \text{other}_1 + \text{other}_2$. Panels (b) and (d) plot average choices separately for the different base values of $\text{other}$ and $\text{B}$.}\label{fig:stakes_and_math}
\end{figure}

Panel (a) of Figure \ref{fig:stakes_and_math} plots the average choice for $\text{self}$ separately depending on the numerical configuration of the math components, i.e., if $\text{self}_1 > \text{other}_1$ or $\text{self}_1 < \text{other}_1$. I observe comparable differences to the Baseline group within the Treatment trials regardless of the relationship between $\text{self}_1$ and $\text{other}_1$. While indeed $\text{other}_1 > \text{self}_1$ (58.95 \%) occurs more frequently than $\text{other}_1 > \text{self}_1$ (41.05 \%), the fact that participants still behave \textit{more} pro-social compared to Baseline in both groups of trials speaks against a purely ``mechanical'' increase in pro-sociality. In the number comparison task (panel c), I observe virtually no difference in behavior between trials where $\text{A}_1 > \text{B}_1$ and $\text{A}_1 < \text{B}_1$.

Another explanation could be that the treatment variation perhaps only works for smaller values of $\text{self}, \text{other}$ where the sums generally contain smaller values. For example, one could expect a stronger treatment effect in trials such as $\text{self}$ = 3.52 (= 1.61 + 1.91) vs $\text{other}$ = 6.55(=4.86 + 1.69) compared to $\text{self}$ = 7.05 (= 4.42 + 2.63) vs $\text{other}$ = 13.1(=10.32 + 2.78) as the components of the sums in the former set of trials are simply smaller (while the ratio between $\text{self}$ and $\text{other}$ remains the same). Thus, the overall treatment effect could be driven by the impact on trials with generally smaller math components which are more likely to be disregarded by participants. Panel (b) of Figure \ref{fig:stakes_and_math} plots the average choice for $\text{self}$ separately for the four different levels of stakes in the trials. Even though there is some difference in behavior between the different stake groups, the \textit{treatment effect} is very similar across different stakes. The same is true for the number comparison task (panel (d)), where I observe similar treatment effects regardless of the value of the number $\text{B}$. Table \ref{tab:reg_stakes} performs a regression analysis akin to the linear probability models in Tables \ref{tab:reg_main} and \ref{tab:reg_number} and shows that the treatment effect does not systematically depend on the general stakes of the trial (or the value of $\text{other}$ and $\text{B}$), and also shows that larger stakes are associated with more choices for $\text{self}$ and $\text{A}$.

\begin{table}[h]
\centering
\caption{Stake-size and choice regression}\label{tab:reg_stakes}
\resizebox{0.8\textwidth}{!}{%
\begin{tabular}[!hptb]{lcc}
\toprule
  & (1) & (2)\\
\midrule
Game Group (1-4) & 0.03749*** & 0.00875***\\
 & (0.00233) & (0.00249)\\
Game Group (1-4) * Treatment Group & -0.00009 & -0.00533\\
 & (0.00330) & (0.00352)\\
Treatment Group & -0.02222** & -0.02400**\\
 & (0.00903) & (0.00965)\\
Intercept & 0.35806*** & 0.36627***\\
 & (0.00638) & (0.00682)\\
\midrule
$N$ & 72000 & 60000\\
Data & Altruism & Number Comp.\\
Clustered Standard Errors & Yes & Yes\\
Unique Obs & 300 & 300\\
$R^2$ & 0.008 & 0.002\\
\bottomrule
\multicolumn{3}{p{0.7\textwidth}}{\textit{Note:} Linear Probability Model. Clustered standard errors (participant-level, ``bias-reduced linearization'' \citep{pustejovskySmallSampleMethodsClusterRobust2018}) in parentheses. * p $<$ 0.1, ** p $<$ 0.05, *** p $<$ 0.01.}
\end{tabular}
}
\end{table}

\subsubsection{Heterogeneous Treatment Effects}\label{sec:het_treatment}

To further investigate the nature of the treatment effect, I analyze heterogeneous treatment effects. To do so, I leverage recent developments in the causal machine learning literature and employ a Causal Forest for estimating heterogeneous treatment effects \autocite{wagerEstimationInferenceHeterogeneous2018}. Causal Forests adapt the logic of tree-based models identifying a split at a given level of a covariate to minimize a loss criterion to the estimation of treatment effects and search for splits that maximize heterogeneity in the estimated conditional average treatment effect. Importantly, these methods are ``honest'', i.e., use a different set of data points to propose and evaluate the splits. I use the CausalForest class implemented in the \texttt{econml} package \autocite{econml}.\footnote{Note that I can define the treatment propensity model -- what usually needs to be estimated from the data -- as a fair coin flip given the exogenous treatment assignment in the experiment.} The main advantage over classical techniques (i.e., interaction terms in OLS) is that they -- constructed using cross-fitting techniques -- are less prone to overfitting and able to pick up other functional forms beyond linear or explicitly pre-specified ones.



\begin{figure}[h]
                \resizebox{1\linewidth}{!}{%
                \centering
	\input{graphs_export/robustness/CATE_graphs.tex}}
	\mycaption[Heterogeneous Treatment Effects]{\quad This plot shows the estimated CATE values for each personal characteristic (sorted by their importance), estimated at each decile of the feature distribution, 95 \% confidence intervals.}\label{fig:het_effects}
\end{figure}
 
Figure \ref{fig:het_effects} shows the estimated CATE values for the different personal characteristics (while holding the remaining characteristics at their median value), sorted in descending order by their importance for the estimated CATE values. The key take-away is that (i) the variation in most personal characteristics does not contribute meaningfully to the CATE estimates, and only participants high on the self-reported GPS Altruism score seem to be slightly less impacted by the treatment compared to lower-scoring individuals. Similarly, participants high on the Need for Cognition scale are less impacted by the treatment, but the confidence intervals are relatively large in both cases. For the remaining features, most variation corresponds to the average treatment effect (ATE). This is true for the remaining ``cognitive'' measures, such as the CRT or BNT performance, which do not indicate systematic treatment heterogeneity. Importantly, further is that personal characteristics that could be related to a tendency to ``exploit'' potential side-effects of the chosen treatment variation, i.e., how strongly their self-reported negative emotions after selfish behavior react to the availability of excuses, the difference between private and social norms in the treatment and the average score on the Machiavellianism scale are not major sources of heterogeneous treatment effects. This is further evidence that the treatment effect -- in addition to leading to \textit{less} selfish choices -- did not invoke motivated ``second-order'' behavior. Overall, the estimation of heterogeneous treatment effects leads to the conclusion that the treatment effect does not operate systematically differently for participants depending on their characteristics and that the present dataset is not large enough to detect minuscule differences in treatment heterogeneity.

\subsubsection{Correlation between Metacognition, RT and Choices}

 Table \ref{tab:meta_cognition_correlations} shows pairwise rank correlation coefficients between the various metacognitive measures and average choices for $\text{self}$, standard deviation (on a game level), and time spent in altruism choices as well as the average correct choices, standard deviation and time spent in the number comparison task. The main finding is that there is a stronger association between choices, inconsistencies and RT with metacognitive measures in the number comparison domain compared to the altruism choices.

 \begin{table}[H]
 \caption{Correlation metacognition altruistic choice and number comparison}\label{tab:meta_cognition_correlations}
 \centering
\resizebox{1\textwidth}{!}{\begin{tabular}[!hptb]{lcccccc}
\toprule
\multicolumn{1}{c}{ } & \multicolumn{3}{c}{Altruism (Avg.)} & \multicolumn{3}{c}{Number Comparison (Avg.)} \\
\cmidrule(l{3pt}r{3pt}){2-4} \cmidrule(l{3pt}r{3pt}){5-7}
  & Choice $\text{self}$ & Std. Deviation  & RT & Choice Correct & Std. Deviation & RT \\
\midrule
\textit{Altruism:} &  &  &  &  &  & \\
Negative Confidence & 0.088 & 0.317*** & 0.080 & -0.127* & 0.109 & 0.048\\
Avg. Attention & -0.215*** & -0.045 & 0.115* & 0.094 & -0.078 & -0.066\\
Precision & -0.195*** & -0.114 & 0.126* & 0.006 & -0.005 & -0.040\\
 &  &  &  &  &  & \\

\textit{Number Comparison:} &  &  &  &  &  & \\
$|\Delta$ Belief Correct$|$ & -0.057 & 0.257*** & 0.198*** & -0.365*** & 0.392*** & 0.433***\\
Belief Correct Confidence & 0.043 & -0.161** & -0.203*** & 0.309*** & -0.285*** & -0.365***\\
$|\Delta$ Belief Time Spent$|$ & 0.074 & 0.107 & 0.125* & -0.077 & 0.040 & 0.209***\\
Belief Time Spent Confidence & 0.033 & 0.003 & -0.121* & 0.025 & -0.031 & -0.151**\\

Precision & 0.020 & -0.017 & -0.113 & 0.254*** & -0.256*** & -0.150*\\
Avg. Attention & 0.008 & -0.131* & -0.093 & 0.223*** & -0.231*** & -0.058\\
\bottomrule
\multicolumn{7}{p{0.9\textwidth}}{\textit{Note:} $p$-values from pairwise rank-correlation tests ($n=300)$. * p $<$ 0.1, ** p $<$ 0.05, *** p $<$ 0.01.}
\end{tabular}

}

\end{table}

Starting with the upper-left quarter of the table, I observe that subjects who decide less often for $\text{self}$ also report higher levels of attention ($\rho = -0.215$) and precision ($\rho = -0.195$), yet there is no apparent correlation of altruistic choices with the confidence measure. Proceeding to the second column, I do observe a positive correlation between the average standard deviation and confidence ($\rho = 0.317$): Participants who report a lower level of confidence are more inconsistent in their altruistic behavior (but not more or less pro-social on average). The average time spent on making altruistic choices does not meaningfully correlate with any of the metacognitive measures. Notably, these correlations also indicate that the metacognitive measures do not replicate the treatment effect: One could expect that ``less metacognitive'' participants also tend to decide more often for $\text{other}$ given the direction of the treatment effect towards fewer choices for $\text{self}$, but this appears not the be the case. Together with the previous fact that there are no treatment differences in the altruism metacognitive measures, this implies that the mechanism through which the treatment effect operates is likely not via impacts on (conscious) metacognition.

The behavioral data from the altruism domain also correlate to some extent with measures of metacognition \textit{across domain} as shown in the lower-left quarter: Participants with a larger $|\Delta$ Belief Correct$|$, i.e., whose beliefs deviate more from their true performance and lower confidence in their belief statements are more inconsistent in their altruism decisions ($\rho = 0.257$; $\rho = -0.161$) and take longer to choose between $\text{self}$ and $\text{other}$ ($\rho = 0.198$; $\rho = -0.203$). This reiterates the argument in Section \ref{sec:altruism_number_comp} that altruism and number comparisons, at least to some extent, are driven by similar processes if elicited in comparable settings.

Turning to the upper-right quarter of Table \ref{tab:meta_cognition_correlations}, I observe no apparent correlations between the altruism metacognitive measures and number comparison behavior. Within-domain, this is different: As shown in the lower-right quarter, the fewer correct choices a participant makes, the more their beliefs deviate from their true performance ($\rho = -0.365$), the less confident they are in their belief estimates ($\rho = 0.309$), the lower the self-reported precision ($\rho = 0.254$) and attention ($\rho = 0.223$). Inconsistency in the number comparison also correlates with belief deviations ($\rho = 0.392$), their confidence ($\rho = -0.285$), as well as self-reported precision ($\rho = -0.256$), and attention ($\rho = -0.231$). Finally, also the time spent is larger the more a participant deviates in their belief statements ($\rho = 0.433$), lower the higher the confidence ($\rho = -0.365$), higher the more a participant deviates in their belief statements of decision time ($\rho = 0.209$) and higher the lower the confidence in these statements ($\rho = -0.151$).

\begin{figure}[H]
        \centering
        \begin{subfigure}{0.49\textwidth}
        \setcounter{subfigure}{0}
            \centering
                \resizebox{1\linewidth}{!}{%
	\input{graphs_export/meta_cognition/beliefs_correct.tex}}
		\centering\footnotesize
		\textbf{(a)} Beliefs (Avg.) Correctly Solved Tasks
        \end{subfigure}
        \hfill
        \begin{subfigure}{0.49 \textwidth}  
            \centering 
                  \resizebox{1\linewidth}{!}{%
	\input{graphs_export/meta_cognition/beliefs_time.tex}}  
		\centering\footnotesize
		\textbf{(b)} Beliefs (Avg.) Time Spent
        \end{subfigure}
        \mycaption[Beliefs Number Comparison]{\quad This plot depicts the number comparison belief data, in correctly solved tasks (a) and the average time spent (b) against either objective counterpart.}\label{fig:beliefs_number_comparison}
\end{figure}

\newpage
\subsection{Experimental Screenshots\label{sec:exp_screenshots}}

\begin{figure}[H]
        \centering
        \begin{subfigure}[b]{0.475\textwidth}
        \setcounter{subfigure}{0}
            \centering
   \includegraphics[width=1\linewidth]{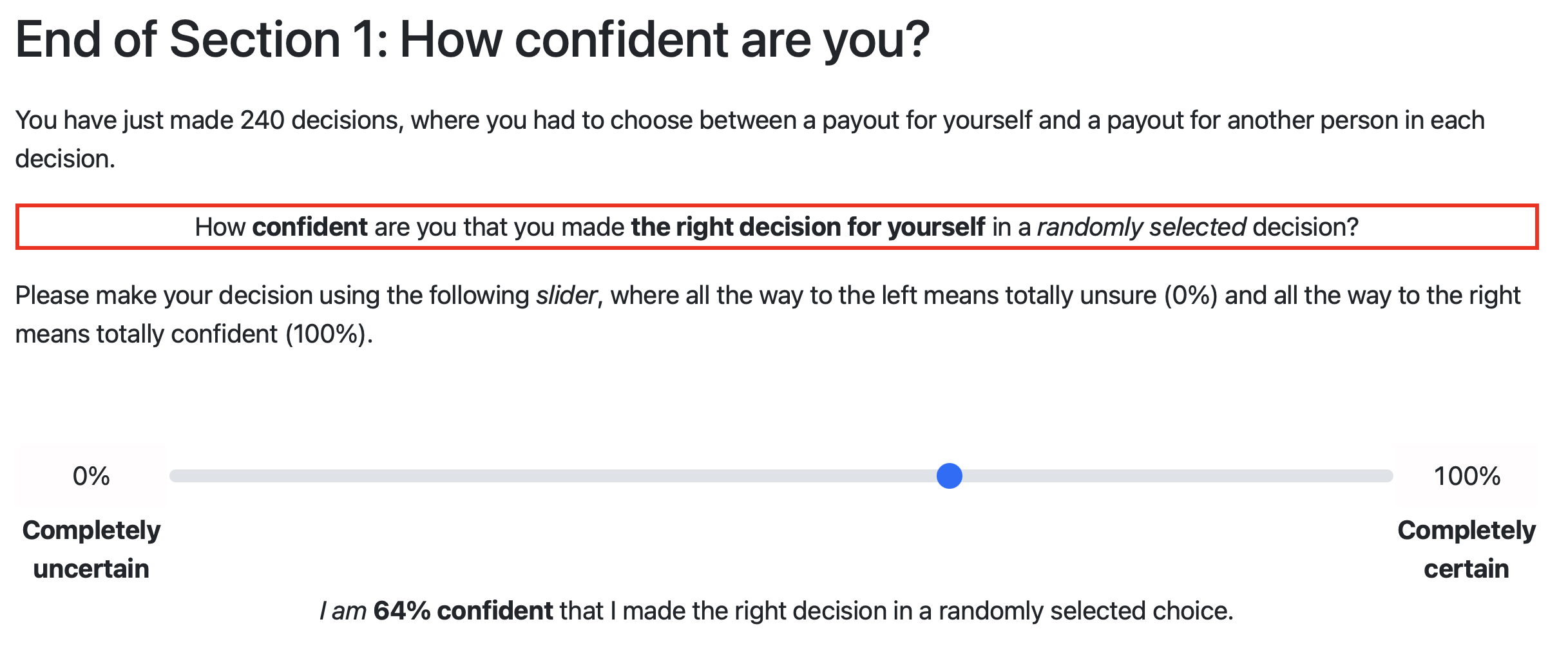}	
		\centering\footnotesize
		\textbf{(a)} Confidence
        \end{subfigure}
        \hfil
        \begin{subfigure}[b]{0.475\textwidth}  
            \centering 
   \includegraphics[width=1\linewidth]{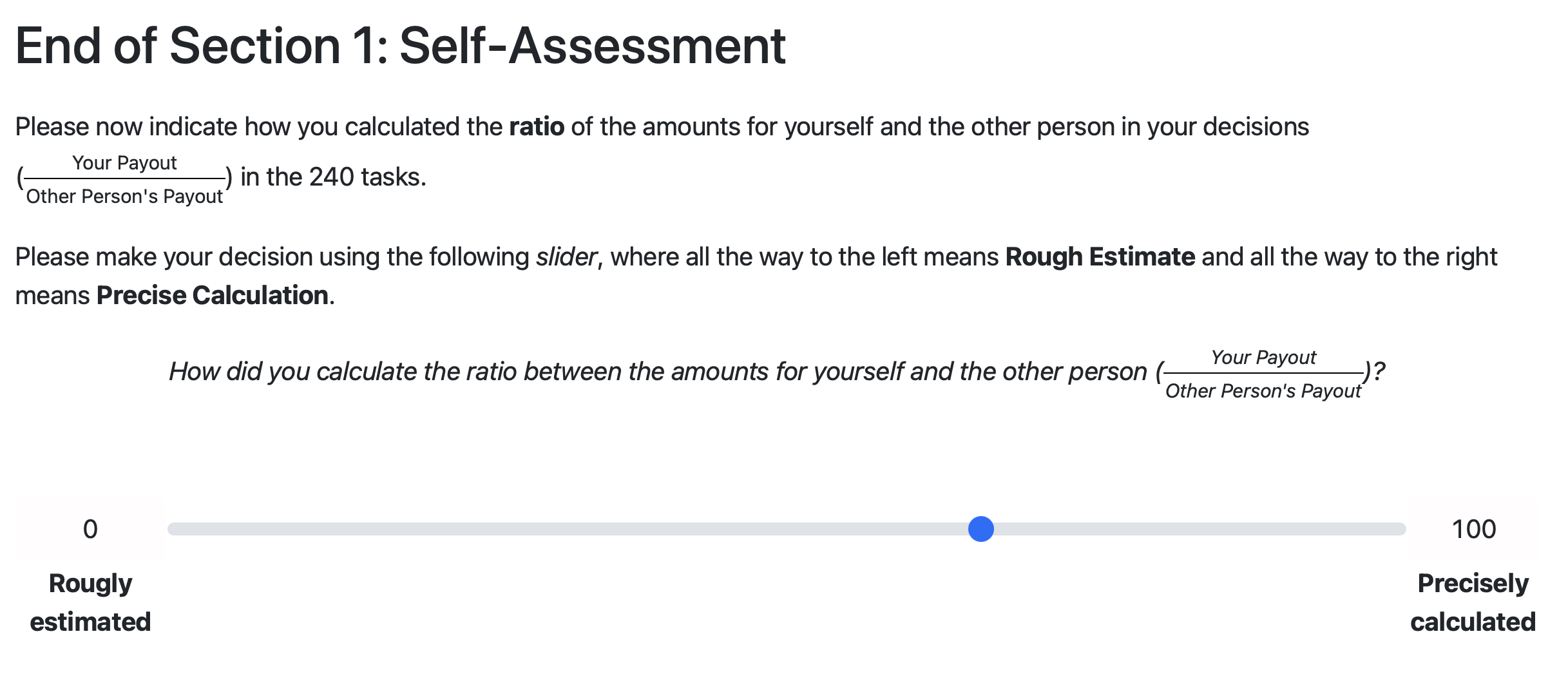}	
		\centering\footnotesize
		\textbf{(b)} Precision
        \end{subfigure}
        \vskip\baselineskip
        \begin{subfigure}[b]{0.475\textwidth}   
            \centering 
   \includegraphics[width=1\linewidth]{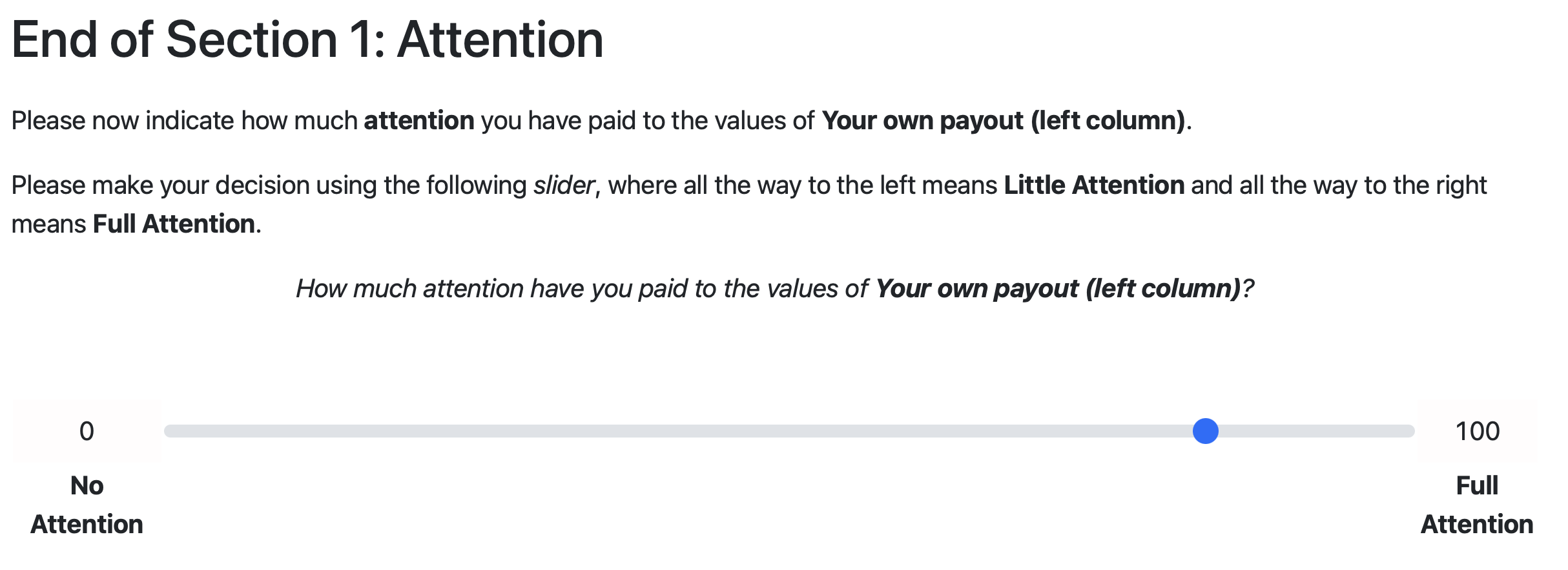}	
		\centering\footnotesize
		\textbf{(c)} Attention $\text{self}$
        \end{subfigure}
        \hfill
        \begin{subfigure}[b]{0.475\textwidth}   
            \centering 
   \includegraphics[width=1\linewidth]{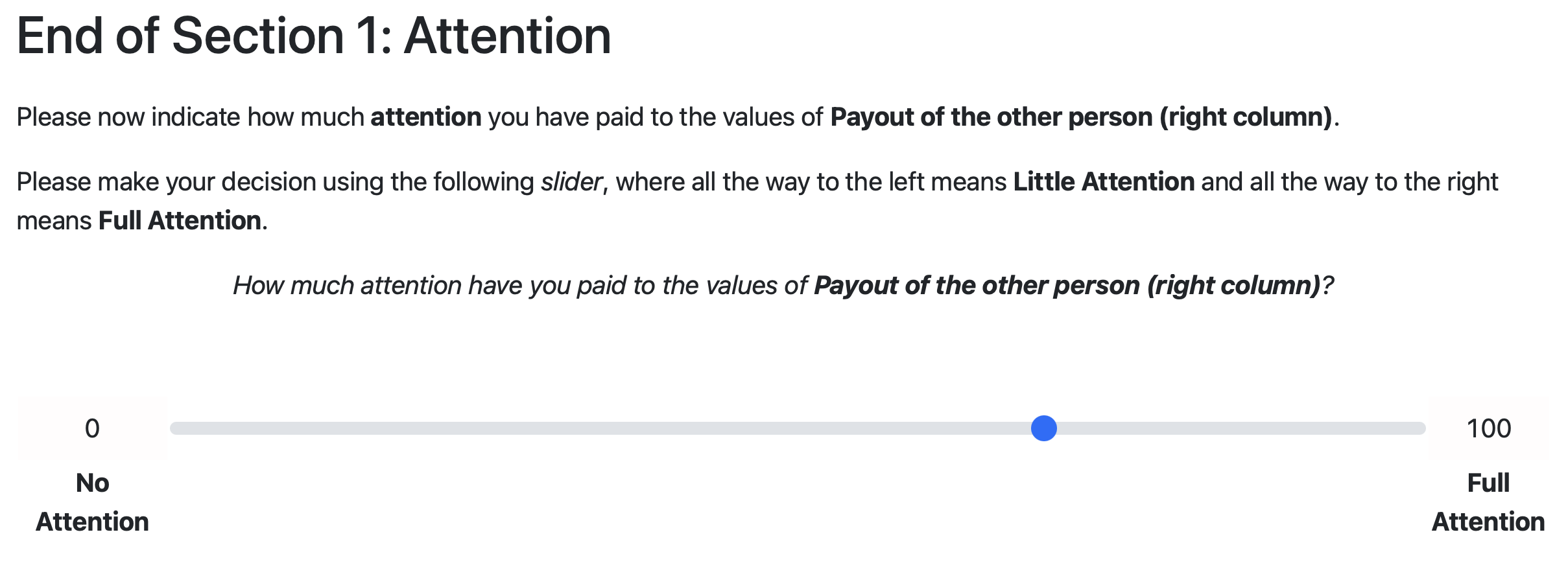}	
		\centering\footnotesize
		\textbf{(d)} Attention $\text{other}$
        \end{subfigure}
        \caption[Screenshots: Metacognition Altruistic Choice]
        {\small \quad Screenshots: Metacognition Altruistic Choice}\label{fig_screenshots_meta_cognition}
    \end{figure}

\begin{figure}[H]
\centering
	\includegraphics[width=1\linewidth]{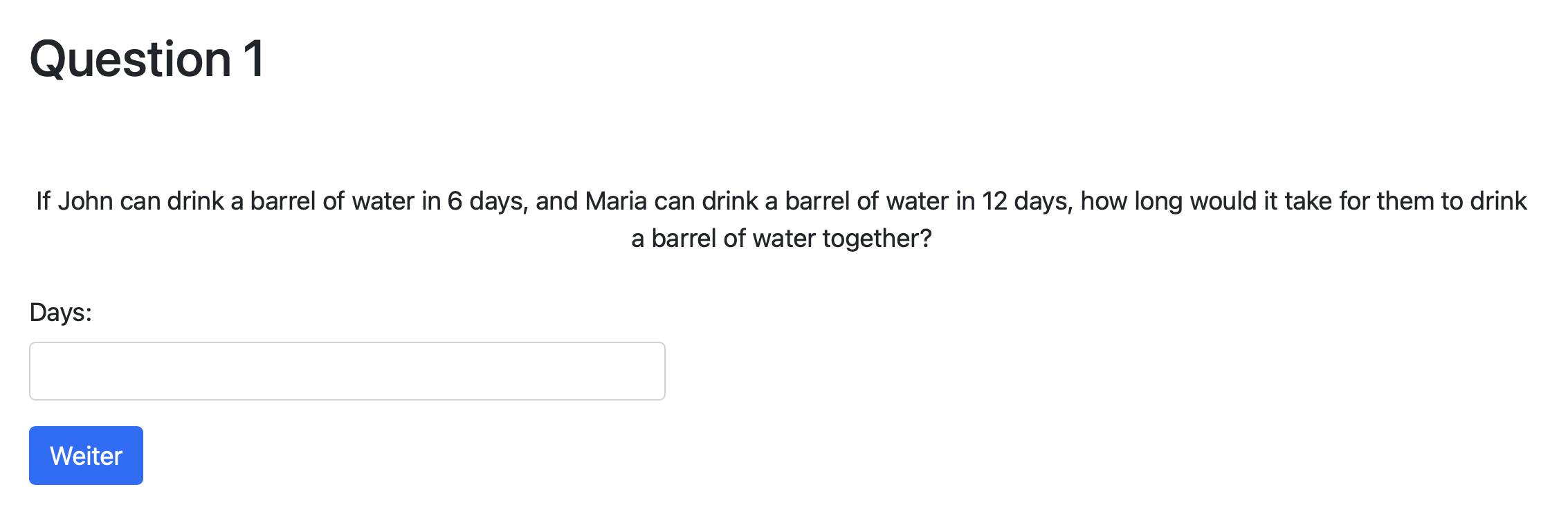}

\caption{Screenshot: CRT4 Question}\label{screenshot:crt}
\end{figure}

\begin{figure}[H]
        \centering
        \begin{subfigure}[b]{0.475\textwidth}
        \setcounter{subfigure}{0}
            \centering
   \includegraphics[width=1\linewidth]{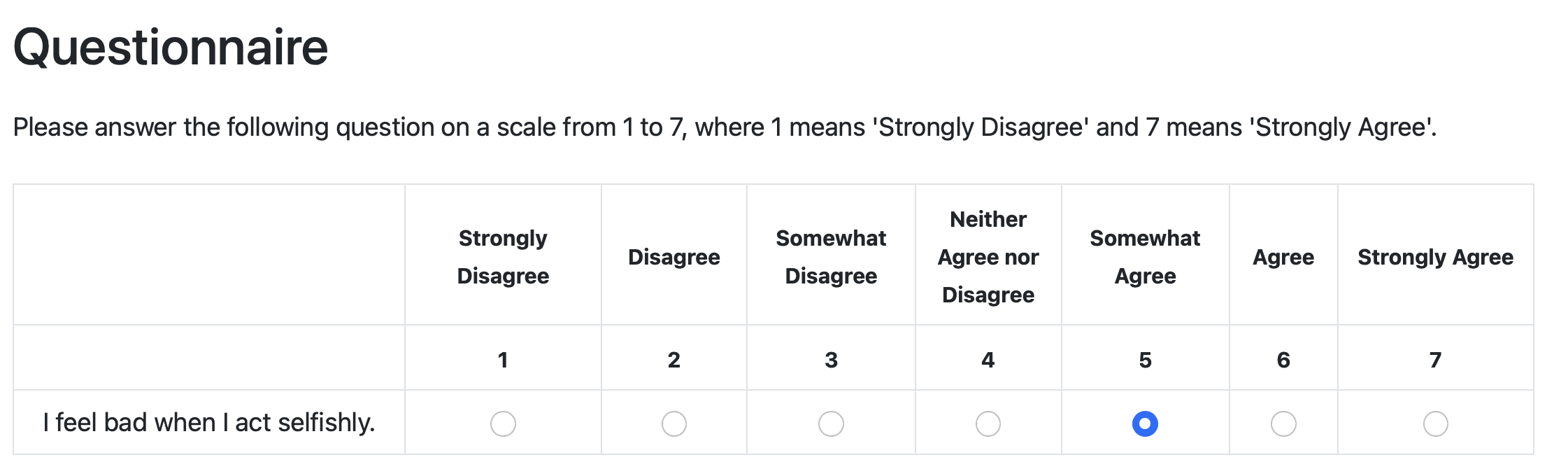}
		\centering\footnotesize
		\textbf{(a)} No Excuses
        \end{subfigure}
        \hfil
        \begin{subfigure}[b]{0.4\textwidth}  
            \centering 
   \includegraphics[width=1\linewidth]{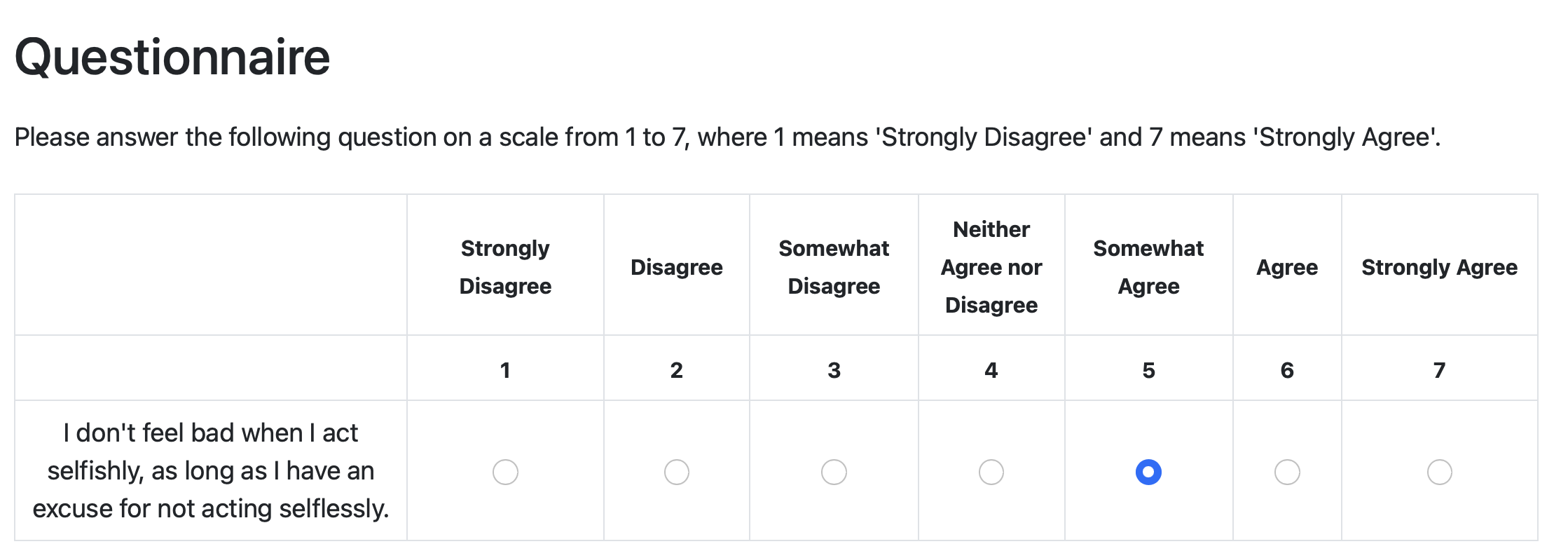}		\centering\footnotesize
		\textbf{(b)} Excuses (reverse formulated)
        \end{subfigure}
        \vskip\baselineskip
        \caption[Screenshots: Excuse-Taking Questions]
        {\small Screenshots: Excuse-Taking Questions \quad Survey Questions inspired by \textcite{lepperExcuseBasedProcrastination2024}. \textbf{(a)} No excuses \textbf{(b)} Excuses. Order in which questions appear is randomized.

        }\label{screenshot:survey_excuses}
    \end{figure}

    \begin{figure}[H]
        \centering
        \begin{subfigure}[b]{0.475\textwidth}
        \setcounter{subfigure}{0}
            \centering
   \includegraphics[width=1\linewidth]{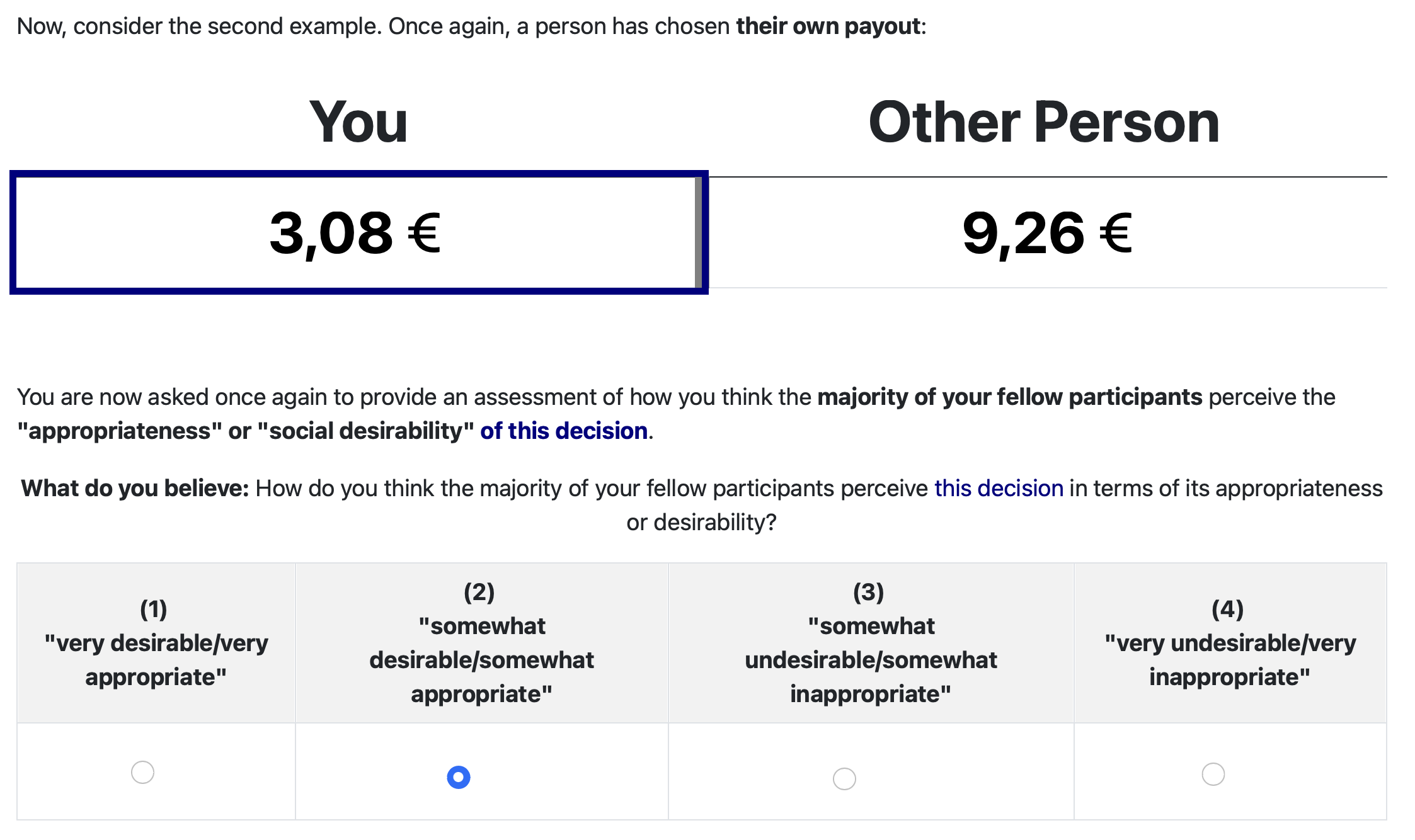}
		\centering\footnotesize
		\textbf{(a)} Baseline
        \end{subfigure}
        \hfil
        \begin{subfigure}[b]{0.475\textwidth}  
            \centering 
   \includegraphics[width=1\linewidth]{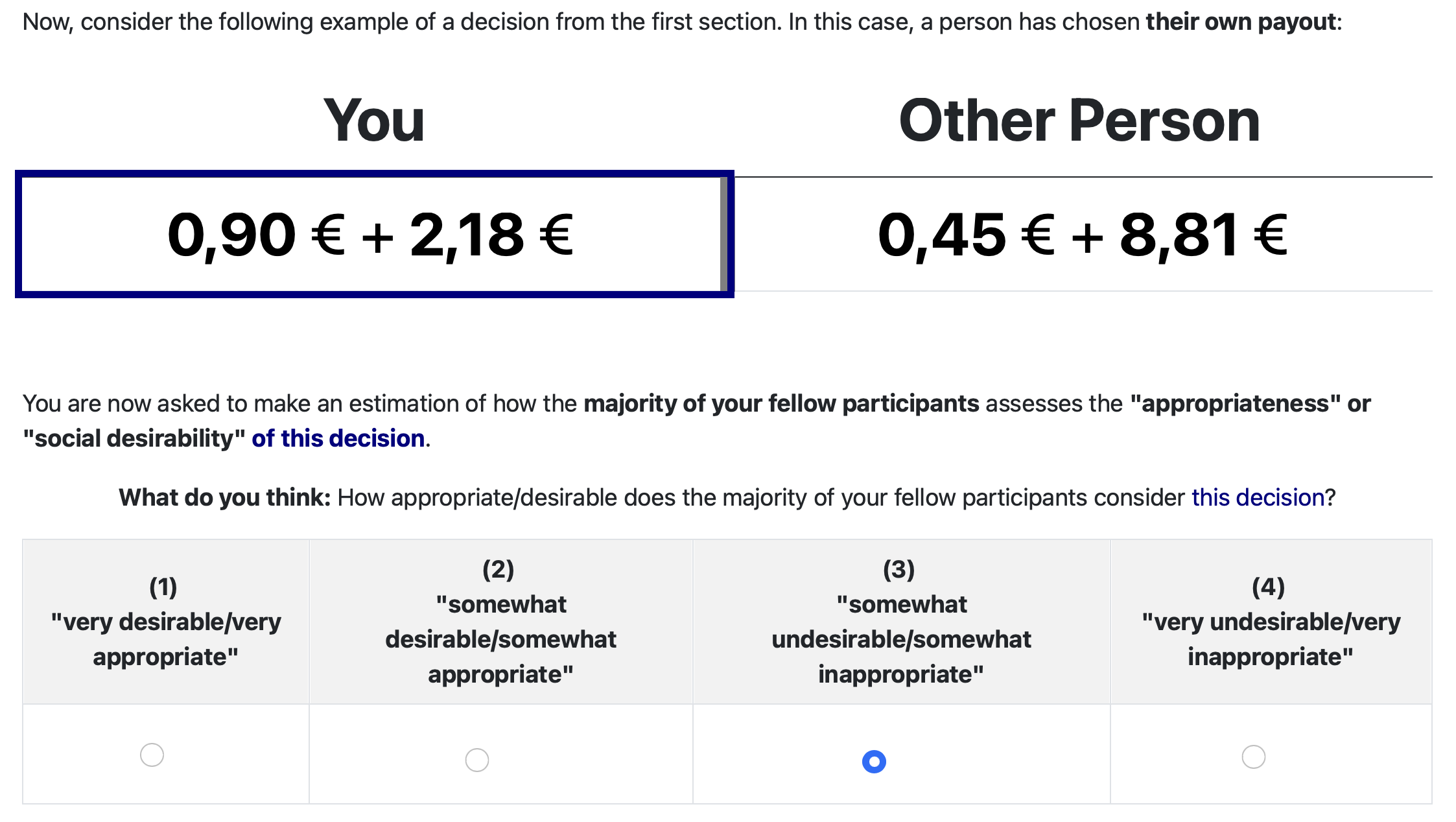}		\centering\footnotesize
		\textbf{(b)} Treatment
        \end{subfigure}
        \vskip\baselineskip
        \caption[Screenshots: Social Norms]
        {\small Screenshots: Social Norms \quad \textbf{(a)} Baseline \textbf{(b)} Treatment. Both variants are shown to all participants, order in which questions appear is randomized}\label{screenshot:social_norms}
    \end{figure}

\begin{figure}[H]
\centering
	\includegraphics[width=1\linewidth]{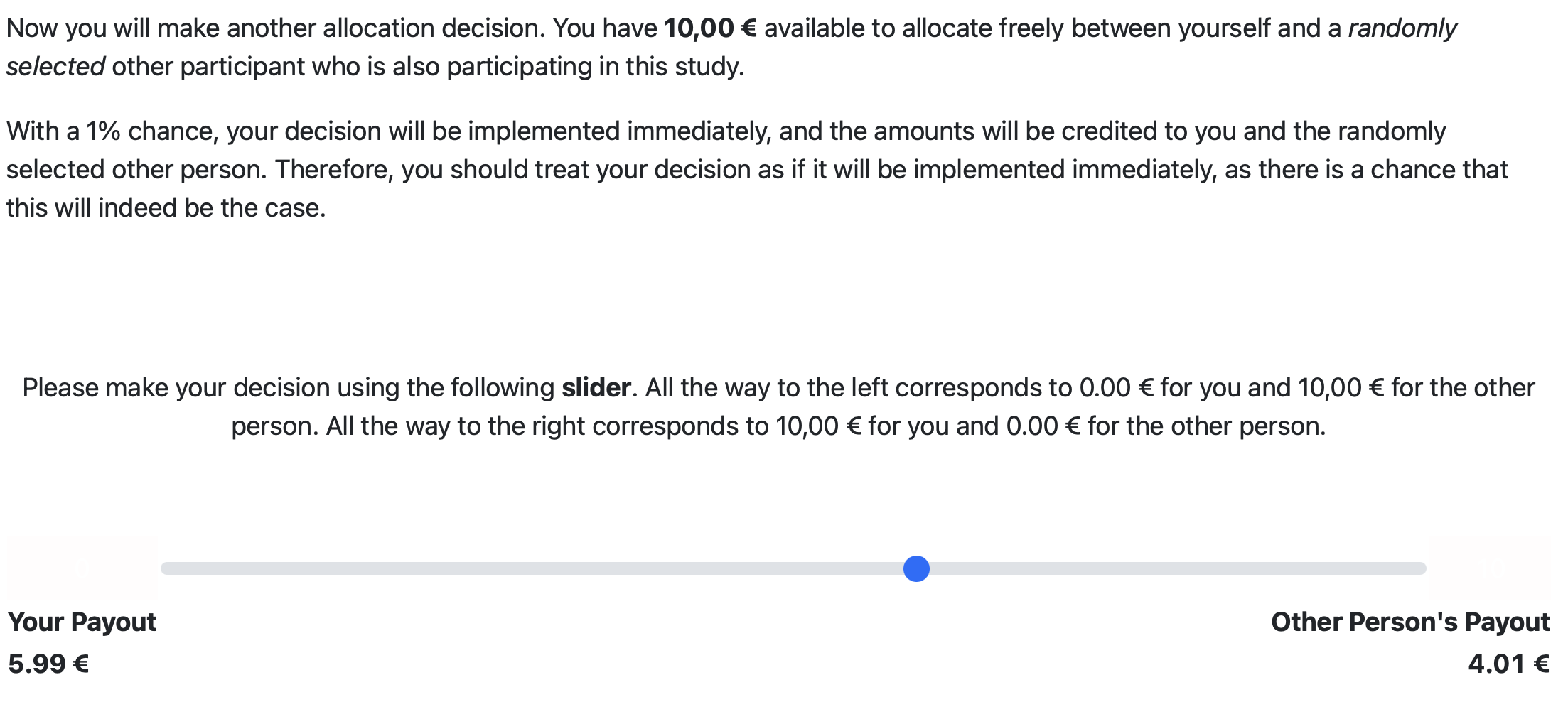}

\caption{Screenshot: Dictator Game}\label{screenshot:simple_dictator}
\end{figure}

\FloatBarrier

\newpage

\phantomsection 

\begin{refcontext} 
\printbibliography[heading=bibintoc]
\end{refcontext}

\end{document}